\documentclass[journal,comsoc]{IEEEtran}
\usepackage[T1]{fontenc}

\usepackage{cite}

\ifCLASSINFOpdf

\else

\fi

\usepackage{booktabs}
\usepackage{footnote}
\usepackage[cmintegrals]{newtxmath}
\usepackage{bbding}
\usepackage{url}
\usepackage[pdftex]{graphicx}
\usepackage{subfigure}
\usepackage{url}
\usepackage{array}
\usepackage{amsmath}
\usepackage{breqn}
\usepackage{mathrsfs}
\usepackage{enumitem}

\RequirePackage[l2tabu, orthodox]{nag}
\usepackage{algpseudocode}
\usepackage{algorithmicx,algorithm}
\usepackage{multirow}
\usepackage{balance}
\usepackage{verbatim}
\usepackage{pifont}

\usepackage{enumerate}
\usepackage{amsmath}
\usepackage{graphicx}

\usepackage[cmintegrals]{newtxmath}
\usepackage{enumitem}

\begin{document}
\title{Internet of Intelligence: A Survey on the Enabling Technologies, Applications, and Challenges}
\author{Qinqin~Tang, F. Richard Yu,~\IEEEmembership{Fellow,~IEEE}, Renchao~Xie,~\IEEEmembership{Senior Member,~IEEE}, Azzedine~Boukerche,~\IEEEmembership{Fellow,~IEEE}, Tao~Huang,~\IEEEmembership{Senior Member,~IEEE}, Yunjie~Liu
	\thanks{This work was supported by the National Natural Science Foundation of China under Grant No. 62171046, the Natural Science Foundation of Beijing under Grant No.4212004 and No. L201002, and the Open Research Fund from Guangdong Laboratory of Artificial Intelligence and Digital Economy (SZ) under Grant No. GML-KF-22-24. (\textit{Corresponding author: Renchao Xie.})}
	\thanks{Qinqin Tang is with Beijing University of Posts and Telecommunications, Beijing, 100876, P.R. China, and also with the Guangdong Laboratory of Artificial Intelligence and Digital Economy (SZ), Shenzhen, 518107, P.R. China  (email: qqtang@bupt.edu.cn).}
	\thanks{F. Richard Yu is with the School of Information Technology, Carleton University, Ottawa, ON K1S 5B6, Canada (email: richard.yu@carleton.ca).}
	\thanks{Renchao~Xie, Tao~Huang and Yunjie~Liu are with the State Key Laboratory of Networking and Switching Technology, Beijing University of Posts and Telecommunications, Beijing, 100876, P.R. China. They are also with the Purple Mountain Laboratories, Nanjing, 211111, P.R. China (e-mail: Renchao\_xie@bupt.edu.cn; htao@bupt.edu.cn; liuyj@chinaunicom.cn).}
	\thanks{Azzedine Boukerche is with the School of Electrical Engineering and Computer Science, University of Ottawa, Ottawa, ON K1N 6N5, Canada (e-mail: boukerch@uottawa.ca).
}
}

\maketitle

\begin{abstract}
The Internet of intelligence is conceived as an emerging networking paradigm, which will make intelligence as easy to obtain as information. This paper provides an overview of the Internet of intelligence, focusing on motivations, architecture, enabling technologies, applications, and existing challenges. This can provide a good foundation for those who are interested to gain insights into the concept of the Internet of intelligence and the key enablers of this emerging networking paradigm.
Specifically, this paper starts by investigating the evolution of networking paradigms and artificial intelligence (AI), based on which we present the motivations of the Internet of intelligence by demonstrating that networking needs intelligence and intelligence needs networking. We then present the layered architecture to characterize the Internet of intelligence systems and discuss the enabling technologies of each layer. Moreover, we discuss the critical applications and their integration with the Internet of intelligence paradigm. Finally, some technical challenges and open issues are summarized to fully exploit the benefits of the Internet of intelligence.

\end{abstract}


\IEEEpeerreviewmaketitle

\section{Introduction \label{sect_1}}

\IEEEPARstart{W}{ith} decades of research and development, the Internet has become an essential information infrastructure in today's world to facilitate economic development and social progress. According to Cisco's white paper, by 2023, the total number of Internet users will reach 5.3 billion, and networked devices will exceed 29.3 billion \cite{Cisco}. Through enabling information networking among people and machines, the Internet can instantly transmit information to people thousands of miles away, realizing global informatization. Despite these benefits, there are still some new challenges in the post-Internet era: i) Information explosion: information overload may occur due to the ever-increasing information in the Internet, which will lead to difficulty in decision-making \cite{9121730}; ii) Fake information: more and more fake information and other lousy information spread rapidly and widely through the Internet, which destroys the harmonious environment of the Internet and harms the economy and society \cite{9233435}; iii) Human-in-the-loop: the explosive trend of the Internet has brought unprecedented challenges of scale, complexity, dynamics and cost to the current ``human-in-the-loop'' network operation and management \cite{8970163}; iv) Limitation: the Internet exhibits limitations in solving existing challenges, including developing reliable, cost-effective autonomous systems (for example, autonomous driving).

To cope with these challenges, networking issues need to be considered in a larger timescale. Networking is for obtaining not just information but also matter and energy. Reviewing the development history of human society, we can find that cooperation is the heart. As a social species, human beings rely on cooperation to achieve survival and prosperity. Consequently, in modern history, to promote human cooperation in the socio-economic system, humans have developed technologies to enable networking for matter (grid of transportation), for energy (grid of energy), and for information (the Internet) \cite{9448012}. In every evolution of the networking paradigm, we can move ``something'' (such as matter, energy, or information) to reduce the disparity of ``something'', thereby facilitating human cooperation through sharing ``something''. In such a large time scale, we can observe the evolution pattern of networking paradigm: the new networking paradigm is based on the existing paradigm, but provides a higher level of abstraction.

\begin{figure}[t]
	\centering
	\includegraphics[width=2.8in]{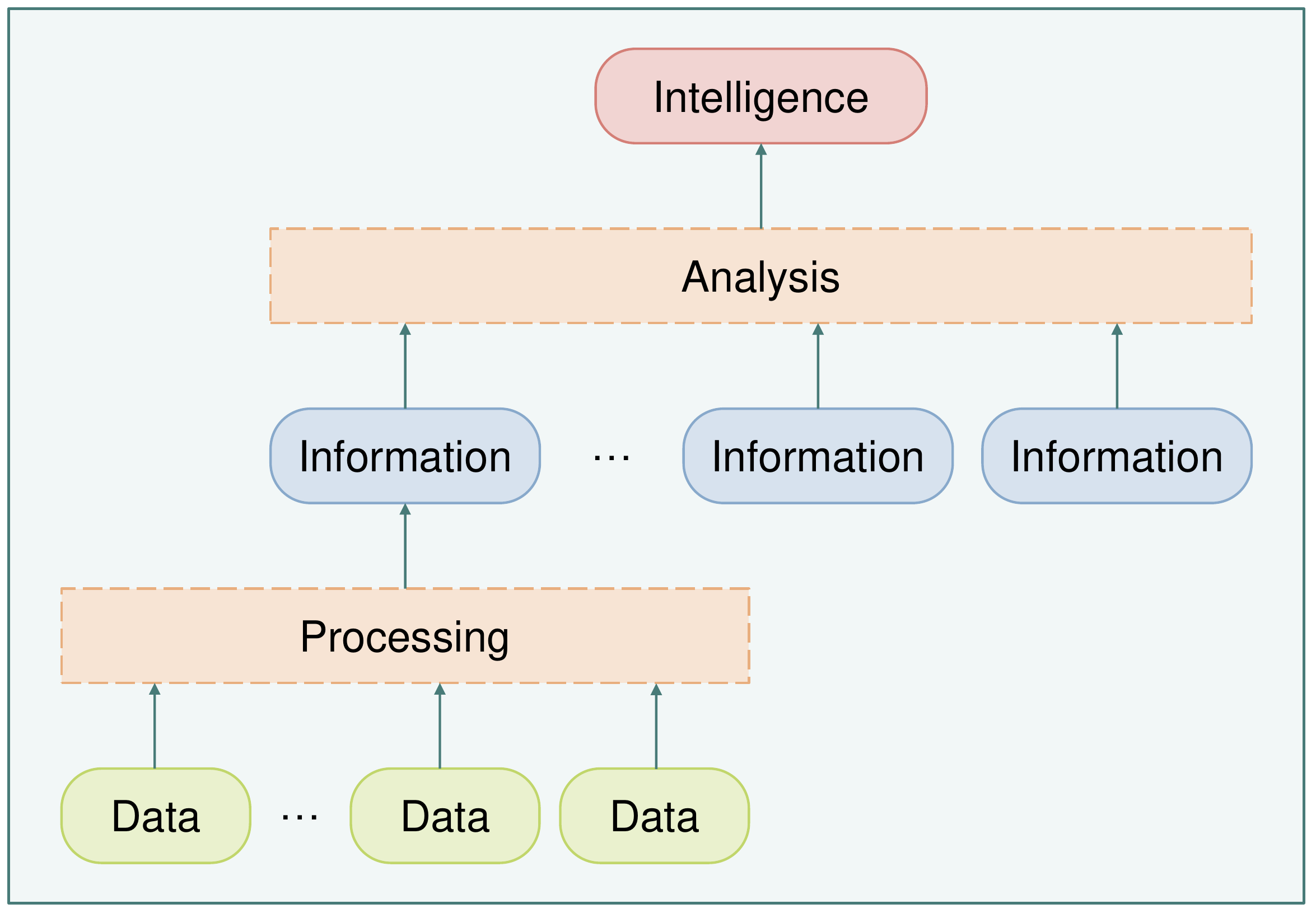}
	\caption{The relationship of data, information, and intelligence.}
	\label{Fig_intelligence}
\end{figure}

The future networking paradigm is inseparable from intelligence to solve the problems existing in the current information networking paradigm. Fig. \ref{Fig_intelligence} shows a conceptual relationship among data, information, and intelligence. Intelligence is the further abstraction and concentration of information \cite{zins2007conceptual}. Although much progress has been made in developing artificial intelligence (AI) \cite{9395691,jackson2019introduction,8734252}, it is far from human learning, which requires much fewer data sets and is more flexible in adapting to new environments. According to the Big History Project \cite{BigHistory}, a decisive feature of human learning is collective learning. Through collective learning, human beings can preserve intelligence, share it with each other, and pass it on to the next generation. This feature enables humans to adapt to the new environment by sharing ideas about dealing with the surrounding environment, thus allowing them to play a leading role in the biosphere. It is difficult to do this in the current AI systems. As a result, the future development of intelligence is also inseparable from networking. Through intelligence networking, distributed intelligence, intelligence storage, and intelligence sharing can be realized, further blurring the boundary between AI and human intelligence, significantly improving training efficiency, and more effectively simulating the real-world environment. Therefore, the future networking needs to shift from an information-based architecture to an intelligence-based architecture \cite{9277903,8685777}. The Internet of intelligence (intelligence networking) is conceived as an emerging networking paradigm, making intelligence easy to obtain, such as matter, energy, and information. 

The Internet of intelligence has attracted great attention from both industry and academia \cite{9430902,9003305,9321476}. To enable researchers to have a more straightforward and deeper understanding of intelligence networking paradigm and to deliver insightful guidelines to fully exploit the profound benefits of the Internet of intelligence, a systematic and comprehensive survey of the Internet of intelligence is provided in this paper. To the best of our knowledge, this paper is the first survey on the Internet of intelligence that covers the motivations, architecture, enabling technologies, applications, and challenges. The key contributions of this paper can be summarized as follows:

\begin{itemize} 
	
	\item We investigate the networking paradigm evolution and the artificial intelligence evolution, based on which we present the motivations of the Internet of intelligence by demonstrating that networking needs intelligence and intelligence needs networking.
	
	\item We propose the layered architecture of the Internet of intelligence, and recent advances of the enabling technologies in each layer are discussed to enable researchers to quickly get up to speed with the key enablers of the Internet of intelligence.
	
	\item We provide an overview of the critical applications and their integration with the Internet of intelligence paradigm.
	
	\item Some technical challenges and open research issues that need to be discussed for future research are summarized to fully exploit the benefits of the Internet of intelligence.
	
\end{itemize}

\begin{figure*}[t]
	\centering
	\includegraphics[width=5.5in]{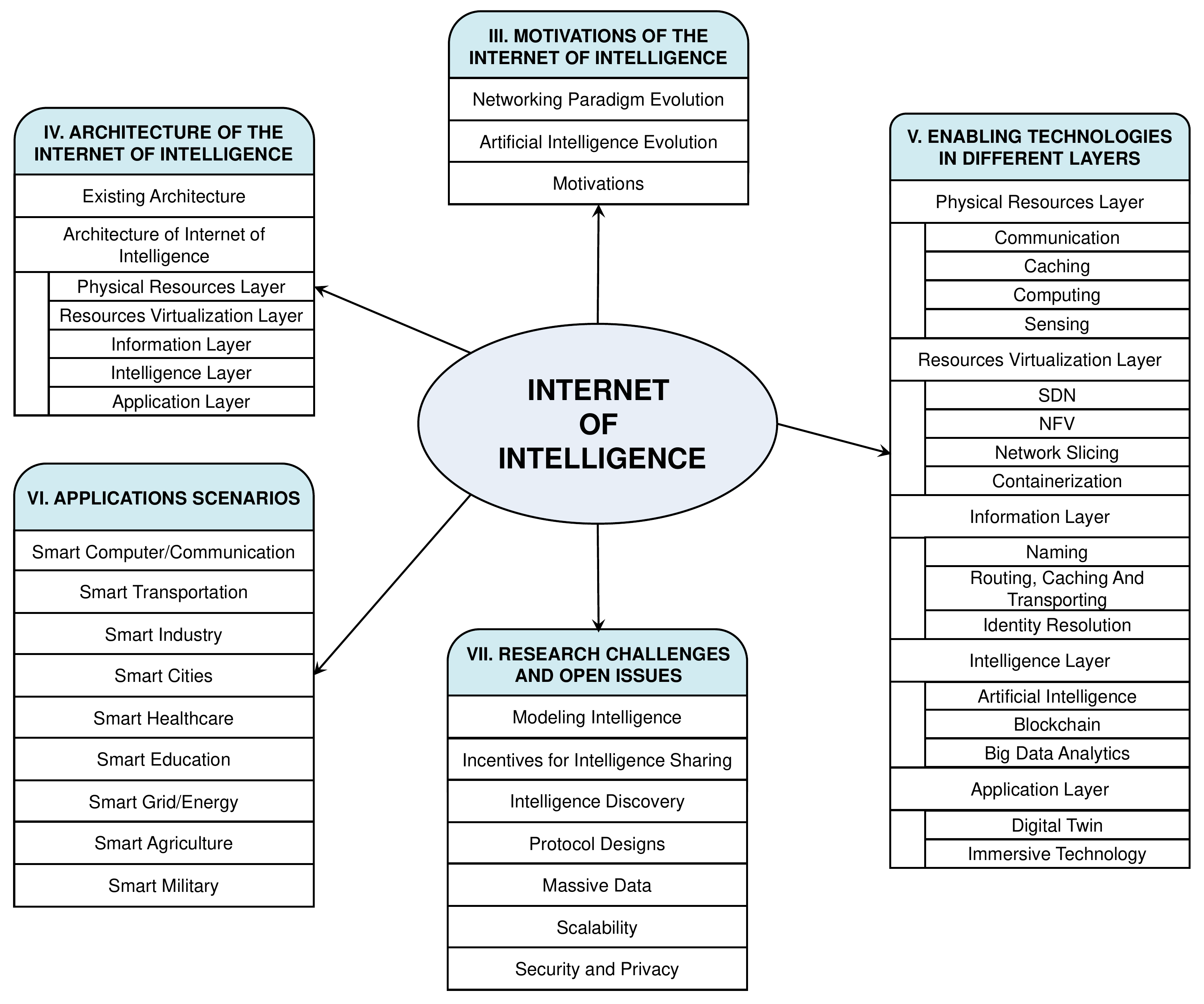}
	\caption{Road map of this article.}
	\label{Fig_roadmap}
\end{figure*}

The paper organization is as follows. Section \ref{sect_2} provides a review of the related work. In Section \ref{sect_3}, we briefly introduce the evolution of networking paradigms and artificial intelligence, and then present the motivations of the Internet of intelligence. Section \ref{sect_4} provides the overall architecture of the Internet of intelligence, which is composed of five different layers: physical resources layer, resources virtualization layer, information layer, intelligence layer, and application layer. Section \ref{sect_5} focuses on the enabling technologies in each of the five layers mentioned above in the presented architecture. In Section \ref{sect_6}, we present the critical applications of the Internet of intelligence, and Section \ref{sect_7} summarizes some technical challenges and open research issues. Finally, the paper is concluded in Section \ref{sect_8}. To facilitate reading, the roadmap of this paper is visualized as Fig. \ref{Fig_roadmap}.

\section{Related Work \label{sect_2}} 
In this section, we summarize some recent surveys related to our work to help readers explore related fields.

AI has gained unprecedented attention from academia and industry. Due to its universality, researchers have started to apply related methods in various fields, such as computer/communication networks, smart cities, transportation, healthcare, industry, grid/energy, etc.
Sun \emph{et al.} in \cite{9146540} discussed the protection of privacy in sixth-generation (6G) by secure machine learning (ML) structures on the one hand, and analyzed the potential for privacy violations in 6G due to attacks or misuse of ML on the other hand. 
Feriani \emph{et al.} in \cite{9372298} outlined the application of model-free, model-based single-agent reinforcement learning, and cooperative multi-agent reinforcement learning frameworks and algorithms for wireless communications.
In \cite{9279299}, Magaia \emph{et al.} investigated how deep learning techniques can be used to enhance the efficiency of the industrial Internet of things (IIoT) and achieve more privacy and security in smart cities.
In \cite{7000557}, Rigas \emph{et al.} presented a systematic survey to study AI technologies for energy-efficient electric vehicle routing and charging point selection, as well as for integrating electric vehicles into the smart grid.
In\cite{9266587}, Wan \emph{et al.} provided a survey to explore the impact of AI on the field of custom smart manufacturing and discussed the challenges associated with smart manufacturing devices due to the introduction of AI.
In \cite{8926369}, Tang \emph{et al.} studied the unique challenges of existing vehicular communication, networking and security, and outlined the corresponding ML-based solutions.
Qadri \emph{et al.} in \cite{8993839} surveyed the application of ML in the healthcare Internet of things (H-IoT) to provide personalized healthcare.

To further enhance the performance of AI, many advanced technologies, such as software-defined networking (SDN), network slicing, edge computing, blockchain, various immersive technologies, etc., can be used to enable AI.
Xie \emph{et al.} in \cite{8444669} investigated how SDN can make the application of ML easier by taking advantage of its centralized control of logic, global view of the network, software-based traffic analysis, and dynamic updating of forwarding rules.
In \cite{9586568}, Li \emph{et al.} examined how network slicing can be used to jointly manage data and general network resources for AI services.
Considering the higher data upload latency and cost as well as the reliability and privacy issues due to AI training in the cloud, Wang \emph{et al.} in \cite{8976180} investigated the realization of edge intelligence through the adoption of edge computing and surveyed the application, inference, training and optimization methods for edge intelligence.
Liu \emph{et al.} in \cite{9007406} analyzed how blockchain can benefit ML, including data and model sharing, security and privacy, and trusted decision-making.
Alsamhi \emph{et al.} in \cite{9635590} focused on the synergy of federated learning (FL) and blockchain for green sustainable edge intelligence.
The survey in \cite{9566732} proposed to use blockchain to enhance the security of decentralized FL in terms of privacy protection and tamper-proof data.
In \cite{9690157}, Kim \emph{et al.} surveyed the application of augmented reality (AR) technology in AI education for non-engineering majors.

However, there are no comprehensive surveys that study AI from the perspective of networking. While there are some works on distributed learning \cite{8976180,9007406,9635590,9566732}, most of them focus on distributed architecture design or optimal algorithm design rather than intelligence networking in a comprehensive way.
By networking the intelligence, distributed intelligence, intelligence storage and intelligence sharing can be realized, thus further blurring the boundary between AI and human intelligence, greatly improving the efficiency of training and simulating the real environment more effectively.
This motivates us to present a systematic survey on the Internet of intelligence, so we review it from multiple perspectives, including motivations, architecture, enabling technologies, applications, and challenges to provide insightful guidelines and up-to-date viewpoints for researchers.

\section{Motivations\label{sect_3}}

In this section, we look back on the networking paradigm evolution and the artificial intelligence evolution.
On this basis, we present the motivations of the Internet of intelligence.

\subsection{Networking Paradigm Evolution}
\begin{figure}[t]
	\centering
	\includegraphics[width=3.0in]{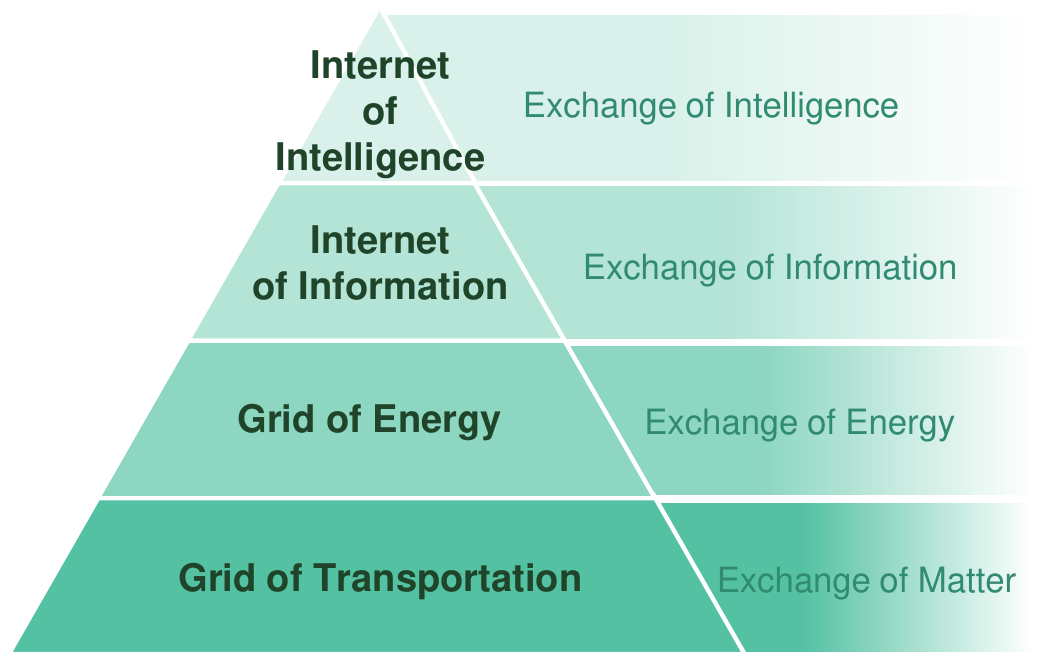}
	\caption{The evolution of networking paradigms.}
	\label{Fig_networking}
\end{figure}

To explore the motivations of the Internet of intelligence, we need to take a look at the development of human society on a larger time scale. It can be observed that cooperation is at the core of human society. As a social species, human beings rely on cooperation to achieve survival and prosperity \cite{harari2014sapiens}. 
It is believed that the reason why humans can dominate the world is because we are the only animals that can cooperate flexibly in large numbers. 
In modern history, to facilitate cooperation in socio-economic systems, humans invented technologies enabling networking for matter (grid of transportation), for energy (grid of energy), and for information (the Internet), as shown in Fig. \ref{Fig_networking}. 
In the following, we will briefly review these three networking technologies.

\subsubsection{Grid of Transportation}
Transportation technologies are essential to facilitate human cooperation. The development and survival of human beings and other organisms are inseparable from matter and energy.
The primary motivation for transportation is to move matter from one location to another, which is matter networking. 
In the initial stage of the grid of transportation, humans transport on foot. Later, they used animals and build transport ships. In the 18th and 19th centuries, plenty of innovative transportation technologies were invented, such as bicycles, cars, trains and trams.
Nowadays, airplanes, high-speed rails, maglev trains, spacecraft, etc., are all powerful transportation technologies.

\subsubsection{Grid of Energy}

The grid of energy is another disruptive technology, which is the foundation of human survival and the cornerstone of human prosperity. Through the grid of energy, electric energy is able to be transmitted from power plants to families and enterprises so that people can easily obtain energy to power appliances, light buildings, and cool houses \cite{kudo2009forecasting}.

Energy is a measure of the system's ability to cause changes in the energy grid. According to the first law of thermodynamics, energy is conserved and can neither be created nor disappear. It can be transferred from one location to another or from one form to another. During the transfer process, the sum of energy does not change. There is a close relationship between energy and matter.
For instance, kinetic energy, which describes the energy of an object due to its mechanical motion, is an important category of energy.
The theorem of kinetic energy can be expressed as: $0.5mv^2 = 0.5m{(d/t)}^2$, where $m$ is the mass, $v$ is speed, $d$ is distance, and $t$ is time.
Accordingly, kinetic energy can be considered as the speed of matter moving in a process.

\subsubsection{The Internet of Information}

The invention of the Internet (of information) aims to move information from one location to another to reduce the disparity in information.
Through the realization of information networking, human cooperation has been promoted to a new level.
The connection between information and energy was first mentioned in the ``demon'' experiment led by Maxwell in 1867, which reduced the system's entropy by converting information (i.e., the position and velocity of each particle) into energy. 
This experiment inspired discussion about the relationship between thermodynamics and information theory \cite{cottet2017observing}.
In 1948, Shannon proposed the concept of ``information entropy'', which solved the quantitative measurement of information and obtained the entropy formula in the same form as in thermodynamics \cite{bratianu2020thermodynamic}. Thermodynamic entropy is a state parameter describing the spontaneous diffusion process of energy. It can be expressed by $\mathrm{d}S=\partial Q/T$, where $\mathrm{d}S$ is the change in entropy, $\partial Q$ is the transferred energy, and $T$ is the temperature. 
Similarly, information entropy measures the state of information, which means the minimum energy required for information to move from one state to another.

\subsection{Artificial Intelligence Evolution}

\begin{figure*}[t]
	\centering
	\includegraphics[width=5.2in]{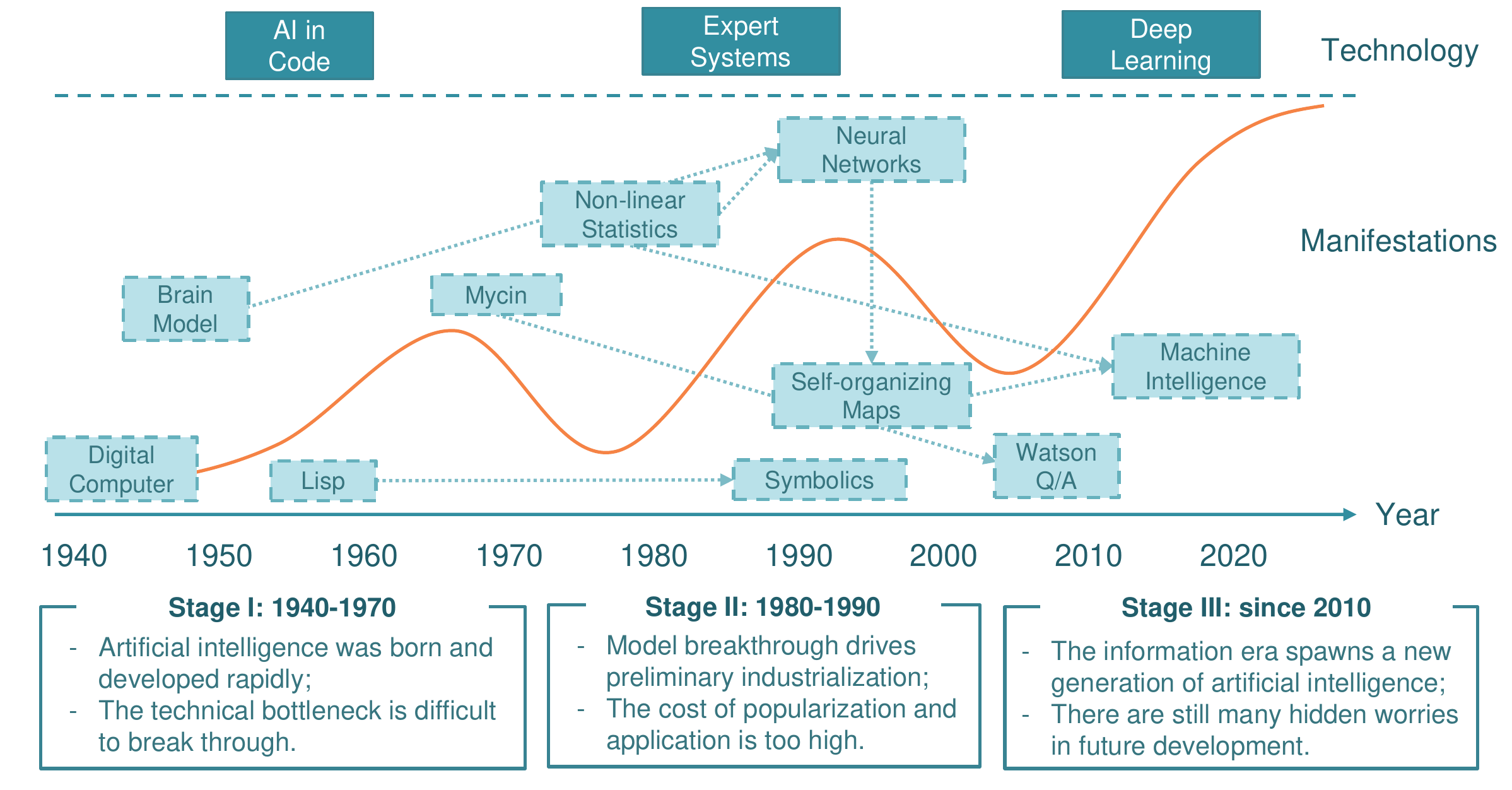}
	\caption{The evolution of artificial intelligence.}
	\label{Fig_AIevolution}
\end{figure*}

As a branch of computer science, AI has undergone more than 70 years of development since its birth. Fig. \ref{Fig_AIevolution} illustrates a schematic diagram of the development process of AI. In the past decades, it has made a lot of progress and achievements but also experienced setbacks and frustrations. Even if the current AI is closer to the level of human intelligence than at any period in history, this automation is still far from human intelligence in the strict sense. In the following, we describe several stages of AI evolution.

\subsubsection{1940-1970}
At this stage, research work focuses on traditional AI problems, such as methodological inference based mainly on logic and heuristic algorithms \cite{zhuang2017challenges}. Specifically, in 1943, W. McCulloch and W. Pitts exploited mathematical and computer models of biological neurons. N. Wienner pioneered cybernetics in 1948 to study the control and communication in animals and machines. At the beginning of 1950, the architecture of modern computers was developed by J. Von Neumann and A. Turing. Moreover, Turing, in the same year, predicted the development of thinking machines. In 1956, J. McCarthy of Stanford University, M. Minsky of Massachusetts Institute of Technology, and other scholars first created the concept of ``artificial intelligence'' at Dartmouth College in the United States. AI is defined as the ability of machines to think and learn in a manner similar to human beings.
The following two decades are the golden age of AI. A successful case was the Eliza computer program developed by J. Weizenbaum in 1966, a natural language processing tool that can simulate a conversation with human beings.
Another example was the General Problem Solver program developed by H. Simon \emph{et al.}, which can automatically solve certain types of problems, such as the Tower of Hanoi.
Thanks to these inspiring success stories, AI has gained unprecedented attention. 
However, since the large amount of capital and workforce investment did not receive the expected results, AI ushered in its first winter from 1974 to 1980.
In particular, due to Minsky's strong criticism of perceptrons, the connectionism (or neural networks) faction has been depressed for almost a decade.

\subsubsection{1980-1990}
With the emergence of the first batch of microprocessors in the late 1970s, AI once again attracted attention and was closely related to expert systems \cite{jaakkola2019artificial}. Expert systems refer to systems built for specific purposes in which intelligence is embedded in tools and knowledge about its operations is publicly available in the system specifications. 
Expert systems are widely applied in industrial fields. A well-known case is that IBM's Deep Blue chess program defeated world champion G. Kasparov in 1997.
Expert systems are most suitable for static problems but not for real-time dynamic issues. 
An expert system defines intelligence in a narrow sense as abstract reasoning, which is far from the ability to simulate the world's complexity. 
Moreover, its development and maintenance are complicated.
In the 1990s, with the failure of the development of Japanese intelligent (fifth-generation) computers and the decline of Cyc, an encyclopedia of human knowledge led by Stanford University, AI entered the cold winter again.

\subsubsection{Since 2010}

Since 2010, AI has taken off again due to major public successes.
For example, in 2011, IBM's Watson defeated two champions in Jeopardy and won the competition.
In 2016, Google's AlphaGo vanquished world champion Lee Sedol, and her predecessor AlphaGo Zero.
In 2020, AlphaFold successfully predicted the shape of proteins so that they can perform life's tasks.
Different from the expert system paradigm, in today's wave of AI, the critical element is the learning ability of the system. 
Two of these technologies play an essential role, i.e., neural networks and deep learning \cite{7932863}. The purpose of the neural network is to build a model similar to the structure of the human brain structure and calculation. Neural networks use rules based on ``what-if'' and teach the network through examples (supervised learning) to learn the nonlinear dependencies between variables. Deep learning originated from the work of G. Hinton in the 1980s. Its learning algorithms are based on nonlinear statistics, and the learning data is organized in multilayer neural networks. 
The neural network is the cornerstone of deep learning algorithms, and the ``deep'' in deep learning refers to the depth of the layers in a neural network.
Today, neural networks and deep learning form the basis of most AI applications, such as image recognition algorithms used by Facebook, speech recognition algorithms for autonomous driving, etc. However, there is still plenty of challenges to be solved for deep learning \cite{jordan2015machine}. For example, deep learning systems require massive amounts of data to obtain powerful learning performance. However, in practice, due to privacy and resource constraints, the system may not be able to obtain high-quality training data. Besides, integration with prior knowledge, interpretability and explainability are also critical issues in deep learning.

\subsection{Motivations of the Internet of Intelligence}

\subsubsection{Networking Needs Intelligence}
As the most crucial information infrastructure, the Internet has experienced leaps and bounds in the past few decades. It becomes jam-packed with billions of websites, active users, and connected devices \cite{9199785}. This explosive trend of the Internet has brought challenges such as information overload, fake information, ``human-in-the-loop'', and the design of trustworthy, cost-effective autonomous systems. Therefore, networking needs intelligence. Intelligence is the further abstraction and concentration of information. Intelligence technologies such as AI are usually used to extract intelligence in the network environment from various information sources.
By injecting intelligence into the network, useful information can be effectively extracted from the massive amount of Internet information. The intelligence can be stored, transmitted and shared, thereby alleviating information redundancy in the network \cite{yu2019internet}. The adoption of blockchain technology can also effectively enhance network security and privacy issues, thereby facilitating intelligence sharing, decentralized intelligence, collective learning, and decision-making trust \cite{9277903,9321476}. ``Human-on-the-loop'' network operation is expected to enhance the robustness of the network and achieve rapid response to network events and dynamics \cite{8970163}. In addition, the network will have the ability of self-configuration, self-organization, and self-adaptation, thus effectively supporting the development of reliable and cost-effective autonomous systems such as autonomous driving \cite{9003305,9448012}.
Intelligence is not only necessary for technologies but also essential for changing daily life. Disruption begins with the adaptive nature of technologies that ensure the quality of social and economic services and will directly affect human lives. The big data generated by the Internet can be used to predict and formulate solutions to intelligently handle real-time situations.

\subsubsection{Intelligence Needs Networking}

Most existing AI works focus their works on single-agent training, which relies heavily on a great quantity of predefined data sets with the local environment. With the rapid growth of data in the Internet, this centralized AI architecture is constrained by local computing capability and storage capacity, and the generalization ability of the trained model needs to be improved. In order to address these problems, AI algorithms need to explore high-quality data sources to better train models. In addition, many systems in actual scenarios are either too complex to be modeled correctly in a fixed, predefined environment or varied dynamically \cite{dulac2019challenges}.
The concept of AI is defined as the science that enables computers to perform tasks that require intelligence like humans.
However, at present, it is far from human learning, which requires much fewer data sets and is more flexible in adapting to the new environment.
According to the Big History Project, a decisive feature of human learning is collective learning. 
Through collective learning, intelligence can be preserved, shared and passed on by humans \cite{harari2014sapiens}.
That is, collective learning is the ability to efficiently share intelligence so that individual ideas can be stored in the community's collective memory and passed on from generation to generation. 
This feature of collective learning allows humans to adapt to the new environment by sharing ideas about how to deal with the surrounding environment, thus allowing them to play a leading role in the biosphere. 
It is difficult for the current AI systems to do this.
Therefore, the future development of intelligence is also inseparable from networking. Through networking, distributed intelligence, intelligence storage and intelligence sharing can be realized, further blurring the boundary between AI and human intelligence, significantly improving training efficiency, and more effectively simulating the real-world environment.

\subsubsection{The Internet of Intelligence}

From the evolution of the networking paradigm, it can be observed that the three major networking paradigms mentioned enabled us to move ``something'' (such as matter, energy, or information) to reduce the disparity of ``something'', thereby facilitating human cooperation through sharing ``something''. Moreover, the evolution pattern of networking paradigms can also be concluded: the new networking paradigm builds on the existing ones but provides a higher level of abstraction. For instance, energy measures the speed at which matter moves, while information measures the degree of energy transmission. 

The Internet of intelligence (intelligence networking) is conceived as an emerging networking paradigm, making intelligence easy to obtain, such as matter, energy, and information. Intelligence is not equivalent to information, but a higher level of abstraction and concentration of information. It describes the general patterns and relationships obtained by analyzing information sets. Intelligence can be algorithms that aggregate information into intelligent fragments, i.e., intelligence models, or abstract information obtained through application models, i.e., intelligence samples \cite{zins2007conceptual}. 
For example, a trained ML model is a kind of intelligence that can return synthetic intelligence from input information \cite{9209668}. The intelligence extracted by advanced technologies can be further utilized by the intelligence combination method, in which existing intelligence is further analyzed/organized to generate new intelligence.
The Internet of intelligence is expected to effectively help solve the challenges in the current socio-economic system and significantly affect human daily life, just like the previous three networking paradigms.

\subsection{Lessons Learned: Summary and Insights}
In this section, we illustrate the motivation for the Internet of intelligence in terms of networking need intelligence and intelligence need networking, respectively.
The explosive growth of the Internet has led to challenges such as information overload, fake information, human-in-the-loop, and limitations.
To meet these challenges, networking needs intelligence.
The introduction of AI in the network can effectively extract useful information from the massive Internet information, thereby alleviating the redundancy of information in the network and making the network self-configuring, self-organizing and self-adaptive.
The adoption of blockchain effectively enhances the security, privacy, reliability and robustness of the network.
Despite the remarkable progress AI has made over the decades, it still falls far short of human learning.
The reason for this phenomenon is the lack of cooperation in current AI systems, which is a defining feature of human learning.
Therefore, the future development of intelligence is also inseparable from networking. Through networking, distributed intelligence, intelligence storage and intelligence sharing can be realized, thus enabling AI to further approach human intelligence.

A review of the past three networking paradigms reveals that the new networking paradigm builds on the existing networking paradigms and allows us to share ``something'' to facilitate cooperation.
Intelligence provides a higher level abstraction of information.
As a result, it is envisioned that the next networking paradigm may be the Internet of intelligence, which will make intelligence easily accessible and effectively address the challenges in current socio-economic systems.

\section{Architecture of Internet of Intelligence\label{sect_4}}

In this section, we first introduce the existing layered architecture of the Internet and describe it in detail. Afterwards, we discuss the architecture of the Internet of intelligence and provide a brief summary and insights at the end of this section.

\subsection{Existing Architecture}

The general principle of layering is widely considered to be one of the key factors in the great success of the Internet \cite{chiang2007layering}. In a layered structure, each layer controls a set of decision variables to achieve a specific function by observing parameters from itself and other layers. Each layer serves the layers above and hides the complexity of the layers below. Layering has the following advantages \cite{yu2019internet}:

\begin{itemize} 
	
    \item Simplicity: Complex tasks are decomposed into discrete subtasks and localized in each layer to simplify the design.
    
	\item Modularity: Each layer is a module that is designed, developed, optimized, managed and maintained independently.
	
	\item Abstract functionality: Each layer can be adjusted without affecting other layers.
	
	\item Reuse: The functionality of each layer can be reused.
	
\end{itemize}

The Internet connects countless computer networks. The current five-layer architecture of computer networks includes physical layer, data link layer, network layer, transport layer, and application layer \cite{9094202,zimmermann1980osi,braden1989requirements}.

\begin{itemize} 
	
	\item \textit{Application layer} is designed to complete specific network applications through the interaction between application processes. The application layer protocol defines rules for communication and interaction of messages between application processes.
	There are many application layer protocols in the Internet, such as domain name system (DNS), hyper text transfer protocol (HTTP), simple mail transfer protocol (SMTP), and so forth. 
	
	\item \textit{Transport layer} is responsible for end-to-end data transmission and data flow. This layer consists of two main protocols: 1) Transmission control protocol (TCP) provides connection-oriented and reliable data transmission services, and 2) User datagram protocol (UDP) provides connectionless and best-effort data transmission services.
	
	\item \textit{Network layer} aims to select appropriate inter-network routing and switching nodes to ensure timely delivery of data. When sending data, the network layer encapsulates TCP segments or UDP datagrams generated by the transport layer into packets for transmission. The network layer protocol used by the Internet is the connectionless Internet protocol (IP) and many routing protocols.
	
	\item \textit{Data link layer} assembles the IP packets passed down from the network layer into frames and transmits them over the link. Each frame includes data and necessary control information (such as synchronization information, address information, error control, etc.).
	
	\item \textit{Physical layer} aims to achieve transparent transmission of bit streams between adjacent computer nodes, shielding the differences between specific transmission media and physical devices as much as possible.
	
\end{itemize}

\begin{figure*}[t]
	\centering
	\includegraphics[width=6.3in]{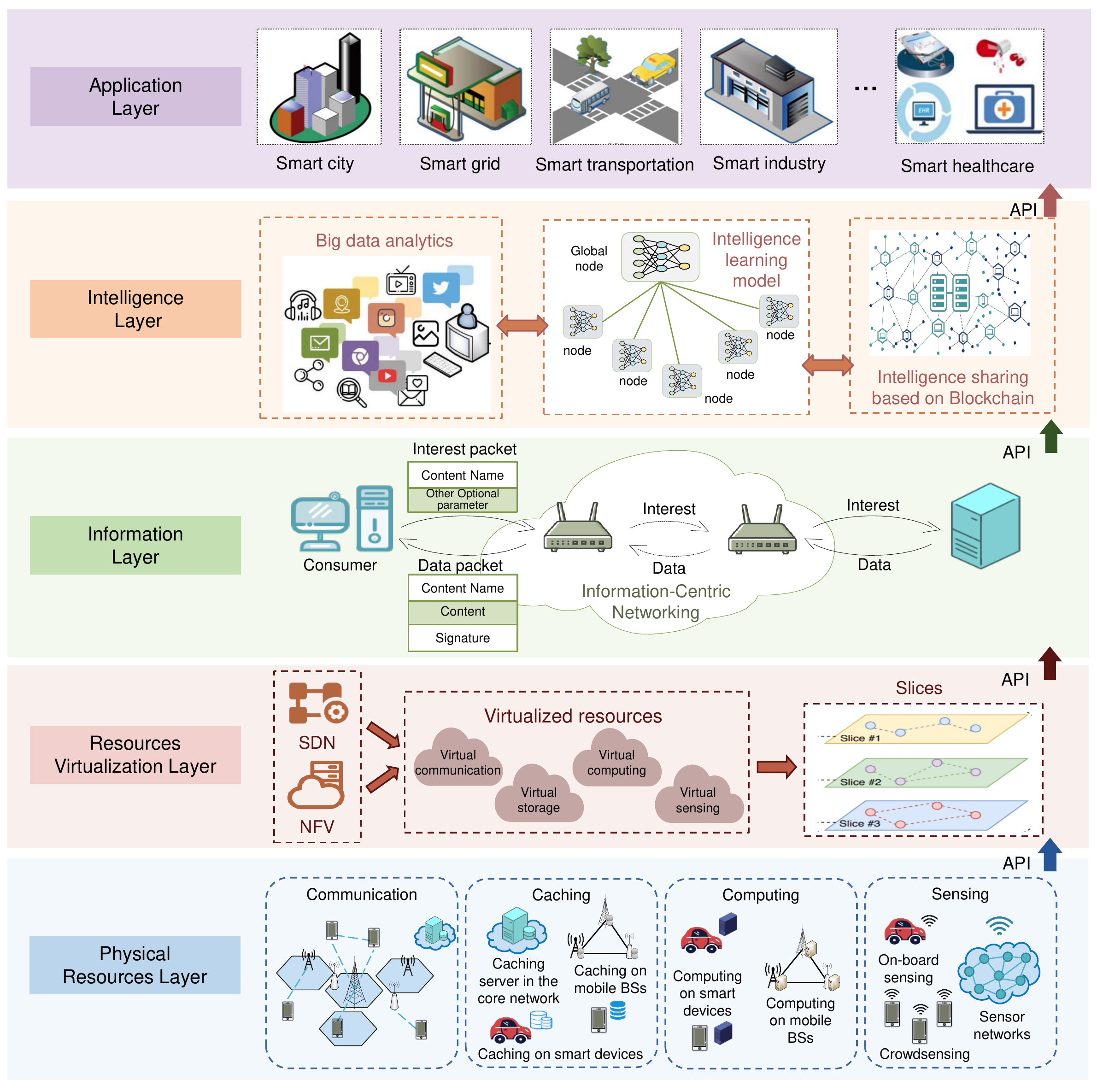}
	\caption{System architecture of the Internet of intelligence.}
	\label{Fig_architecture}
\end{figure*}

\subsection{Architecture of Internet of Intelligence}

Consistent with the design concept of the existing architecture, we also adopt a layered structure to design the Internet of intelligence architecture. 
As illustrated in Fig. \ref{Fig_architecture}, the architecture consists of five different layers: physical resources layer, resources virtualization layer, information layer, intelligence layer, and application layer. 
The physical resources layer includes various underlying physical resources of the Internet of intelligence, such as communication resources, caching resources, computing resources and sensing resources. 
With the latest advances in technologies such as network virtualization, SDN, and so forth, network cloudification is an important development trend for future networks, which also provides multiple benefits for the implementation of the Internet of intelligence.
Driven by this trend, the data transmission of the underlying infrastructure is no longer the focus but the utilization of resources \cite{9293089}.
Therefore, above the physical resources layer is the resources virtualization layer, which abstracts physical infrastructure resources into logical resources for flexible scheduling through various virtualization technologies.
The information layer corresponds to the network layer in the existing architecture, which supports intelligent data processing to extract useful information and transfers it to the intelligence layer. The intelligence layer then merges the information and develops comprehensive intelligence using technologies such as AI, blockchain and big data analytics. The application layer implements dynamic deployment and management of applications through various standardized interfaces. The details of the architecture are summarized as follows.

\subsubsection{Physical Resources Layer}

The physical resources layer is mainly composed of various infrastructure resources of the Internet of intelligence, including communication resources, caching resources, computing resources, and sensing resources. For instance, communication resources include various communication and network resources from radio access networks (RANs), transmission networks, core networks, and so forth. The caching resources and computing resources are composed of various types of entities that support content storage and data computing, such as smart devices, edge computing servers, base stations (BSs), and data centers. 
Sensing resources consist of various sensing devices that sense the surrounding environment and collect data in real time.
These resources from distributed infrastructure are usually ubiquitous and heterogeneous. Therefore, it is necessary to abstract the physical infrastructure resources into logical resources through various virtualization technologies at the resources virtualization layer to form a shared resources pool to provide support for different applications of the Internet of intelligence \cite{9474932}.

\subsubsection{Resources Virtualization Layer}
In this layer, heterogeneous and ubiquitous communication, caching, computing and sensing resources are abstracted and pooled. Resources virtualization is the abstraction of resources using appropriate technologies. Resource abstraction is to represent resources based on attributes that match pre-defined selection criteria while hiding or ignoring aspects that are not related to such criteria, trying to simplify the use and management of the resources in a helpful way. The resources to be virtualized can be physical or virtualized, which support recursive modes with different abstraction layers. 
Through the introduction of the pooling hypervisor, various physical resources can be perceived from the physical resource layer, and the scattered resources can be aggregated into communication pools, caching pools, computing pools and sensing pools \cite{9277903}. 
Some advanced technologies such as SDN, NFV and containerization can be effectively adopted to realize resource virtualization and provide on-demand virtual resources for various applications. Virtualized resources can customize services on demand through slicing technology to achieve effective resource sharing.

\subsubsection{Information Layer}

The primary responsibilities of the information layer are to process and analyze the massive raw data generated from various devices in the Internet of intelligence, and infer useful information from it to pass to the intelligence layer.
In recent years, this layer has attracted a lot of attention since it facilitates the development and integration of services.
A large amount of duplicate and redundant content and inefficient transmission in the information layer pose challenges to the current TCP/IP-based network architecture. 
Therefore, information-centric networking (ICN) is emerging as an alternative network architecture for this layer. 
It is characterized by name-based data retrieval and in-network data caching, allowing users to retrieve data from nearby copyholders. 
ICN separates the content from the location and provides services such as storage and multi-party communication through the publish/subscribe paradigm, so that people's attention has shifted to the rapid acquisition of information regardless of the location of information storage \cite{8410516,6563278,9277899}. By converting raw data into useful information, this layer introduces self-awareness to the physical components of the Internet of intelligence systems.

\subsubsection{Intelligence Layer}
Intelligence is an essential characteristic of the future networking paradigm. The Internet of intelligence is driven by intelligence to achieve self-configuration, self-optimization, self-organization and self-repair, ultimately improving feasibility. Therefore, the intelligence layer is expected to convert the information from the information layer into intelligence and provide intelligent and adaptive management and control on the Internet of intelligence through intelligent decision-making. Using technologies such as big data analytics and AI, functional modules such as intelligence discovery, intelligence sharing, intelligence storage, intelligence delivery, and intelligent service application programming interfaces (APIs) are implemented at the service level. At the data level, the intelligence layer performs further intelligence processing on the valuable information from the information layer to realize networked intelligence computing. At the control level, it dynamically configures various resources according to the real-time situation of the network and provides customized services for upper-layer applications. As an intermediate layer, based on technologies such as blockchain, the intelligence layer can simultaneously provide authentication, authorization, and accounting (AAA) services for users who subscribe to smart services to ensure their privacy.

\subsubsection{Application Layer}
The application layer aims to provide application-specific services to users according to their diverse requirements and to evaluate the provided services before feeding back the evaluation results to the intelligence process. Intelligent programming and management can be realized to support higher-level smart applications such as smart cities, smart industries, smart transportations, smart grids and smart healthcare, and achieve overall management of them. This layer also manages all activities of smart devices and infrastructures in the Internet of intelligence through intelligent technologies to achieve network self-organization capabilities. In addition, this layer is responsible for evaluating service performance, which may involve many considerations and factors, such as the quality of service (QoS), the quality of experience (QoE), the quality of collected data, and the quality of acquiring intelligence. The cost dimension measures of resource efficiency, such as computing efficiency, energy efficiency and storage efficiency, also need to be taken into consideration to improve intelligent resource management and smart service provisioning.

\subsection{Lessons Learned: Summary and Insights}
The contribution of this section is to present the general architecture of the Internet of intelligence. Layered architectures offer the advantages of simplification, modularity, abstract functionality, and reuse.
Driven by these advantages, the Internet is implemented based on the 5-layer architecture of TCP/IP, including the physical layer, data link layer, network layer, transport layer, and application layer. Due to the consistency with the design concept of the existing architecture, the architecture of the Internet of intelligence is also designed based on layering, which consists of five layers: physical resources layer, resources virtualization layer, information layer, intelligence layer, and application layer. However, there are still many issues to be dealt with in the proposed architecture if it is to be actually applied in practice. For example, how to integrate the proposed architecture with other existing protocols and standards, and how to evaluate the intelligence correctly and efficiently. Besides, in practice, the total number of networks, devices, tasks and users is increasing dramatically, which poses a great challenge for optimization. Only when these issues are reasonably addressed can the proposed architecture of the Internet of intelligence be implemented and applied in a practical environment. The authors hope that this section will inspire readers to contribute more to the development of the Internet of intelligence.

\section{Enabling Technologies in Different Layers \label{sect_5}}

The realization of the Internet of intelligence largely depends on a variety of enabling technologies. Therefore, this section focuses on the enabling technologies of each of the five layers of the above architecture to help in-depth understand the functionalities of the Internet of intelligence.

\subsection{Physical Resources Layer}

The physical resources layer is mainly composed of various infrastructure resources of the Internet of intelligence, including communication resources, caching resources, computing resources, and sensing resources.
In the following, a brief overview of the current communication, caching, computing, and sensing technologies that support the Internet of intelligence is given.

\subsubsection{Communication}

The communication and networking technology of the fifth-generation (5G) network is a key driving enabler of the Internet of intelligence, which can bring users with high service quality and experience such as enhanced mobile broadband (eMBB), ultra-reliable and low-latency communications (uRLLC), and massive machine type communications (mMTC) \cite{9269939,7744805}. One of the significant advancements in 5G comes from the ability to virtualize and redistribute network functions. Dynamic network slicing dramatically improves the flexibility, data rate, and delay performance of the network, thereby providing support for the emergence of various Internet of intelligence services \cite{8683970}. SDN and NFV technologies can further increase the flexibility of slicing \cite{9363323}.

In 5G, to provide better services and reduce capital expenditures and operating expenses, cloud radio access network (C-RAN) with the characteristics of centralized processing, collaborative wireless, real-time cloud computing and energy-saving infrastructure has attracted widespread attention. Meanwhile, to improve network coverage and link capacity to cope with explosive data traffic growth, heterogeneous network (HetNet) architecture with different types of cells has been proposed as a promising technology \cite{8088524}. Long term evolution-advanced (LTE-A) band below 6 GHz is too crowded for 5G. An effective way to achieve a higher data rate is to use massive frequency resources available in the millimeter wave (mmWave) range (30 - 300 GHz) \cite{8454666}. To tackle the issue of high propagation loss in mmWave communication, a practical solution is to use the short wavelength of millimeter wave to design a smaller but more compact large-scale multiple-input multiple-output (MIMO) \cite{7454778}. MmWave communication assisted by MIMO plays an important role in fundamentally solving the problem of 5G spectrum congestion and supporting high data rates.

Improving the efficiency of spectrum utilization is another feasible method to address the scarcity of spectrum. Device-to-device (D2D) communication enables two devices to transmit data directly and can flexibly reuse radio resources, thereby improving spectrum efficiency and reducing the workload of core network data processing \cite{8088537}. Besides, ICN is also a candidate technology worth considering \cite{7926924}. Due to the widespread distribution of mobile devices and the ever-increasing storage capacity, the caching capability prevalent in mobile devices should not be ignored.

\begin{figure}[t]
	\centering
	\includegraphics[width=3.3in]{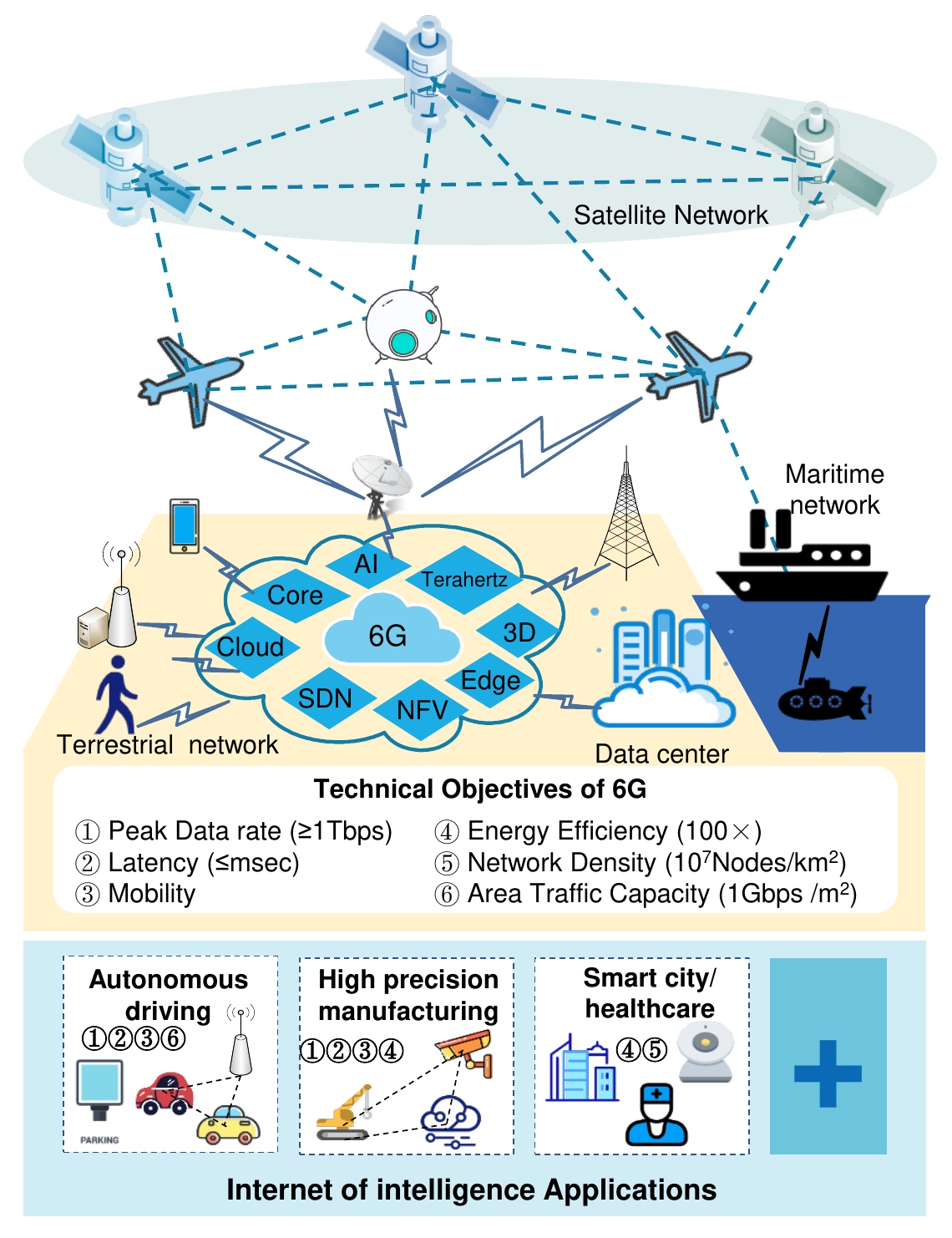}
	\caption{The architecture of 6G-enabled Internet of intelligence.}
	\label{Fig_communication}
\end{figure}

To better satisfy the strict requirements of the Internet of intelligence, such as ultra-high data rate, ultra-low latency, ultra-high reliability, and seamless connection, the next 6G network is conceived by industry and academia \cite{9403380}. A simple architecture of 6G-enabled Internet of intelligence is illustrated in Fig. \ref{Fig_communication}. The architecture of the 6G will be implemented mainly under the support of the 5G architecture, such as the architectures of SDN, NFV, and network slicing. However, compared with 5G, the development of 6G is capable of the characteristics of large dimension, high complexity, dynamics, and heterogeneity. 

AI has powerful learning ability, reasoning ability, as well as intelligent recognition ability and can realize completely autonomous services that execute human intelligence \cite{9237460}. The application of AI in 6G enables its architecture to perform adaptive learning without manual intervention, thereby effectively supporting various Internet of intelligence services \cite{9205981}. Ubiquitous AI services are provided from the core of 6G to terminals, such as network management, resource scheduling, network maintenance, etc.

The design of the three-dimensional (3D) communication infrastructure that combines terrestrial, aerial radio access points and mobile edge hosts enables 6G to provide cloud functions on demand \cite{8792135}. This idea can provide cloud capabilities under controllable delay constraints by managing large aerial platforms. When the request changes greatly over time and space, this strategy is more cost-effective than current methods that rely on a fixed infrastructure.

The endless pursuit of higher data rates has led to the exploration of higher frequency bands and corresponding communication solutions in 6G. The terahertz (THz) frequency band (0.1-10 THz) has been promoted as a key technology to meet this requirement.
Due to the huge unallocated bandwidth, it can support data transmission rates above 10 Gbps, and has better confidentiality and anti-interference capabilities.
Consequently, the use of terahertz can effectively alleviate the increasingly tight spectrum resources and the current capacity limitations of communication systems \cite{9269931}.

The exchange of large amounts of data, information and intelligence in the Internet of intelligence poses considerable challenges to communication networks in terms of security, privacy, trust, and multitude. Innovative cryptographic technologies are used in 6G to achieve effective integration between AI and privacy. Moreover, decentralized authentication is another critical problem in the Internet of intelligence. Technologies like distributed ledgers in blockchain are expected to play an important role in distributed authentication \cite{8496756}.

\subsubsection{Caching}

With the development of the Internet of intelligence, the amount of data exchanged within the network will increase significantly, and the backhaul may become a bottleneck. By storing the content of the Internet of intelligence at infrastructures such as access points and edge computing servers, caching techniques can achieve the reuse of cached content, thus effectively reducing the use of backhaul. 
The caching paradigm's requirements for content distribution and retrieval-based usage patterns will lead to challenges in network mobility, availability, and scalability. 
Toward this end, content delivery networks (CDNs) and peer-to-peer networks (P2P) have been proposed to endow the network with content distribution and retrieval capabilities by using existing storage and computing infrastructures \cite{8060515}.

In CDN, multiple service locations (service nodes) cache content in local storage. The load balancer optimally distributes user requests to these service nodes and provides users with a duplicate of the content cached in these service nodes. Generally, when assigning backup content source locations, requests will be redirected to the nearest service nodes through the domain name server of the CDN to improve the response speed \cite{6195531}. The selection of service nodes to cache the content replication will be determined based on continuous monitoring of data traffic and load balancing in the network \cite{6782765}.

P2P networks eliminate the need to use central servers, allowing all users to communicate and share content. Each user can act as a cache node/server in P2P networks. In a P2P network, a single server (also called a seeder) distributes a large content to a large number of interested peers through the Internet. Instead of uploading the content to each peer, the seeder first divides the content into data packets called blocks and then intelligently distributes these blocks so that the participating peers can exchange them with each other \cite{9240921}. P2P-based content distribution dramatically reduces the download time of each peer and reduces the pressure on a single server.

Besides, information-centric networking (ICN) can also effectively realize information distribution and content retrieval by running common protocols at the network layer \cite{8240926}. According to the caching location, caching mechanisms of ICN can be classified into two categories: i) on-path caching: On-path caching involves the transmission path, so it is usually aggregated with the ICN forwarding mechanism; ii) Off-path caching: off-path caching only cares about content storage and delivery, and does not consider the delivery path.

\subsubsection{Computing}

\begin{figure}[t]
	\centering
	\includegraphics[width=2.8in]{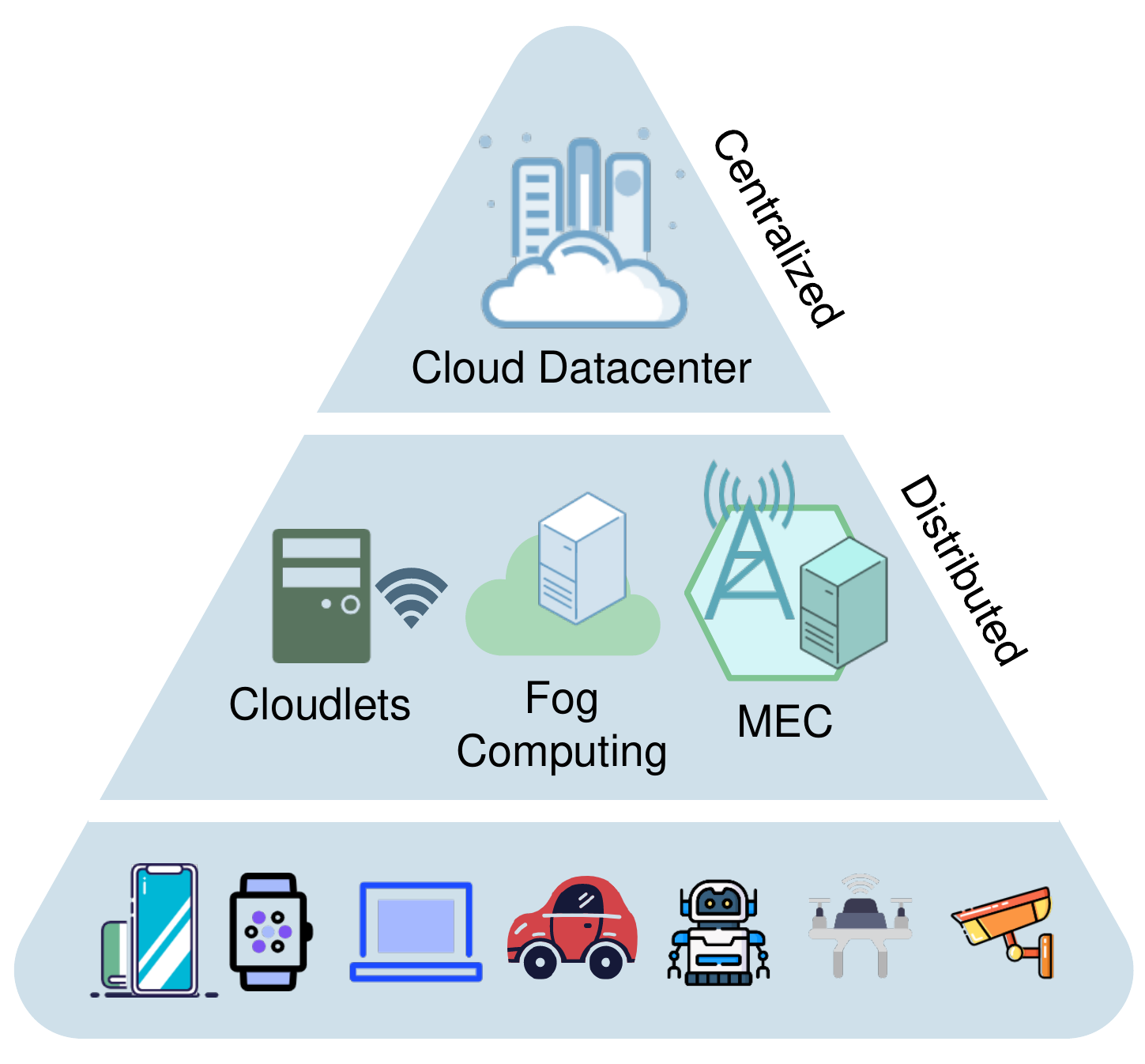}
	\caption{Computing technologies in the Internet of intelligence.}
	\label{Fig_computing}
\end{figure}

The development of applications such as autonomous driving, smart cities, and energy Internet poses huge challenges to the computing capabilities of the Internet of intelligence. Due to limitations of size and battery life, smart mobile devices are often unable to meet such demands \cite{8012473}. By offloading those applications to powerful cloud centers, cloud computing is an effective approach to solve the above challenges \cite{8352664}.

However, since cloud centers are usually built far away from smart mobile devices, the low-latency requirements of some delay-sensitive applications in the Internet of intelligence may not be satisfied by cloud computing. Besides, it is sometimes not feasible and economical to migrate masses of computation tasks over long distances. To cope with such problem, edge computing has been proposed to enable users to be served by proximity servers with abundant computing resources \cite{8981923}. Edge computing aims to migrate cloud computing platforms (including computing, storage, and network resources) to the network edge, and strive to achieve the deep integration among traditional mobile communication networks, the Internet and the Internet of intelligence, thereby exploring the inherent abilities of the network, reducing the end-to-end latency, and improving the user experience \cite{9048610,9562560}. Fig. \ref{Fig_computing} shows the computing technologies that enable the Internet of intelligence.

Currently, the concept of edge computing first appeared in the form of Cloudlets, which enabled users to have superior computing and storage performance, and later paved the way for more complex fog computing and multi-access edge computing (MEC) \cite{9063670}.

Cloudlet is originally proposed by the open edge computing (OEC) project initiated by Carnegie Mellon University \cite{7954011}. 
It can be regarded as a node with trusted Internet connection and rich resources, providing computing, storage and access services for neighboring users.
The instantiation of cloudlet is based on the soft state that depends on virtual machines (VMs).
Using VMs, cloudlet offers a temporary environment isolated from another host software for individual users. As a micro data center in a box, cloudlet provides end-users with access to deploy and manage their own VMs via Wi-Fi \cite{7450563}.

Fog computing was first proposed by Cisco in 2012 \cite{cv2019fog}. 
It is a distributed computing paradigm in which network entities with heterogeneous computing and storage capabilities are placed in the vicinity of the RAN to connect users to the cloud or Internet. 
Fog computing is a supplement to cloud computing rather than a substitute. The main purpose of fog computing is to offer low-latency services in response to real-time applications \cite{yi2015survey,8981923,9075084}. When involving applications that require massive computation or permanent storage, fog merely acts as a gateway or router to redirect data to the cloud center.

MEC was proposed by the European telecommunications standards institute (ETSI) in 2014 \cite{ETSI}. 
It gathers service providers, mobile operators, over-the-top (OTT) participants and subscribers to provide them with a sustainable business model. 
The gist of MEC is to offer cloud capabilities within the RAN in proximity to mobile users \cite{peng2020multi}. Exactly as the objective of other edge computing methods, MEC provides accelerated content and services by improving the responsiveness of the edge.

Fog computing and MEC conceive an open platform that provides similar functionalities. Fog computing focuses on applications (mainly the IoT) utilize the platform set that collectively assists users, while MEC concentrates on application-related enhancements in information computing and caching. Moreover, cloudlets or Fog nodes are mainly managed by individuals and can be deployed in any suitable location, but the MEC server is owned by the mobile operator and requires to be placed in the vicinity of BSs to offer access to users through the RAN.

\subsubsection{Sensing}

In the Internet of intelligence era, a wide variety of sensing devices are deployed to provide meaningful sensing data records, such as environmental sensing devices, color sensing devices, flame sensing devices, motion sensing devices, cameras, etc.
With the help of identification, positioning, detection, imaging and other technologies, these sensing devices can sense the environment in real-time, thus enabling pervasive sensing and providing intelligent decision support for the Internet of intelligence \cite{7879243,9200330}.

Radio frequency identification (RFID) is a non-contact sensing technology that can effectively identify and track objects in a non-contact manner.
It is developed from radar technology in the 1940s and obtains relevant information about physical objects through radio frequency signals.
In RFID, electronic tags are used to mark an entity object, and the data of the entity object is received through RFID readers \cite{lahiri2005rfid}. 
RFID has a strong anti-interference ability so that it can not only identify solid objects moving at high speed but also identify the target of each solid object simultaneously.
RFID offers many advantages over other technologies, including fast scanning, non-contact reading, reusability, large storage, low cost, and security \cite{tan2014survey,cika2012active}.
Due to these advantages, RFID can be effectively applied in the Internet of intelligence for identifying and tracking objects and exchanging information.

On the other hand, to sense real-world physical parameters related to the surrounding environment, wireless sensor networks (WSNs) play a vital role in the Internet of intelligence \cite{9433518}. WSNs can monitor and track the status of devices and transmit the status data to control centers or receiver nodes via multiple hops \cite{lin2011towards}. Therefore, WSNs can be considered a further bridge between the real world and the cyber world \cite{7060643}.
Compared to other technologies, WSNs offer advantages such as dynamic reconfiguration, scalability, reliability, low power consumption, low cost, and small size \cite{8396218}. All these benefits facilitate the integration of WSNs into various domains with different requirements.

With the rapid development of the Internet of intelligence, the popularity of various smart devices makes their built-in sensing devices (such as accelerometers, cameras, microphones, etc.) attract much attention, and the data collected by these sensing devices can be analyzed to extract a great deal of useful intelligence \cite{8689081}.
Crowdsensing, which combines the idea of crowdsourcing and the sensing ability of smart devices, is a new mode for data acquisition \cite{8703469}. In this model, large-scale users collect sensing data through their own smart devices and upload them to the server. The service provider records and processes the sensing data and uses the collected data to provide daily services to users.
Due to the prevalence of smart devices and the flexibility of user movement, crowdsensing effectively solves the problems of high maintenance costs and limited coverage of traditional sensing networks \cite{8703108}. Therefore, it can be widely used in the Internet of intelligence.

\begin{figure}[t]
	\centering
	\includegraphics[width=3.2in]{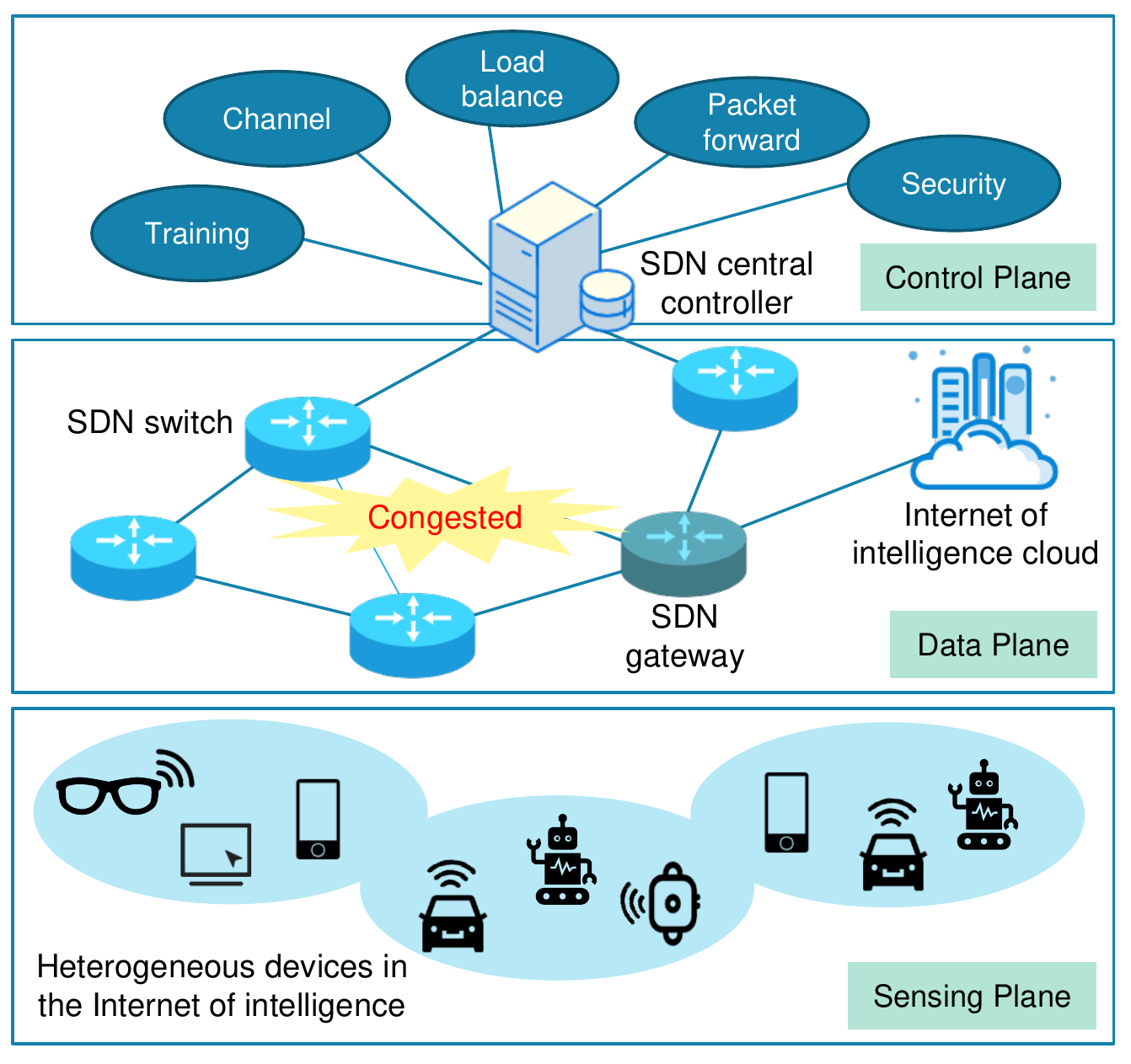}
	\caption{Overview on the SDN technology.}
	\label{Fig_SDN}
\end{figure}

\subsection{Resources Virtualization Layer}

The resources virtualization layer aims to virtualize the heterogeneous and ubiquitous communication, caching, computing and sensing resources in the Internet of intelligence to fully utilize these resources. The following will introduce four supporting resources virtualization layer technologies: SDN, NFV, network slicing, and containerization.

\subsubsection{Software-Defined Networking}

The considerable growth of network scale and diversity has created the heterogeneity of the network and has generated massive data that need to be analyzed, extracted and processed.
Consequently, the Internet of intelligence will face challenges in flexibility, interoperability, and reconfiguration. 
To meet these challenges, SDN can effectively simplify the network performance by separating the control plane from the data plane, and play an essential role in network management, network virtualization, resource utilization and security \cite{8017556,8770301,8025644}.

\begin{figure}[t]
	\centering
	\includegraphics[width=3.2in]{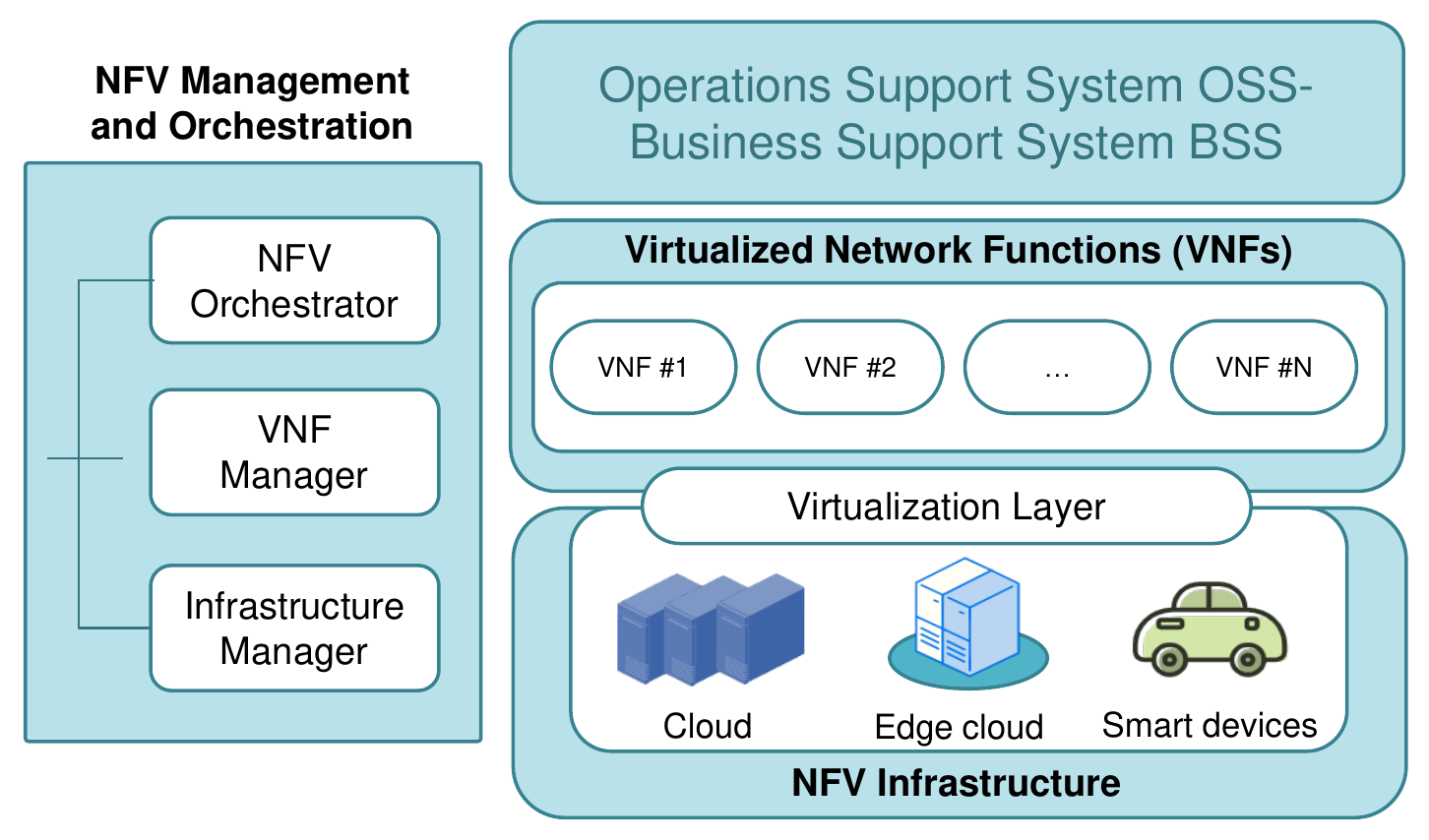}
	\caption{Network function virtualization in the Internet of intelligence.}
	\label{Fig_NFV}
\end{figure}

\begin{figure*}[t]
	\centering
	\includegraphics[width=5.5in]{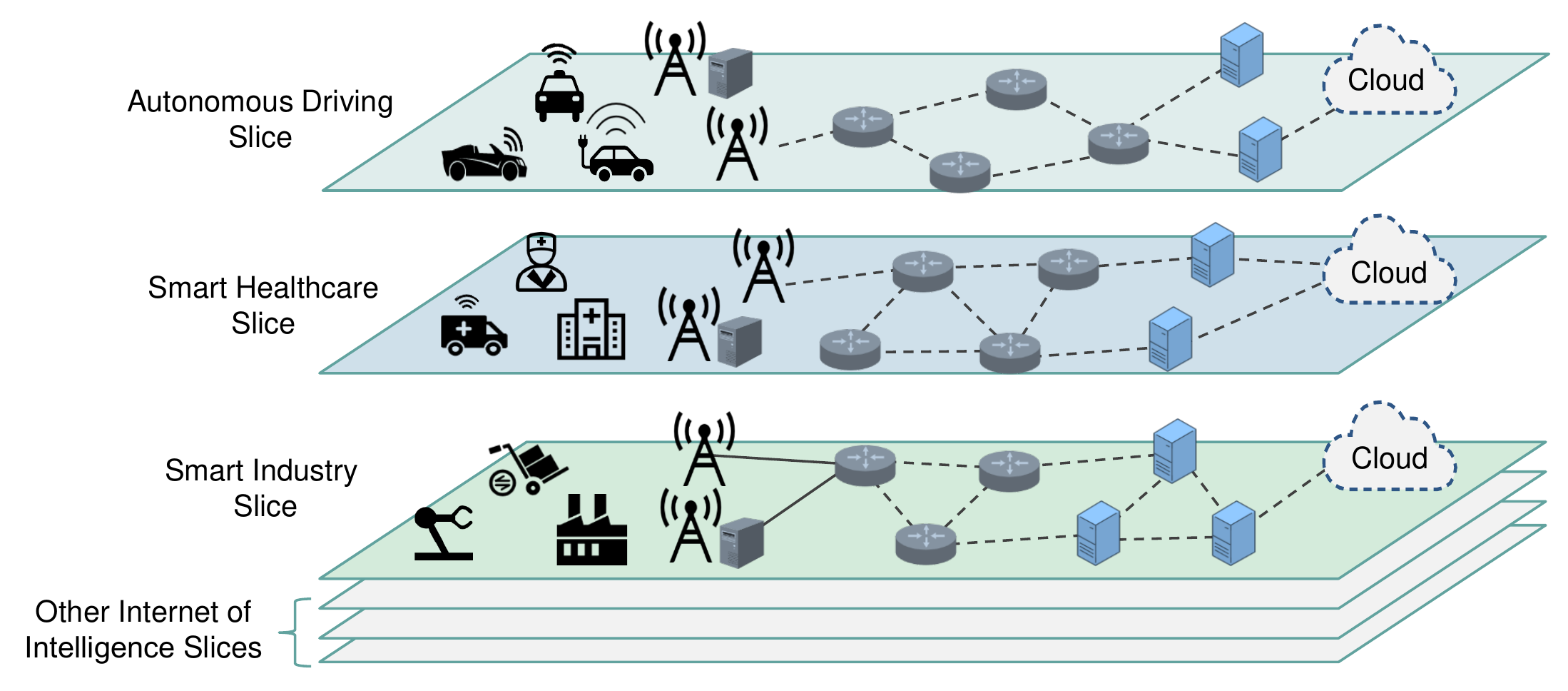}
	\caption{Use of network slicing in the Internet of intelligence.}
	\label{Fig_NetworkSlicing}
\end{figure*}

As shown in Fig. \ref{Fig_SDN}, the architecture of SDN in the Internet of intelligence includes three planes, i.e., sensing plane, data plane, and control plane \cite{8277323}. The sensing plane comprises heterogeneous intelligent devices in the Internet of intelligence, which are mainly used to sense large amounts of data. 
The data plane adopts distributed working mode and transmits data by following a flow table.
In the Internet of intelligence, routers, switches, circuits, etc., are equipped with flow tables for data transmission. The control plane is a centralized unit that mainly configures the network according to the requirements of the Internet of intelligence applications, such as network paths, routing protocols, and network policies. In SDN, OpenFlow has become a standard protocol for communication between software controllers and data planes \cite{OF}. OpenFlow conveniently exposes the network elements and hardware functions to controllers to realize the programmability of data planes. SDN provides two types of vertical APIs, i,e., northbound API and southbound API, to support the communication between applications and controllers and the communication between controllers and network devices, respectively. SDN also provides two types of horizontal APIs, eastbound API and westbound API, to support the communication between SDN controllers. In addition, SDN offers hardware reusability through reconfigurable features, updates the software of various modules and fixes errors through policy updates, and provides a quality-aware framework for the Internet of intelligence by implementing network logic, thus further expanding network capacity and coverage.

In this way, SDN has excellent flexibility and a high degree of programmability, which can support the cost-effective and dynamic network configuration of the Internet of intelligence and provide a basis for the service customization of Internet of intelligence applications. Besides, efficient capacity sharing, secure routing and mobility management can also be realized by the introduction of SDN into the Internet of intelligence.

\subsubsection{Network Functions Virtualization}

NFV is a network concept that uses virtualization technology to manage core network functions through software-based methods \cite{7243304}. Therefore, in the resources virtualization layer, NFV has gradually become one of the critical drivers for the Internet of intelligence to improve computing capabilities to meet the growing network demands. As shown in Fig. \ref{Fig_NFV}, the NFV architecture and orchestration framework is composed of three key components, namely virtualized network functions (VNF), NFV infrastructure (NFVI), and NFV management and orchestration (NFV MANO) \cite{ETSINFV}. VNF is the software implementation of network functions. NFVI includes hardware and software components that provide the network environment for deploying VNF, such as central processing unit (CPU), storage, virtualization, etc. NFV MANO is responsible for managing and orchestrating the physical and virtual resources of NFVI and the life cycle management of VNFs. 
By integrating network functions into the cloud platform, NFV can greatly reduce network operating costs.
In addition, it also provides flexibility in both the data plane and the control plane by allocating resources on-demand according to service requirements. Specifically, VNFs allow multiple instances to be collocated on the same hardware or the same virtualized environment (such as VM) \cite{7243304}.

In the Internet of intelligence, NFV is a key enabling for handling virtualized instances related to specific services and shows good performance in achieving service migration, flexibility and scalability. 
For example, by adding additional specific resources or software instances, resources can be easily expanded for popular applications in the Internet of intelligence. The Internet of intelligence services can benefit from the dynamic characteristics of NFV: i) support the deployment of portable functions on an operable distributed virtual network; ii) migrate independent blocks of services to different cloud environments in various networks; iii) slice the virtual network resources of specific applications, and iv) share configurable resource pools for on-demand access \cite{7931566}.

SDN and NFV are developing towards network virtualization and automation, which are two closely related technologies \cite{9351537,9477144,8945310}. Accordingly, combining NFV with SDN to jointly optimize network functions and resources is an increasing trend. On the one hand, SDN can assist NFV in managing dynamic resource management and orchestrating intelligent functions. On the other hand, SDN can dynamically create virtual environments for specific service chains by implementing NFV \cite{7350211}. With the aid of SDN and NFV, the Internet of intelligence can realize the dynamic scheduling, management, and utilization of communication, cache and computing resources to improve the system performance.

\subsubsection{Network Slicing}

The Internet of intelligence enables various types of connections and services, which need performance and security guarantees. Moreover, different Internet of intelligence applications have different requirements for services. For instance, smart factory applications need to ensure a packet loss rate of less than $10^{-9}$ and an end-to-end delay range of 250 $\mu$s to 10 ms. Smart vehicular applications require a reliability level of more than 99.99\% and an end-to-end delay of less than 1 ms \cite{7842415}. Network slicing can effectively meet such diverse requirements of different Internet of intelligence applications by creating multiple logical network segments on a common shared physical infrastructure \cite{wu2021ai}. A multi-tenant environment is introduced in network slicing to support flexible provisioning of network resources and dynamic allocation of network functions \cite{9382385}. Fig. \ref{Fig_NetworkSlicing} illustrates the use of network slicing in the Internet of intelligence.

Network slicing improves resource utilization efficiency by dynamically adjusting network resources between slices.
The autonomous system is a feasible way to achieve such dynamic adjustment and enhance the scalability of Internet of intelligence applications \cite{8636831}. The rapid growth in intelligent devices has also attracted many attackers into the Internet of intelligence. Intelligent devices in the Internet of intelligence are vulnerable to attack due to their limited resources, which will cause them to generate distributed denial of service (DDoS) attacks on the network \cite{7971869}. By using network slicing to isolate various applications, the severity of such attacks can be minimized, and the high security and privacy of the Internet of intelligence can be achieved. In addition, the dynamic allocation of idle resources to the attacked slices can also guarantee the QoS during the attack to a certain extent.

To meet the differentiated service requirements of various applications of the Internet of intelligence and realize service customization, the combination of network slicing with NFV and SDN is essential \cite{7926921,8954683,9076109}. Tight coordination for VNF allocation and service provisioning can be supported at the edge/cloud, thus realizing flexible service control and enabling service differentiation customization.

\subsubsection{Containerization}

Containers provide a lightweight virtualization solution for the portable operation of the Internet of intelligence services. Due to its advantages of being quickly packaged and deployed to various interconnected intelligent platforms, the Internet of intelligence can benefit from containers. In traditional computing platform virtualization, the creation of VMs provides the effect of physical resources running its own operating system (OS). Hardware virtualization occurs on the host, and the VM is the guest machine. In contrast, the abstraction in containers occurs at the OS level \cite{7931566}. Containers partition the resources of physical machines to create multiple isolated instances smaller than VMs \cite{6096132,barham2003xen}. As a result, multiple containers can run in a single OS, which can be quickly deployed on CPU, memory, and disk with performance close to the native machine \cite{6498558}.

Containers have additional advantages over VMs in the context of the Internet of intelligence: i) Containers build and create an application through image layering and expansion, and then store it as an image in a repository, which helps achieve packaging, delivery, coordination, and rapid deployment; ii) Container APIs provide the lifecycle management of containers, such as creating, defining, composing, distributing, and running containers; iii) Networking performed through port mapping simplifies the linking of containers; iv) Data storage support is provided by connecting one or more containers to the ongoing service. 
The superiority of the container is more obvious in mobile environments. In contrast, VMs can support heavier applications or bundled applications associated with specific third parties to ensure higher security. Docker is an outstanding container solution in edge environments \cite{7377291}. Containers, particularly Docker containers, are drinkable and can efficiently run inside a VM. They can be ported from one VM to another VM or even to bare metal without spending a lot of effort to facilitate the migration \cite{combe2016docker}.

\subsection{Information Layer}

The information layer aims to process the massive raw data in the Internet of intelligence, analyze the valuable information, and transfer it to the intelligence layer. In this layer, the information itself is more valuable than its source since it describes intelligent events. ICN can build an information/content-centric information layer architecture based on the requirements of the Internet of intelligence.
ICN mainly includes two types of packets: information packets and data packets \cite{mccarthy2021qosa}. Information packets are used to record the path of the user's request to facilitate the content publisher to return response data, and data packets are the content that the publisher responds to the user's request. Both the information packet and the data packet contain three tables: content store (CS), pending interest table (PIT), and forwarding information base (FIB) \cite{9350832,7996572}. CS is used as a temporary cache of content to meet future requests for stored content. PIT tracks unmet interests (more precisely, interests that have been reposted but have not yet received data). Finally, FIB maintains a record of each suitable interface that can access the name prefix.

\begin{figure}[t]
	\centering
	\includegraphics[width=3.0in]{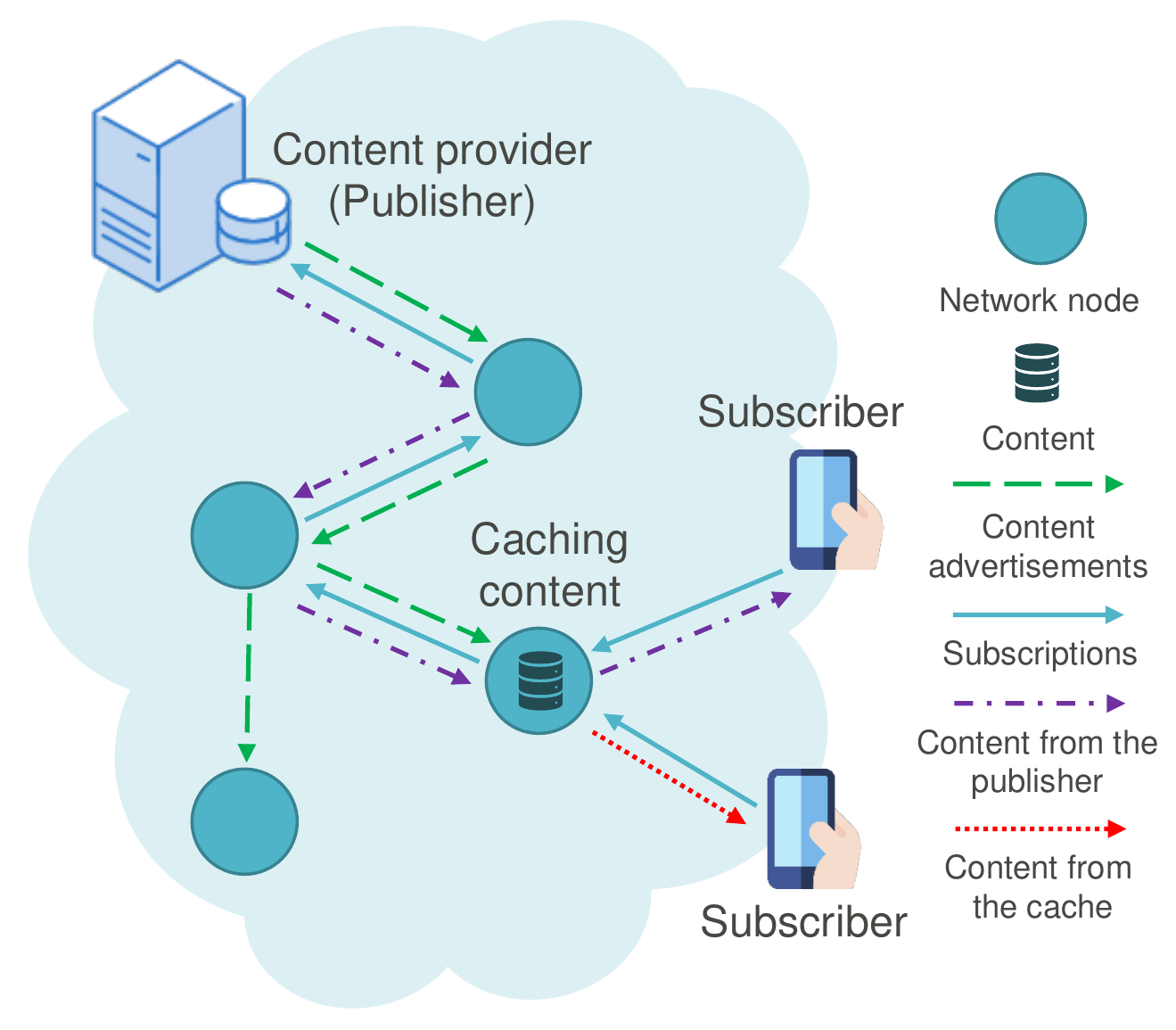}
	\caption{The basic information-centric networking model.}
	\label{Fig_ICN}
\end{figure}

Fig. \ref{Fig_ICN} shows the ICN model. Unlike the IP architecture, the main focus of ICN is to disseminate, find, and deliver information, instead of concentrating on the accessibility of end hosts and maintaining the conversation between them. Users in ICN can request content without knowing where the content is located. The communication follows the receiver-driven principle, and the data follows the reverse path \cite{7113228}. Therefore, the connection establishment in ICN is determined by matching the requested content rather than the endpoint providing the content.

In this subsection, we mainly introduce the key enabling technologies of the information layer supported by ICN, including naming, routing, caching, transport, and identity resolution.

\subsubsection{Naming}

ICN decouples information from location and uses content-based names rather than locations to route ICN nodes. Therefore, information naming is a critical issue in ICN because it is the foundation of caching and replication in the network. During network transmission, the same value can be used to access and retrieve naming information. In addition, the value of the name's aggregation and verifiability varies in different ICN designs. There are two primary naming schemes for ICN: flat naming and hierarchical naming \cite{6563278}.

Flat naming is mainly adopted by ICN architectures like MobilityFirst \cite{li2014comparative}, DONA \cite{misra2017accconf}, and NetInf \cite{potys2015netinf}. Flat names are generated by the encrypted hash of the content or its subcomponents or its attributes. Such a naming process ensures location independence, global uniqueness and application independence. 
Flat naming can effectively avoid location identification binding, so the information data is more dynamic. However, this approach has encountered difficulties in supporting the scalability of network routing. In addition, flat names may encounter bottlenecks in readability, and thus an additional system is needed to classify readable and unreadable names.

The hierarchical naming is utilized in the two well-known ICN architectures, named data networking (NDN) \cite{zhang2010named} and content centric networking (CCN) \cite{perino2011reality}. It uses a hierarchical structure to assign content names in a human-readable way. The aggregate information name can easily match the uniform resource locator (URL) and can be deployed in the current network environment. Hierarchical naming can realize the aggregation of data and the compatibility of the current system. Moreover, it minimizes routing information by aggregating file names and improving the backbone network's core routing performance. However, the variable component name length may make the content name difficult to remember \cite{7437030}.

In the information layer of the Internet of intelligence, to effectively realize the collection and analysis of information, the content naming mechanism needs to support: location independence, global unique information retrieval, human readability, push support, and security. Naming content through a hybrid naming scheme that combines the characteristics of the above two naming schemes is a feasible solution \cite{8254350}. Specifically, it adds a sub-name in the content name to push and pull communication modes. Meanwhile, the hash value of the content is appended to enhance security and ensure the integrity of the content and its provider. Furthermore, some intelligence layer technologies, such as AI, can be introduced to solve the scalability problem in the naming scheme \cite{8740792}.

\subsubsection{Routing, caching and transport}

In addition to the naming mechanism, some other vital technologies that ICN supports for the Internet of intelligence are its routing, caching, and transport mechanisms. ICN has a more straightforward mechanism of content discovery and data transmission than the current IP-based network, which can effectively improve content acquisition and distribution in the Internet of intelligence. 
Since the inter-network caching in the ICN can satisfy requests from users, session-based communication and continuous/timely availability are not necessary.
ICN uses name-based routing, where the content name is adopted to forward requests in a hop-by-hop manner to discover content along the way and pass data back to the requester \cite{8624354}. 
Different physical interfaces can be supported simultaneously, such as cellular, WiFi, short-distance, which can effectively achieve load balancing \cite{8885909}. In addition, by aggregating the same interests in the PIT, ICN can improve content retrieval for multiple users, track future data delivery, and support multicast data delivery locally.

Since the content in ICN is decoupled from the location, in-network caching can be applied to retrieve content from the most convenient cache node to improve content retrieval and availability \cite{8240926}. Efficient cache replacement strategies are needed for cache nodes to cache popular and most commonly used content to improve caching performance and network resource utilization. Due to the application of in-network caching, the user's request for certain content may be fulfilled at the CS of the intermediate nodes. This request may not be forwarded to other nodes in such cases, which effectively reduces data waiting time and decreases network costs \cite{8975984}. The content can be cached in any cache node to serve subsequent requests. In addition, large content can be divided into packet-sized blocks. The cache node can cache part of the content or the entire content.

The transport issue is an important issue that affects the resource allocation of the Internet of intelligence. It directly affects the transmission efficiency of ICN and indirectly affects the performance of the Internet of intelligence. In the traditional end-to-end network congestion control mechanisms, the network can be considered as a ``black box'', and the congestion control is completed by the end-user. The congestion control in ICN is based on the user-driven transmission model, which is mainly implemented by controlling the rate at which the requester sends interest \cite{7590427}. 
Each routing node on the path can preserve data locally and return the data to requesters. In small and medium-sized Internet of intelligence networks, internal information sharing can effectively achieve load balancing. Besides, some ML algorithms can be used to predict delay and dynamically select forwarding paths to effectively improve congestion control performance \cite{8290944}.

\begin{figure}[t]
	\centering
	\includegraphics[width=3.0in]{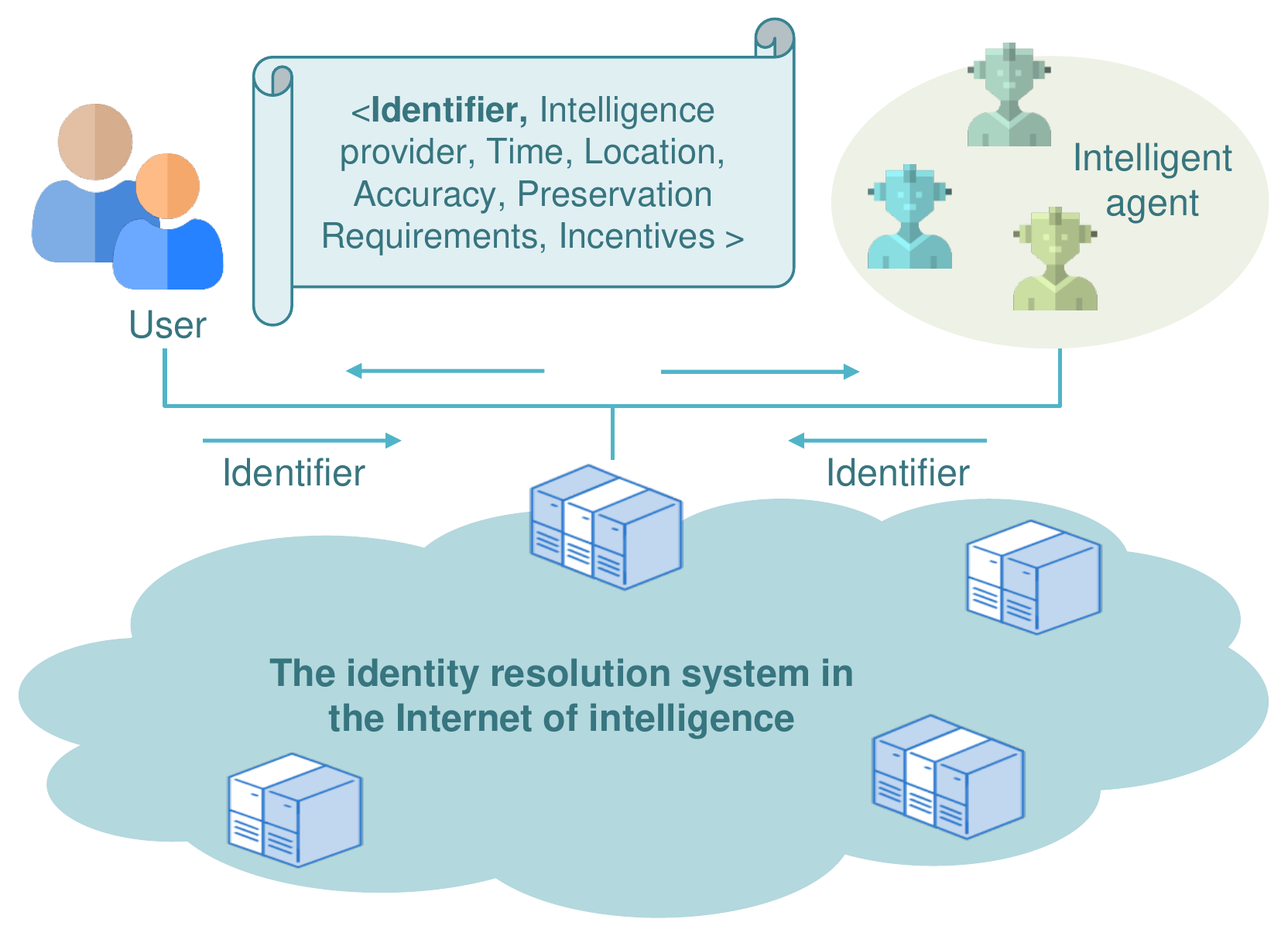}
	\caption{Application of identity resolution system in the Internet of intelligence.}
	\label{Fig_IS}
\end{figure}

\subsubsection{Identity Resolution}

Identity resolution is a key technique in the information layer of the Internet of intelligence. Similar to the domain name system (DNS) in the Internet, it plays the role of entrance in the Internet of intelligence. 
DNS maps domain names and IP addresses to each other, allowing people to access the Internet more conveniently without having to remember complex IP addresses. The input of DNS is the domain name, and the output is its IP address. In contrast, the input of identity resolution is the identifier, and the output is the mapping information of the identifier. Through the identity resolution system, scattered intelligence can be associated, as shown in Fig. \ref{Fig_IS}. The user only needs to provide a unique identifier to obtain the traceability information of intelligence.

\renewcommand{\arraystretch}{1.5}
\begin{table*}[t]
	\caption{Comparison of Different AI Approaches}
	\begin{center}
		\begin{tabular}{m{2.0cm}<{\centering}|m{2.1cm}<{\centering}|m{1.4cm}<{\centering}|m{2.1cm}<{\centering}|m{1.8cm}<{\centering}|m{2.5cm}<{\centering}|m{2.6cm}<{\centering}} 
			\hline
			\hline
			\textbf{Approach} & \textbf{Driving Factors} & \textbf{Required Data} & \textbf{Training/Learning} & \textbf{Calculation Process} & \textbf{Advantages} & \textbf{Disadvantages}\\ 
			\hline	
			Symbolic AI & Knowledge-driven & Large & Human knowledge and experience; Logical inference & Transparent and understandable audit trail & High interpretable; Fast detection speed.
			& Weak generalization; Poor robustness; Low accuracy.\\ 
			\hline
			Connectionist AI & Data-driven & Small & Machine learning; Deep learning & ``Black box'' & Strong generalization; High accuracy; Strong robustness. & High computational complexity; Poor interpretability; Over-reliance on data.\\ 
			\hline
			Hybrid AI & Combining knowledge-driven and data-driven methods & Medium & Combining the advantages of symbolic and connectionist AI & Understandable for humans & High interpretable; Strong robustness. & Potential privacy and trust issues.\\ 
			\hline
			\hline
		\end{tabular}
	\end{center}
	\label{tb2}
\end{table*}

The construction of an identity resolution system can bring plenty of benefits to the Internet of intelligence.
Firstly, a unified method for object recognition, addressing, and understanding for applications can be provided by the identity resolution system to meet the data homogeneity requirements in big data analytics \cite{7571188,van2019self}. Secondly, the procedure of service providers can be simplified by the construction of the identity protocol layer. The absence of the identity protocol layer will shift the responsibility of identity verification to service providers, leading to inefficiency. Thirdly, the identity resolution system can achieve better management of various network devices \cite{9296359}. It can identify various subjects such as terminals, network equipment, service resources and data, thus realizing the joint optimization of multiple systems and the unified scheduling of resources. Fourthly, the identity resolution can also facilitate intercommunication between multiple isolated systems \cite{8015935}.

A large amount of information will be generated with the further application of identity resolution in the Internet of intelligence. The analysis of this information can improve the performance of Internet of intelligence services. Therefore, such information will then be transmitted to the intelligence layer and further processed by intelligent technologies such as AI to extract its intelligence through continuous learning, thereby improving the performance of specific systems and making the Internet of intelligence services smarter.

\subsection{Intelligence Layer}
In the intelligence layer, the intelligence behind the massive information needs to be analyzed and extracted. Some state-of-the-art technologies such as AI, blockchain and big data analytics fit well in this goal.

\subsubsection{Artificial Intelligence}


In the Internet of intelligence, AI is a powerful technology used to extract valuable intelligence from a large amount of information. AI can be divided into two categories according to the method of modeling high-level cognitive processes such as human decision-making, i.e., symbolic AI approach and connectionist AI approach. We have already introduced the evolution history of AI in Section II. In this subsection, we will discuss the classification and realization of current AI in more depth and make an outlook on the development of AI in the future.

\textit{- Symbolic AI Approach:} Symbolic AI is the general term for all advanced symbolic representation methods based on deductive reasoning, logical inference and search algorithms in AI research \cite{bolander2019human}. In symbolic AI, knowledge about environments is not obtained from mathematical models through optimization techniques but is hard-coded through formal languages \cite{santoro2021symbolic}. Symbol processing uses the rules or operations on the symbol set to encode comprehension. This set of rules is called an expert system, which uses a series of ``if-then'' statements to follow a top-down approach to formalize and reconstruct human intelligence.
Symbolic AI acquires skills through education and uses logical inference to solve problems \cite{martin2018symbolic}. The reasoning process is open and transparent and can be read and understood by people. All the knowledge learned by AI comes from human experience and logic, and this process does not require a lot of data. The modern implementation of symbolic AI is represented by ontologies, domain conceptualized formal representations and knowledge graphs (KG), large-scale network representations of entities, and relationships related to specific domains \cite{futia2020integration}. 
By integrating and linking data from various types of representations, KGs are exploited to capture knowledge within domains \cite{ehrlinger2016towards,seeliger2019semantic}. Compared with deep learning models, the KG is less flexible and robust, but it is developed to be interpretable.

Since it does not entirely rely on massive data, the advantage of symbolic AI is that it can train AI models on a single event, and logical inference can avoid problems caused by omissions or errors in data. Furthermore, symbolic AI can provide a transparent and understandable audit trail. Humans can understand using AI to solve problems step by step and know how AI obtains the final result. However, the symbolic AI system has the drawbacks of weak generalization and poor robustness. It is limited to a system built for one task and cannot be easily generalized to other tasks due to the artificial environment of fixed symbols and rule sets. In addition, the system may fail if one hypothesis or rule does not hold since it may break all other rules.

\textit{- Connectionist AI Approach:} The purpose of connectionist AI is to model intelligence by simulating neural networks in our brains. These computational neural networks construct the path between input and output through a series of interconnected units.
ML forms the core of contemporary connectionist AI \cite{poersch2005new}. The basis of ML is data. It uses algorithms to analyze large amounts of data, explores the mathematical and logical relationship between input and output through continuous learning, and then judges and predicts what is happening in the environment to make it intelligent \cite{8444669}. In a broader sense, ML methods can be divided into the following categories \cite{9138370}:
(i) Supervised learning: In such algorithms, a labeled data set containing inputs and known outputs is provided for training optimal models based on the relationships contained in the data set; 
(ii) Unsupervised learning: In contrast to the above-mentioned supervised learning approaches, this kind of algorithm is realized through unlabeled training data. The algorithm is designed to learn or search for hidden insights in unlabeled data;
(iii) Deep neural networks (DNN): DNN performs complex calculations and processes high-dimensional input data through multiple hidden layers. Compared with the traditional iterative optimization method, deep learning algorithms can better approach the optimal solution while significantly reducing the complexity;
(iv) Reinforcement learning: Agents in reinforcement learning perform some actions and learn from the feedback after interacting with the environment. The trade-off between exploration and exploitation is a unique feature that distinguishes it from other types of learning algorithms;
(v) Transfer learning: Transfer learning focuses on storing the knowledge acquired in solving a problem and applying it to different but related issues. It improves the learning process through the knowledge transfer between the source domain and the target domain.

A series of algorithms based on connectionist AI can well learn from examples, and perform much better than symbolic AI in many fields. For instance, AlphaGo is an implementation of connectionist AI. The system finds out the corresponding logic and abstract concepts on its own through a large number of chess records and then predicts the placement of human chess pieces and their corresponding strategies when playing against humans.
The connectionist AI system is robust to change. Even if one neuron or calculation is removed, the system can still work normally due to all other neurons. In addition, the neural network has a stronger generalization ability to different problems.
However, they often lack interpretability. Humans cannot understand how the system obtains a decision and why the system believes that this decision is the best path. The computing process of connectionist AI is a ``black box'' for humans.
Another disadvantage of connectionist AI is that it relies too much on large amounts of data. Without the collection of preliminary data, it is difficult for connectionist AIs to acquire capabilities through training.

\textit{- Hybrid AI Approach:} As mentioned above, symbolic AI is based on creating explicit (symbolic) models, while connectionist AI is based on learning implicit models. In the real world, humans seem to combine implicit and explicit models when solving problems. For example, when humans throw snowballs, we generally use the learned implicit model to predict where the snowballs will fall, consistent with the connectionist AI. In contrast, when we play chess, we use explicit symbol (language) models to reason about our possible sequence of actions, which is consistent with symbolic AI.
Therefore, future AI for the Internet of intelligence should be a combination of symbolic and connectionist AI. It can not only find out the internal laws based on large amounts of data but also need to learn human experience, professional knowledge and logical reasoning ability. In addition, the process and results of its calculations should be interpreted to establish a high degree of trust relationship with humans. Even if the data is missing or the information has errors, the AI system can solve the problem smoothly.

Such a hybrid approach can perceive the image as a fixed symbol through connectionist approaches, and the upper-level reasoning constructs the rule operation through the symbol \cite{babbar2018connectionist}. For example, how to deliver and share intelligence is a major challenge for the Internet of intelligence. Currently, the network model learned through ML is very large, which leads to the low efficiency of intelligence delivery. Therefore, we can leverage the KG technique to simplify knowledge into symbols, which can greatly improve the efficiency of intelligence delivery and better enable the Internet of intelligence.
Although the current two frameworks have their advantages and disadvantages, combining the two makes it possible to realize true AI for future Internet of intelligence. Table \ref{tb2} summarizes these three types of AI approaches.

\subsubsection{Blockchain}

\begin{figure}[t]
	\centering
	\includegraphics[width=3.0in]{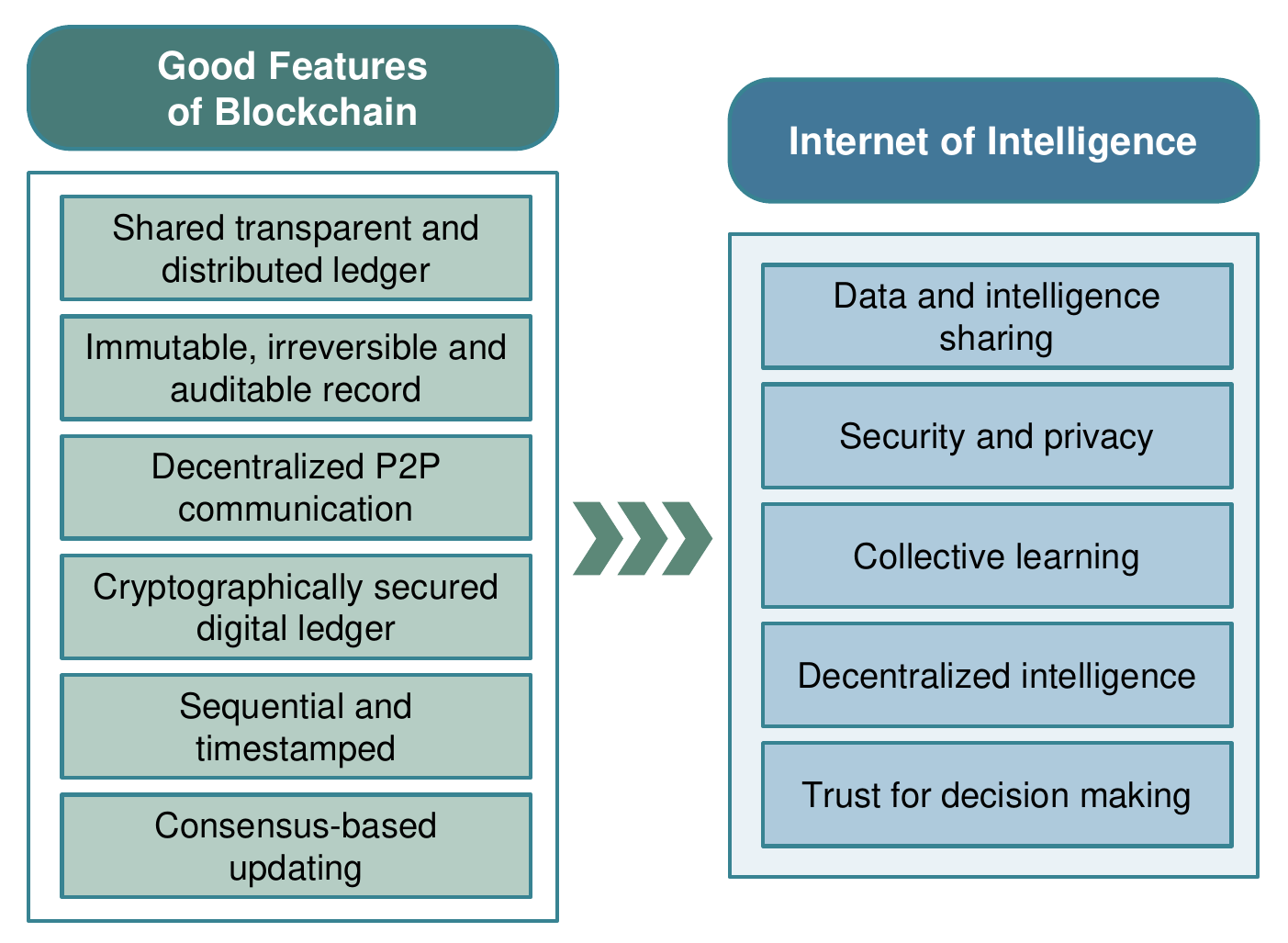}
	\caption{Some good features of blockchain that can enable the Internet of Intelligence.}
	\label{Fig_BlockchainIntel}
\end{figure}

With the large-scale application and deployment of the Internet of intelligence in the future, many data sensing and intelligence sharing records (transactions) will be generated in a distributed manner. A critical challenge on the Internet of intelligence is the management of transaction records and equipment. Besides, due to the trust and privacy issues, the trustworthiness of data and intelligence sharing is a bottleneck for the Internet of intelligence. As a universal, secure and verifiable distributed ledger technology, blockchain is also a vital enabler of the intelligence layer to cope with the low efficiency of intelligence sharing effectively \cite{8993839}. The incentive mechanism embedded in blockchain can encourage distributed parties to share intelligence. In Fig. \ref{Fig_BlockchainIntel}, the good features of blockchain that can enable the Internet of intelligence are illustrated, including intelligence sharing, security and privacy, decentralized intelligence, collective learning and trust issues for decision-making. Thanks to these features of blockchain, provenance can be enabled on the Internet of intelligence, and the trustworthiness of the network can be improved \cite{9448012}.

The key mechanism for the blockchain to realize the secure data management in the Internet of intelligence is that each participant in the network can hold a copy of the blockchain as a ledger to record all transactions (such as intelligence sharing, data sensing, etc.) among the Internet of intelligence devices.
Every transaction on the blockchain is verified according to one-way encrypted hash functions and stored in a distributed ledger \cite{9165552}. Any Internet of intelligence participants cannot arbitrarily change this ledger, and they must reach an agreement on the current state and changes of the blockchain ledger \cite{8977441}. Without the general agreement of blockchain peers, unauthorized changes to the blockchain ledger will be detected and blocked. In order to reach such an agreement, consensus mechanisms are adopted by the blockchain peers. Commonly used consensus mechanisms include proof of stake (PoS), proof of work (PoW), and directed acyclic graph (DAG), and so forth \cite{8629877,8436042,9681977}. Relying on the consensus mechanism and the chain structure of the blockchain, the Internet of intelligence system can use the blockchain to store, access and protect historical transactions to exploit these data further.

\begin{figure}[t]
	\centering
	\includegraphics[width=3.4in]{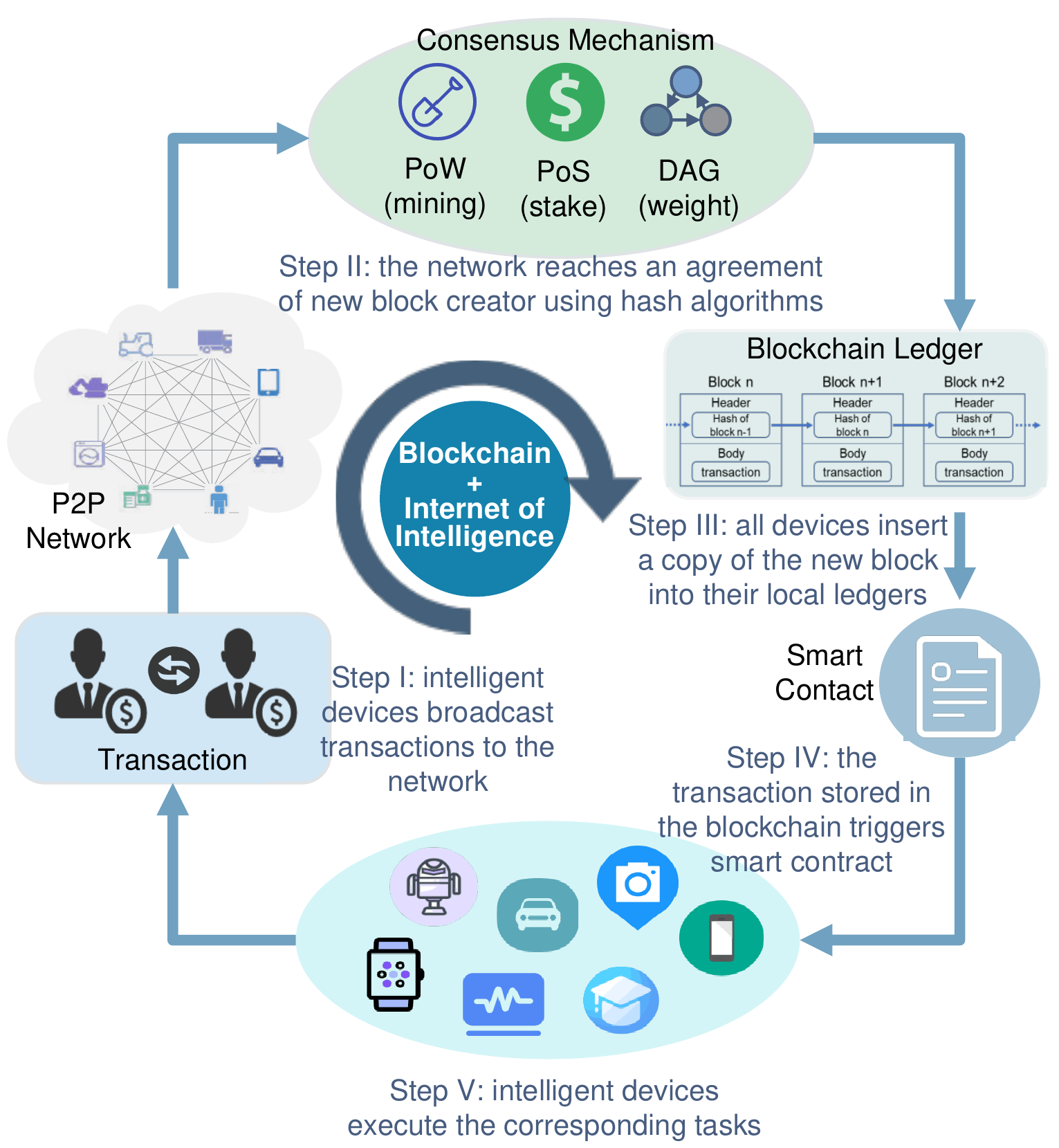}
	\caption{An example of implementing blockchain in the Internet of intelligence system.}
	\label{Fig_Blockchain}
\end{figure}

Fig. \ref{Fig_Blockchain} presents the significant steps of operating a blockchain-enabled Internet of intelligence system: i) Intelligent devices generate transactions (or records of important information) and broadcast them to the network; ii) Intelligent devices verify the information and transactions received, and the network uses hash algorithms to reach an agreement of new block creator; iii) All intelligent devices insert the identical copy of the new block into their local ledgers; vi) Transactions stored in the blockchain ledger trigger smart contracts in intelligent devices; v) Intelligent devices execute the corresponding tasks.

The deployment of blockchain can enhance the Internet of intelligence in secure data management and system scalability. Besides, data transmission and processing efficiency in the Internet of intelligence can be greatly improved due to its distributed features. The recorded historical transaction data can also be used for further system decision-making, system status rollback and data audit of the Internet of intelligence system.
However, there are some challenges in integrating blockchain into the Internet of intelligence system. The deployment of blockchain needs to consider the unique characteristics of the Internet of intelligence. Blockchain needs to offer differentiated data management services for various Internet of intelligence devices \cite{8977452}. Accordingly, it is unrealistic to deploy the blockchain functionalities on each device. Besides, the blockchain and non-blockchain sub-networks should be strategically organized by the blockchain-enabled Internet of intelligence system. The architecture and deployment of blockchain as a ledger need to take into account the integration with other enabling technologies in the Internet of intelligence. For instance, edge/cloud computing resources can be leveraged to execute blockchain consensus tasks for resource-constrained devices in the Internet of intelligence through computation offloading. In such cases, some issues such as the availability of edge/cloud resources and the security of edge/cloud environments need to be considered.

\subsubsection{Big Data Analytics}

Big data analytics can be effectively used to extract the knowledge behind the massive information, and is becoming a key contributor to enriching the intelligence of the Internet of intelligence. 
Big data usually refers to data sets that are larger than the commonly used software tools can capture, manage, and process in a tolerable time \cite{snijders2012big}. The defining characteristics of big data are the 5Vs, i.e., volume, velocity, variety, value and veracity. Specifically, volume refers to the vastness of data, velocity describes the speed and method of data arrival, variety is the diversity of data sources and types, value describes the low value density of the data, and veracity refers to quality of big data.

\begin{figure}[t]
	\centering
	\includegraphics[width=3.2in]{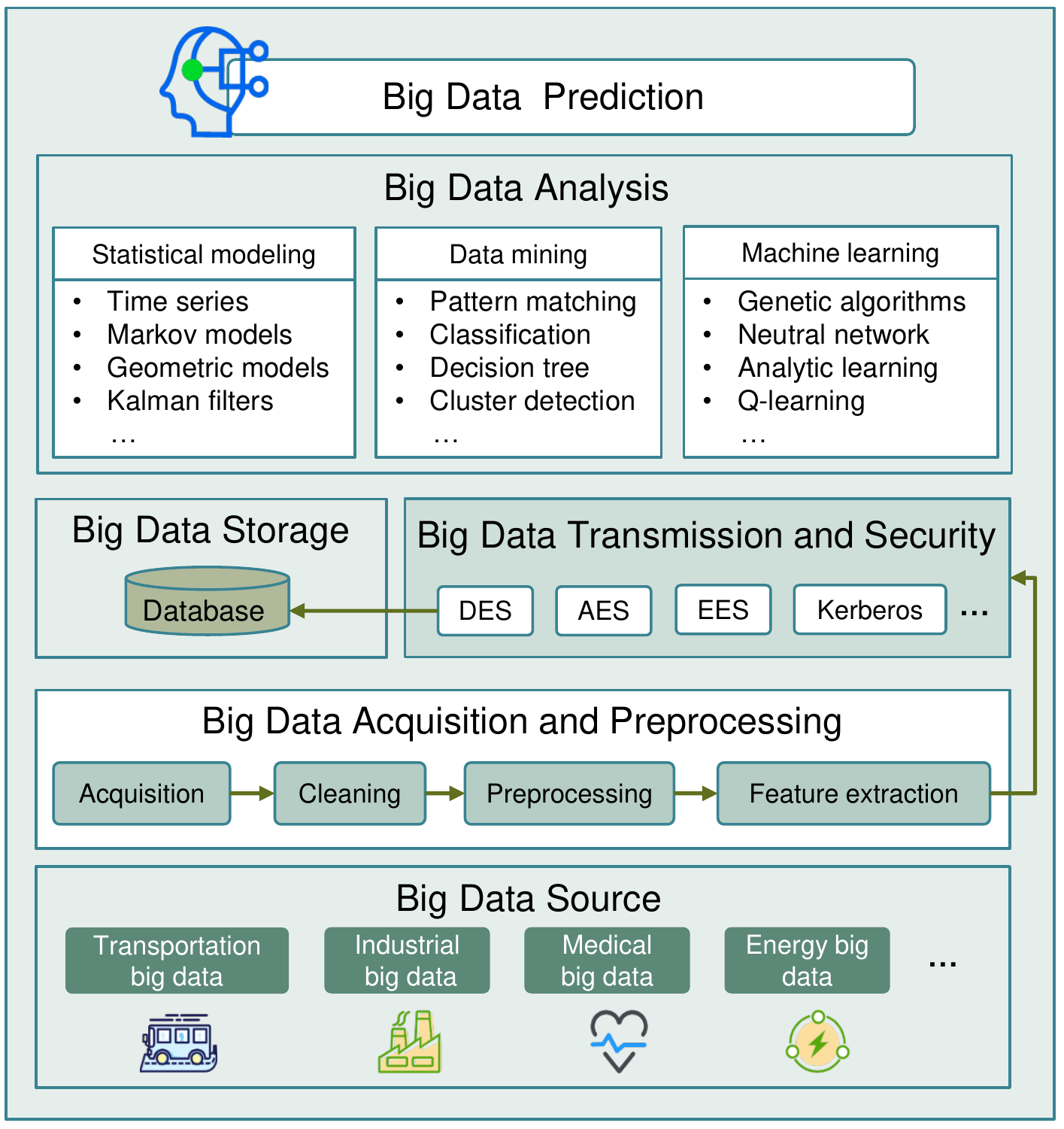}
	\caption{Big data analytics in the Internet of intelligence.}
	\label{Fig_BigData}
\end{figure}

Specific technologies and analysis methods are needed to transform the information assets represented by big data into value \cite{7892976}. Big data in the Internet of intelligence requires better knowledge representation, more profound domain knowledge, and more explicit analysis of system functions to support more informed decision-making \cite{li2016learning,obitko2013big}. Traditional big data analytics technologies may not meet the requirements of the Internet of intelligence. Fig. \ref{Fig_BigData} illustrates the big data analytics in the Internet of intelligence, and the main data processing steps are discussed in the following.

\begin{itemize} 
\item Data Acquisition and Preprocessing: Appropriate data is essential because the analysis cannot be performed without data. At this stage, a large amount of data is collected through masses of sensors and actuators deployed in cities, transportation, energy, hospitals, factories and other departments in the Internet of intelligence. Then, data cleaning and preprocessing are applied to improve data quality and extract important features.

\item Data Transmission and Security: The security and reliability of data transmission need to be ensured during data transmission to prevent data from being sniffed by third parties. Many encryption methods such as advanced encryption standard (AES), data encryption standard (DES), escrowed encryption standard (EES), and authentication methods such as Kerberos authentication protocol \cite{miller1988kerberos} can be applied to solve insecure communication problems in the Internet of intelligence.

\item Data Storage: The extracted feature needed to be stored as data sets. The final and intermediate results of the Internet of intelligence big data and indexes can be stored based on NoSQL databases such as MongoDB, Neo4j, FlockDB, etc \cite{mathew2015novel}. In addition, various storage mechanisms (including HDFS, HBase, and Hive) can be used to store and maintain processing results \cite{vora2011hadoop}. This is also based on a NoSQL database.

\item Data Analysis: Up to the minute analytical tools such as stochastic modeling, data mining and ML can be used for big data analysis in the Internet of intelligence. For example, estimating the intensity of content requests requires cluster analysis and probabilistic preference inference on social-related data \cite{8535091}.

\item Data Prediction: The prediction needs to refer to human intelligence, which establishes meaningful connections between different data sets for specific purposes based on the results obtained from the original data.
\end{itemize}

\subsection{Application Layer}
Digital twins and various immersive technologies such as VR, AR, and mixed reality (MR) are effective enabling technologies at the application layer of the Internet of intelligence.

\subsubsection{Digital Twin}

The digital twin has become an expected enabling technology at the application layer to promote the digital conversion and intelligent evolution of the Internet of intelligence. 
Digital twins build high-fidelity digital-virtual models of physical objects to simulate their behaviors and describe their operating states, thereby providing a solution for realizing cyber-physical integration \cite{negri2017review,9372842}. Advanced computing and analysis functions in cyberspace have also opened up a bright perspective for it.

\begin{figure}[t]
	\centering
	\includegraphics[width=3.2in]{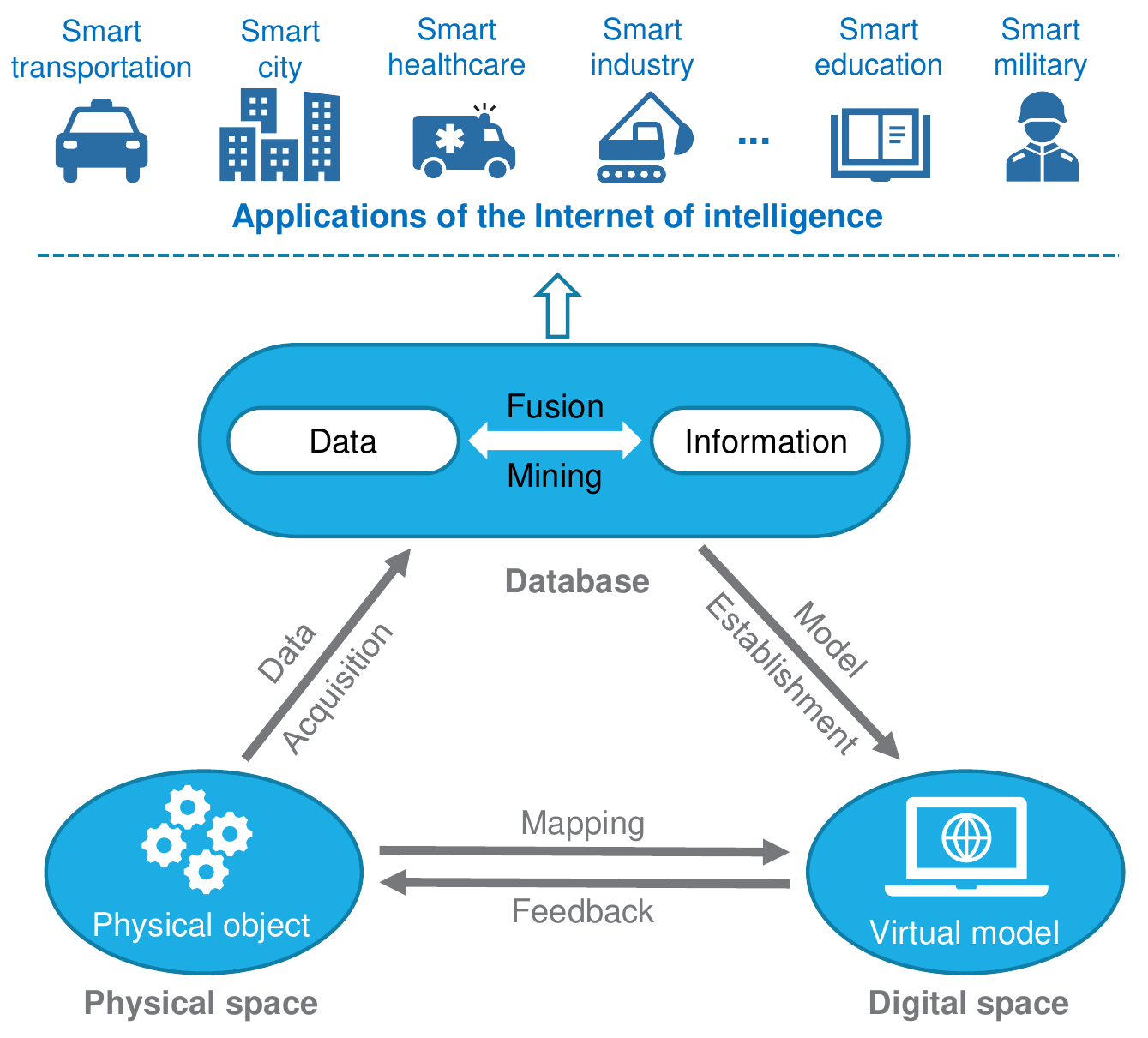}
	\caption{Digital twin in the Internet of intelligence.}
	\label{Fig_DigitalTwin}
\end{figure}

The basic principle of the digital twin is illustrated in Fig. \ref{Fig_DigitalTwin}. In the physical space, sensors and monitors collect data from physical objects in real time and store it in the database. 
Through data mining and various fusion processing, useful information can be extracted, and deeper intelligence can be exploited to dynamically construct virtual models in digital space. 
The interaction between reality and virtuality is realized by forwarding mapping to the virtual and backward reactions to physical objects. Virtual models will generate the corresponding target results according to the different requirements of various applications and then feedback the matching policies to the physical space \cite{9356524}.

Digital twin technology has three main characteristics: i) Full life cycle: the digital twin should run through the entire life cycle of the physical object, and can even go beyond its life cycle and persist in the digital space; ii) Real-time/quasi-real-time: the physical object and the twin are not completely independent, and a completely real-time or quasi-real-time mapping relationship should be established; iii) Bidirectional: the data flow between entity object and twin is bidirectional.
Based on the idea of digital twin, a series of successful applications in the Internet of intelligence have been demonstrated in areas such as smart cities, manufacturing, and electric power. It has implemented various services, including 3D visualization, intelligent manufacturing, life cycle management, state prediction, and so forth \cite{9120192,9345490,9502678}.
In general, through the digital twin technology, the Internet of intelligence can obtain massive data and build a powerful digital twin, thereby simulating the operation rules of entity objects and helping to put them into practice in the real world. Furthermore, the digital twin can also be integrated with other application layer technologies to better serve human beings.

\subsubsection{Immersive technology}

Immersive technologies such as VR, AR, and MR allow people to experience remote scenes immersively without leaving home, which have become critical enablers in the application layer of the Internet of intelligence.
VR can present ultra-high-resolution video content through a display close to human eyes or arrange a multi-projection environment in specially designed rooms to provide users with immersive scene feeling \cite{9112752}. Moreover, by detecting the user's posture and movement combined with wearable sensors or external sensors, real-time interaction between human and virtual worlds can be realized by VR to obtain deep immersion \cite{8197481}.

AR technology is dedicated to integrating the digital world into people's perception of the real world. Reality is the main focus in AR. Virtual information is presented in reality, and real and virtual objects are combined and run simultaneously \cite{7435333,6915663,7999155}. The synchronized actions of real and virtual objects can provide users with an appropriate depth of perception. For example, displaying text and image information on special lenses helps users obtain current environmental information in a timely manner. Furthermore, in addition to visual information, it can also provide perceptual information such as hearing, touch, and smell. 

MR combines VR and AR to allows people to interact more naturally with the virtual world depend on the information collected by the Internet of intelligence \cite{9133121}.
In general, the ability of VR creates the next generation of immersive experiences, while MR and AR are expected to improve the user experience. The essential difference between VR, AR, and MR lies in the proportion of mixing digital content with reality \cite{8329628}. 
AR focuses on real elements, while MR pays more attention to virtual elements.
AR or MR devices do not need to cover the surrounding world but will cover the digital layer to the user's current view. Moreover, to ``feel real'', 3D models of the environment need to be constructed by AR or MR systems to place the virtual object in the correct position and deal with the occlusion. In contrast, VR refers to a complete virtual simulation experience. The only connection with the external real world in VR is the input from the VR system to the user's senses, which helps to increase the credibility of the illusion living in the virtual replicated location.

Immersive technologies with the Internet of intelligence provide great contributions to facilitate indispensable services such as smart education, telemedicine and entertainment activities. Besides, with the expansion of the Internet of intelligence, immersive technologies will become destructive technologies and can be used in more interesting ways to improve the interaction between humans and machines in smart environments.

\renewcommand{\arraystretch}{1.5}
\begin{table*}[!htb]
	\caption{Summary of Enabling Technologies for the Internet of intelligence}
	\begin{center}
		\begin{tabular}{m{1.6cm}<{\centering}|m{1.80cm}<{\centering}|m{2.5cm}<{\centering}|m{3.9cm}<{\centering}|m{4.1cm}<{\centering}|m{1.5cm}<{\centering}} 
			\hline
			\hline
			\textbf{Layer} & \textbf{Technology} & \textbf{Description} & \textbf{Advantages} & \textbf{Potential role in the Internet of intelligence} & \textbf{References}\\ 
			\hline	
			\multirow{7}{*}{\shortstack{Physical \\ resources \\ layer}} & Communication & 5G; 6G &
			\begin{itemize}[leftmargin=*,parsep=\parskip,topsep=2pt]
				\item Ultra-high data rate
				\item Ultra-low latency
				\item Ultra-high reliability
				\item Seamless connection
				\vspace{-1.35em}
			\end{itemize}
			& Provide ultra-high QoS and QoE for communication services & [51]-[65] \\
			\cline{2-6}
			& Caching & CDN; P2P; ICN & 
			\begin{itemize}[leftmargin=*,parsep=\parskip,topsep=2pt]
				\item Data traffic reduction
				\item Low backhaul usage
				\item Low latency
				\item Load balancing
				\vspace{-1.35em}
			\end{itemize}
			& Achieve the reuse of cached content; Provide information distribution and content retrieval capabilities & [66]-[70] \\
			\cline{2-6}
			& Computing & Cloud computing; Cloudlet; Fog computing; MEC &
			\begin{itemize}[leftmargin=*,parsep=\parskip,topsep=2pt]
				\item Real-time computation
				\item Energy efficiency
				\item High QoS
				\vspace{-1.35em}
			\end{itemize}
			& Provide rich and nearby computing resources to the Internet of intelligence & [71]-[83]  \\
			\cline{2-6}
			& Sensing & RFID; WSNs; Crowdsensing &
			\begin{itemize}[leftmargin=*,parsep=\parskip,topsep=2pt]
				\item Real-time and pervasive sensing
				\item High efficiency and accuracy
				\item Low cost
				\vspace{-1.35em}
			\end{itemize}
			& Enable real-time and pervasive sensing of the Internet of intelligence environment & [84]-[95]  \\
			\cline{2-6}
			\hline	
			\multirow{10}{*}{\shortstack{Resources \\ virtualization \\ layer}} & Software-defined networking & - &
			\begin{itemize}[leftmargin=*,parsep=\parskip,topsep=2pt]
				\item Control and data plane isolation
				\item Great flexibility
				\item High programmability
				\vspace{-1.35em}
			\end{itemize}
			& Support cost-effective and dynamic network configurations & [96]-[100] \\
			\cline{2-6}
			& Network functions virtualization & -&
			\begin{itemize}[leftmargin=*,parsep=\parskip,topsep=2pt]
				\item High service flexibility
				\item Strong scalability
				\item Flexible migration
				\vspace{-1.35em}
			\end{itemize}
			& Provide dynamic management and orchestration; Optimize network functions and resources & [101]-[107]  \\
			\cline{2-6}
			& Network slicing & - &
			\begin{itemize}[leftmargin=*,parsep=\parskip,topsep=2pt]
				\item Flexible provisioning of network resources
				\item Multi-tenant environment
				\item High security and privacy
				\vspace{-1.35em}
			\end{itemize}
			& Support flexible provisioning of network resources and dynamic allocation of network functions & [108]-[115]  \\
			\cline{2-6}
			& Containerization & - &
			\begin{itemize}[leftmargin=*,parsep=\parskip,topsep=2pt]
				\item Lightweight
				\item Rapid deployment
				\item High security
				\vspace{-1.35em}
			\end{itemize}
			& Provide a lightweight virtualization solution for the portable operation of services & [103], [116]-[120] \\
			\cline{2-6}
			\hline	
			\multirow{6}{*}{\shortstack{Information \\ layer}} & Naming & - &
			\begin{itemize}[leftmargin=*,parsep=\parskip,topsep=2pt]
				\item Location independence
				\item Unique retrieval
				\vspace{-1.35em}
			\end{itemize}
			& Provide content-based names instead of locations to route nodes & [49], [125]-[132] \\
			\cline{2-6}
			& Routing, caching and transport & Routing mechanism; Caching mechanism; Transport mechanism & 
			\begin{itemize}[leftmargin=*,parsep=\parskip,topsep=2pt]
				\item Location independence
				\item High efficiency
				\vspace{-1.35em}
			\end{itemize}
			& Provide efficient routing, caching and transport for information & [133]-[137]\\
			\cline{2-6}
			& Identity resolution & - &
			\begin{itemize}[leftmargin=*,parsep=\parskip,topsep=2pt]
				\item Efficient management
				\item Convenience
				\vspace{-1.35em}
			\end{itemize}
			& Provide unified identification, addressing and understanding for intelligence & [138]-[141]  \\
			\cline{2-6}
			\hline	
			\multirow{6}{*}{\shortstack{Intelligence \\ layer}} & Artificial intelligence& Symbolic AI; Connectionist AI; Hybrid AI &
			\begin{itemize}[leftmargin=*,parsep=\parskip,topsep=2pt]
				\item Intelligent computing
				\item High efficiency
				\item Labor saving
				\vspace{-1.35em}
			\end{itemize}
			& Extract intelligence from various information sources & [142]-[150] \\
			\cline{2-6}
			& Blockchain & - &
			\begin{itemize}[leftmargin=*,parsep=\parskip,topsep=2pt]
				\item Decentralized intelligence
				\item Trusted decision-making
				\item Incentives
				\vspace{-1.35em}
			\end{itemize}
			& Provide a secure and reliable intelligence transaction guarantee for intelligent participants & [5], [22], [151]-[156] \\
			\cline{2-6}
			& Big data analytics & - & 
			\begin{itemize}[leftmargin=*,parsep=\parskip,topsep=2pt]
				\item Massive data processing and analysis
				\vspace{-1.35em}
			\end{itemize}
			& Extract the knowledge behind the collected information & [157]-[164] \\
			\cline{2-6}
			\hline	
			\multirow{4}{*}{\shortstack{Application \\ layer}} & Digital twin & - &
			\begin{itemize}[leftmargin=*,parsep=\parskip,topsep=2pt]
				\item Digitization
				\item Efficiency improvement
				\vspace{-1.35em}
			\end{itemize}
			&Achieve cyber-physical integration and promote digital transformation& [165]-[170] \\
			\cline{2-6}
			& Immersive technologies & VR; AR; MR &
			\begin{itemize}[leftmargin=*,parsep=\parskip,topsep=2pt]
				\item High QoE
				\item Convenience
				\vspace{-1.35em}
			\end{itemize}
			&Improve the interaction between humans and machines in intelligent environments& [171]-[177] \\
			\hline
			\hline
		\end{tabular}
	\end{center}
	\label{tb1}
\end{table*}

\subsection{Lessons Learned: Summary and Insights}

This section discusses some of the enabling technologies that support the Internet of intelligence. These enabling technologies are briefly summarized in Table \ref{tb1}. Key lessons learned from the discussion are as follows.

\begin{itemize} 
	
	\item Enabling technologies for the physical resources layer include various communication, caching, computing, and sensing technologies.
	In terms of communication, 5G/6G can support high QoS, high QoE and high reliability of communication and networking of the Internet of intelligence.
	In terms of caching, CDN and P2P can be applied to effectively improve content distribution and retrieval capabilities of the Internet of intelligence.
	In terms of computing, cloud computing and edge computing technologies such as cloudlets, fog computing, and MEC are effective enablers to respond to the intensive computing challenges posed by diverse Internet of intelligence services.
	In terms of sensing, RFID, WSN and crowdsensing can be used to achieve pervasive sensing through identification, positioning, detection and imaging, providing data support for the Internet of intelligence.
	
	\item Enabling technologies for the resources virtualization layer include SDN, NFV, network slicing and containerization, which enable the flexible utilization of heterogeneous and ubiquitous multidimensional resources in the Internet of intelligence.
	SDN can effectively support cost-effective dynamic network configuration for the Internet of intelligence thanks to its high flexibility and programmability.
	NFV demonstrates excellent performance in achieving service flexibility, scalability and migration for the Internet of intelligence through a software-based approach to managing core network functions.
	In combination with SDN and NFV, network slicing enables service customization by dynamically adjusting network resources between slices.
	Besides, the use of containers provides a lightweight virtualization solution for the portable operation of Internet of intelligence services.
	
	\item The information layer infers useful information from massive data and moves it to the intelligence layer. In this layer, the information itself is more valuable than its source because it describes intelligent events. 
	ICN can help build an information/content-centric  architecture for this layer. Naming, routing, caching and transport are key technologies for the information layer based on ICN. AI-assisted approaches can be applied to further improve their performance. Another core technology of this layer is identity resolution. By constructing identity resolution systems, the convenient management of the network can be effectively achieved, and the interconnection of the Internet of Intelligence can be promoted.

	\item Key enabling technologies for the intelligence layer include AI, blockchain and big data analytics that can be used to extract the intelligence behind massive information.
	AI for the Internet of intelligence combines current symbolic and connectionist AI approaches to make AI computing interpretable, highly generalizable, and robust.
	Blockchain can effectively realize the good characteristics of intelligence networking, thereby providing efficient and reliable guarantees for the large-scale application and deployment of the Internet of intelligence.
	Big data analytics is promising to extract the knowledge behind the collected information. Big data in the Internet of Intelligence requires better knowledge representation, more profound domain knowledge, and more explicit analysis of system functions to support more informed decisions.
	
	\item Effective enabling technologies for the application layer include digital twins and various immersive technologies to provide higher quality services for diverse Internet of intelligence applications.
	Digital twin provides solutions to enable cyber-physical convergence, facilitating digital transformation and intelligence for the Internet of intelligence.
	Immersive technologies such as AR, VR and MR improve human-machine interaction in the Internet of intelligence environment in a more interesting way.

\end{itemize}

\section{Applications Scenarios \label{sect_6}}

In this section, we provide an overview of the primary application scenarios and their integration with the Internet of intelligence paradigm.

\subsection{Smart Computer/Communication Network}

As already introduced in Section \ref{sect_5}, computing and communication are key enabling technologies for the physical resources layer of the Internet of intelligence.
As the Internet of intelligence flourishes, it will in turn support the development of 5G/6G and the future Internet, promoting intelligence in computer/communication networks.
By distributing intelligence and federated learning at the edge of smart computer/communication networks, the Internet of intelligence can break through the limitations of users and data in a single network node or a single domain to improve the versatility and scalability of intelligence models \cite{9651548}.
Meanwhile, the paradigm of intelligence networking can also enable intelligence to be effectively stored, delivered, and shared, thus effectively alleviating information redundancy within the smart computer/communication network.
In addition, the security and privacy issues of smart computer/communication networks can also be effectively enhanced through the application of blockchain technology, thus facilitating intelligence sharing, decentralized intelligence, and trust in decision-making \cite{9448012}.
The smart computer/communication network with the Internet of intelligence has the following characteristics.

\subsubsection{Joint Optimized Computing Efficiency and Accuracy}
In smart computer/communication networks, users inevitably generate similar ML tasks that may require the same type of data and even expect the same training results \cite{9091106}. Since the computing resources of smart computer/communication networks are limited, these large numbers of repeated model training lead to a large number of redundant computations in the network and cause a significant waste of resources.
In addition, the small sample size of data also causes overfitting of the model. As a result, ineffective and meaningless model training is prevalent, leading to low training efficiency and training accuracy.
Currently, many works have proposed solutions to efficiently utilize the limited resources of smart computer/communication networks to further improve model training efficiency or QoS, such as data sharing \cite{9051991,9454587}, gradient compression \cite{8852172,wen2017terngrad}, adaptive global aggregation \cite{8664630}, and so forth. Data sharing can effectively improve the training accuracy of the model and thus providing better QoS. However, it requires the computation of a large number of data samples, and the transmission of shared data also incurs high transmission costs. Gradient compression and adaptive global aggregation can be effective in ensuring learning efficiency but at the expense of training accuracy.
Innovative solutions to jointly optimize these two metrics are urgently needed.
Fortunately, the Internet of intelligence can provide an effective alternative solution.
Distributed computing nodes in smart computer/communication networks can efficiently use their computing resources for various types of model training and share intelligence through the Internet of intelligence during the training process.
In this way, training efficiency and training accuracy can be effectively improved without sampling and training a large amount of data, thereby reducing information redundancy in the network.

\subsubsection{Global Domain Intelligence}
Recent advances in ML algorithms have accelerated the deployment of AI in smart computer/communication networks \cite{9618666,8712527,8926369}, yet most of them use AI techniques to solve specific network problems or to optimize single-user or limited-scale systems.
The rapid growth of smart computer/communication networks requires AI to permeate every corner of the network, from network management to network services \cite{9627726}. This is extremely challenging because smart computer/communication networks will be highly heterogeneous, with network nodes having different communication, computing, caching and sensing resources. 
Meanwhile, network dynamics, such as time-varying channel conditions and spatio-temporal service requirements between network nodes, further complicate decision-making problems \cite{9651548}.
Internet of intelligence will effectively help solve these challenges through intelligence networking.
For network problems with different characteristics and different decision time scales in smart computer/communication networks, appropriate AI techniques can be selected to solve them. Besides, the cooperation between intelligent modules is important for the efficient and flexible implementation of smart computer/communication networks.
The Internet of intelligence enables cooperative decision-making and network control by networking the intelligence among intelligent modules.
In this way, AI models and network resources located at the edge of the network are connected through a distributed architecture that connects AI algorithms to support an ever-expanding AI services, thereby effectively realizing global domain intelligence and global domain network optimization of smart computer/communication networks.

\subsubsection{Improved Network Security and User Privacy}
AI is already an integral part of smart computer/communication networks for analyzing large amounts of data.
Massive data collection makes the privacy issues worthy of attention.
Moreover, AI algorithms lack fairness and transparency. The large amount of training data and its own vulnerabilities make it a prime target for attackers seeking to violate privacy in the age of smart computer/communication networks.
The Internet of intelligence can effectively improve network security and user privacy \cite{9146540}.
Using distributed computing technologies such as edge computing, the Internet of intelligence can sink the computation and training of AI models from the centralized cloud to the edge of smart computer/communication networks, thereby effectively reducing the risk of cloud leakage and providing a degree of privacy protection through local data storage \cite{9479786,peng2021edge}. 
In the distributed computing framework, FL has attracted widespread interest in computer/communication fields, which effectively mitigate data privacy breaches by avoiding data uploads during centralized training \cite{9562560,8994206,9567690}.
Meanwhile, through the application of blockchain technology, the Internet of Intelligence can also make the specific process of ML decision-making more transparent. The blockchain and its ledger can record all data and variables in the ML process and track the reasons for ML decisions, thus improving  the credibility of ML decisions. In addition, blockchain not only facilitates the development of ML in terms of data collection, but also plays an important role in security and privacy protection. Chen \emph{et al.} in \cite{8622598} develop a decentralized blockchain-based privacy-preserving and secure ML system. Shen \emph{et al.} in \cite{8653362} propose a privacy-preserving IoT data training scheme that ensures the privacy of data providers and the confidentiality of ML models by using support vector machines (SVM), blockchain, and homomorphic encryption.

\begin{figure*}[t]
	\centering
	\includegraphics[width=6.5in]{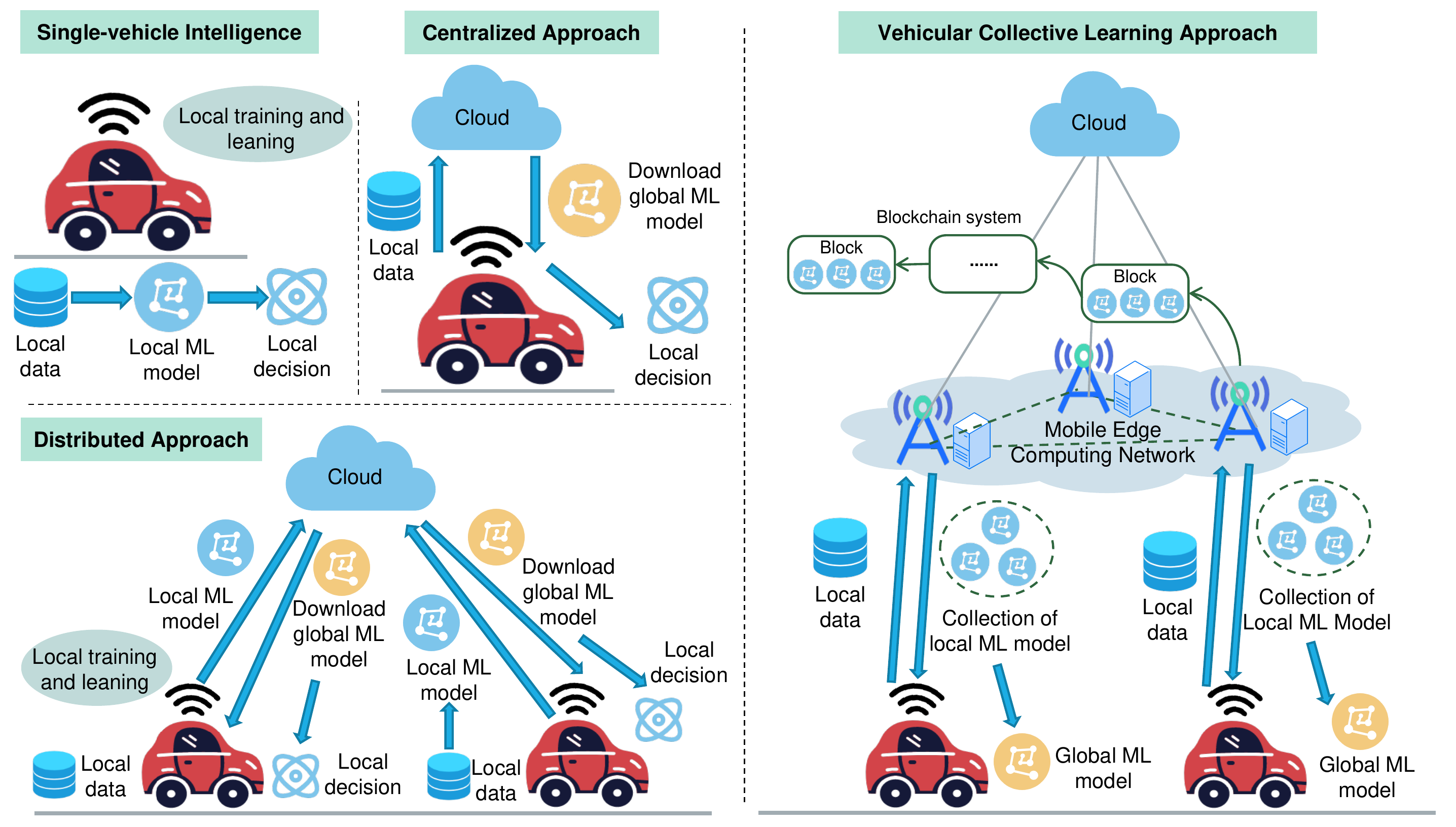}
	\caption{Comparison of single-vehicle intelligence, centralized approach, distributed approach and BCL approach of autonomous driving.}
	\label{Fig_Transportation_1}
\end{figure*}

\subsection{Smart Transportation/Autonomous Driving}

Smart transportation integrates people, vehicles and road infrastructure into a large and rigorous system that manages urban traffic by collecting real-time vehicle and road information and controls vehicle driving mode through end-to-end joint decision-making. The Internet of intelligence can accelerate the transition of smart transportation from connected vehicles to fully automated driving \cite{8032474}. Autonomous driving is a technology that integrates AI, visual computing, radar, monitoring equipment, and positioning systems \cite{9264235}. It not only relies on efficient and real-time data computing but also requires real-time access to road infrastructure and Internet data. Consequently, autonomous driving relies on the entire network to coordinate all vehicles and road infrastructure, rather than being limited by the computing and sensing capabilities of any single vehicle. In autonomous driving, simple information processing and forwarding mechanisms are no longer sufficient. The decision model is based on the environment and is constructed from a larger set of vehicular information. In such cases, the model may reduce the load and delay of autonomous driving since the key content is extracted from a larger set of input information. Therefore, the Internet of intelligence plays a vital role in autonomous driving to obtain intelligence by processing, analyzing and understanding the collected information. Moreover, intelligence sharing between vehicles can provide more accurate location awareness and higher communication efficiency. Following we provide several fields of autonomous driving systems that can be benefited from the Internet of intelligence.

\subsubsection{Vehicular Collective Learning}
In recent years, the application of AI technology to autonomous driving has attracted extensive attention from academia and industry. In an AI-enabled autonomous driving system, the accuracy and efficiency of ML models directly affect the performance of CAVs. To obtain more accurate ML models, CAVs need to collect a large amount of data from onboard sensors \cite{6823640}. However, with the increasing complexity of the onboard sensor system, its security is still unable to be absolutely guaranteed. For example, fatal accidents occurred in Tesla and Uber's CAVs \cite{banks2018driver}.
Currently, there are three approaches to realize AI-enabled autonomous driving, i.e., single-vehicle intelligence approach, centralized approach, and distributed approach (federated learning) \cite{liu2017computer,zeng2021federated}. Due to the characteristics of ML, current autonomous driving systems based on these three approaches face challenges in terms of a single model, independent learning, and privacy and security issues. The major reason for such challenges is ``the lack of trust mechanism and collective intelligence''.
Toward this end, a blockchain-based collective learning (BCL) framework based on the Internet of intelligence is presented in \cite{9003305}. Inspired by the idea of collective intelligence, each autonomous vehicle can share the learned local model as intelligence to improve the efficiency and accuracy of ML. Moreover, adopting the blockchain system can make the intelligence sharing process in collective learning secure, automatic, and transparent. The comparison between the BCL approach and the traditional approaches is shown in Fig. \ref{Fig_Transportation_1}.

\subsubsection{Consumer-Centric Experience}
By analyzing drivers' characteristics and behaviors, such as age, gender, driving style, driving experience, and accident history, the Internet of intelligence can help the autonomous driving system better understand different driving modes and provide a customized experience for drivers. The customization of CAVs is highly dependent on the intelligence of human behavior \cite{8457076}. For instance, the speed of driving is closely related to human characteristics such as age, gender, preference, and emotion. Young drivers tend to drive faster than older ones, and female drivers usually drive more carefully than male drivers. Moreover, some people like to drive on less busy roads, even if it takes longer to reach the destination. Autonomous driving should be able to be customized for different drivers for various scenarios \cite{li2016human,Tesla,8924619}. The Internet of intelligence will pave the way for passenger-centric autonomous driving applications. The intelligence model will be trained based on a large amount of onboard preference information, such as the passenger's background and preferences for speed, in-car entertainment, risk level, etc. The trained intelligence models can be stored in vehicular users, updated in real-time, and shared with other users through the Internet of intelligence. Driven by the Internet of intelligence, human characteristics and motivation will play significant roles in the driving experience.

\subsubsection{Vehicular Security and Privacy}
The dynamic vehicle-to-vehicle (V2V) communication in untrusted vehicle environments and the dependence on centralized network authorization pose security challenges for autonomous driving.
The Internet of intelligence can help build a secure, trustworthy, and decentralized autonomous driving ecosystem. Blockchain with high security can enhance the security and service quality in autonomous driving \cite{9094168}. First, the blockchain can create a secure P2P network to achieve seamless communication between ubiquitous vehicles for model sharing, collaborative trust, and service management \cite{8705921}.
A blockchain-enabled model sharing approach is recently proposed to improve the cross-domain adaptive object detection performance of autonomous driving \cite{8963705}.
Based on blockchain and MEC, a domain-adaptive you-only-look-once (YOLOv2) model is trained across nodes to significantly reduce the domain difference of different object categories, and smart contracts are developed to efficiently perform data storage and model sharing.
In addition, due to the time-varying, delayed and location-related characteristics of autonomous driving, it is a considerable challenge to motivate CAVs to spread intelligence in the Internet of intelligence. Meanwhile, attackers may not only spread false information to confuse the network, but may also bring security vulnerabilities and privacy issues to CAVs. The immutable ledger, cryptocurrency, and asymmetric encryption of the blockchain also promote secure vehicular services and play an essential role in ensuring the security of intelligence transactions and reputation values.
For instance, the authors in \cite{9048619} proposed a blockchain framework for the security and incentives of vehicular content delivery. The reputation model is designed to assess the credibility of CAVs and road side units (RSUs) while incentivizing participants to act honestly. 

\subsection{Smart Industry}

Smart industry refers to the integration of intelligence into the manufacturing process, which is an application of the Internet of intelligence in the manufacturing field.
The Internet of intelligence integrates many advanced communication and automation technologies in smart industry, such as AI, machine-to-machine (M2M) communication and big data analytics, in order to improve its intelligence and connectivity \cite{6714496,zhang2021deep}.
For example, the Internet of intelligence can connect all industrial information and processes on the factory floor and forward them to the industrial cloud data center.
Then, decision makers can utilize this information to create an intelligent and accurate view of the manufacturing process, thereby enhancing their ability to make smarter decisions.
The Internet of intelligence also helps to develop and implement novel smart technologies to accelerate the innovation and transformation of the factory workforce \cite{lasi2014industry}.
For example, the way of intelligence networking can make industrial machines have higher precision, higher efficiency and continuous working ability, thereby saving the cost of the manufacturing process \cite{7004894}.
Therefore, the Internet of intelligence has significant potential in improving overall efficiency, quality control and sustainability.
Then, we introduced several Internet of intelligence implementation cases in the smart industry.

\subsubsection{Industry 4.0}
Industry 4.0 is a new concept in the industrial revolution, which aims to facilitate the automation of manufacturing and industrial operations \cite{lu2017industry}.
Distributed intelligent Industry 4.0 systems can be effectively supported by the Internet of intelligence.
For example, a new type of intrusion detection system based on the Internet of things (IoT) Industry 4.0 collaborative learning model was proposed in \cite{9120761}. 
This model allows the smart filter installed at the IoT gateway to learn information from others by exchanging trained models, significantly improving the accuracy of detecting intrusions and reducing network traffic and information leakage.
In addition, blockchain can also be applied in the Industry 4.0 to provide security for the implementation of intelligence learning. The authors in \cite{9134967} jointly use FL and blockchain to develop a new type of direct-to-consumer (D2C) platform to improve the computational processing efficiency of Industry 4.0 smart manufacturing.
FL can effectively solve the privacy and efficiency issues of Industry 4.0, and blockchain promotes FL for poisoning attack functions by adopting advanced selection methods.

\subsubsection{Robotics}
Robotic can handle manufacturing tasks through automation and programmable capabilities and is an important component of industrial systems. Nowadays, a lot of work has been done on robotics in smart factories. The feasibility of robotics in collecting and processing sensor data of autonomous vehicle navigation systems in the industrial environment is discussed in \cite{8371642,7384376,cardarelli2017cooperative}. The ubiquitous product management, customer maintenance and material handling in the smart factory also adopt this concept. One of the critical challenges of the smart industry is the real-time data processing and data privacy of robotic systems. The Internet of intelligence can effectively cope with these challenges by distributing intelligence to robotic equipment without relying on remote servers for data processing. Robots can perform AI training locally. 
Then, each robot only need to upload gradient parameters to construct a shared model in the cloud without sharing its raw data based on differential privacy technology, thereby effectively reducing network transmission delay.
For instance, a federated imitation learning scheme for cloud robotics is proposed in \cite{9013081,8772088}.
Each robot uploads the model parameters to the cloud for intelligent fusion, and then the fused parameters are returned to the robot for the next round of training.
Through the rounds of training, the cloud can accumulate the vital intelligence of different robots and build accurate intelligence models, and the robots can also benefit from the exchanged intelligence.

\begin{figure}[t]
	\centering
	\includegraphics[width=2.9in]{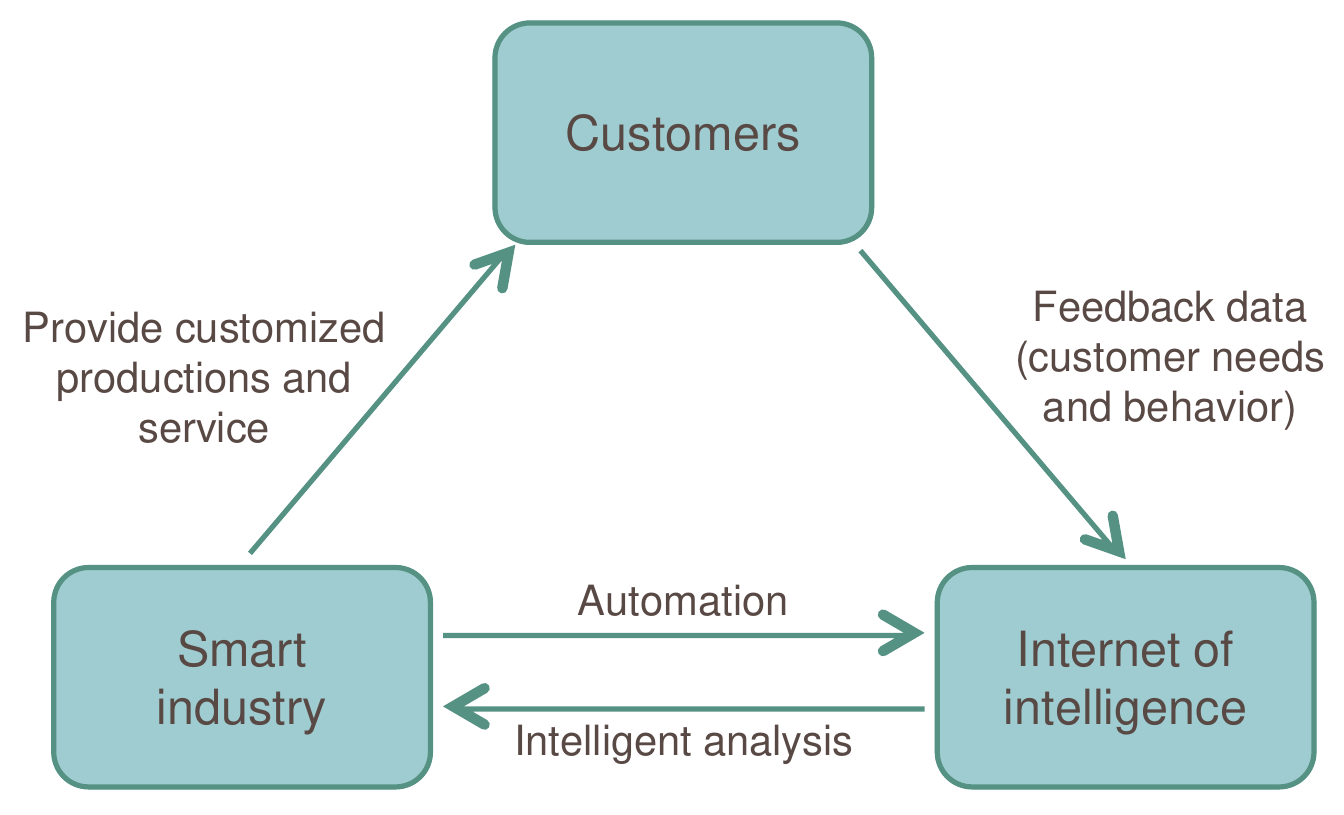}
	\caption{Customized production driven by the Internet of intelligence.}
	\label{Fig_SmartIndustry}
\end{figure}

\subsubsection{Smart Manufacturing}

Smart manufacturing is a broad manufacturing category that benefits the manufacturing industry by adopting cloud manufacturing, IoT technologies, and service-oriented manufacturing. Currently, smart manufacturing is facing challenges in centralized industrial networks and third-party-based permissions, such as flexibility, efficiency, and security.
With the support of distributed AI, blockchain, and edge/cloud computing, the Internet of intelligence has the potential to cope with these challenges in manufacturing systems \cite{zhang2021optimizing}.
The Internet of intelligence can effectively enhance and optimize the manufacturing process and realize intelligent information analysis.
Recently, the authors in \cite{banerjee2017generating} developed a knowledge graph-based digital twin model, which can infer intelligence from large-scale production line data and improve manufacturing process management through semantic relational reasoning.
Based on the blockchain, the authors in \cite{8789508} proposed ManuChain, which is a new iterative bi-level hybrid intelligence model named, to eliminate the imbalance/inconsistency between overall planning and local execution in a personalized manufacturing system. 


\subsubsection{Customized Production}
Advanced industrial intelligence service embodies the integration of automation and intelligent technologies, which can create an intelligence-oriented end-to-end production process to fulfill the requirements of different enterprises and customers \cite{wan2020artificial}.
The Internet of intelligence can realize networked, intelligent and customized industrial design, production and circulation. Employees, design agencies, factories, warehouses and supply chains are integrated into an intelligence networking system, thereby reducing logistics and management costs, and improving industrial production efficiency and profits. As shown in Fig. \ref{Fig_SmartIndustry}, with the support of the Internet of intelligence, the interconnection between people and machines, machines and machines, and services and services can be formed to achieve a high degree of horizontal, vertical, and end-to-end integration to meet the personalized customization needs of users. 

\subsection{Smart Cities}

Smart cities have become a new paradigm, which can dynamically exploit resources in cities from ubiquitous devices to provide citizens with a wide range of services \cite{9105931}. Smart cities involve various components, including ubiquitous intelligent devices, a variety of heterogeneous networks, and data centers with large-scale storage and powerful computing capabilities. Despite the potential vision of smart cities, how to provide intelligent, secure and efficient services for smart cities is still a challenging issue \cite{9163375}. In such cases, by using attractive technologies such as AI, blockchain, digital twin and so forth, the Internet of intelligence can become a promising candidate for empowering smart city services. The Internet of intelligence allows intelligence transmission rather than information transmission in smart cities, thereby providing safer and more effective services and making smart cities smarter. Compared with traditional smart cities, smart cities with the Internet of intelligence has the following characteristics:

\begin{figure}[t]
	\centering
	\includegraphics[width=3.2in]{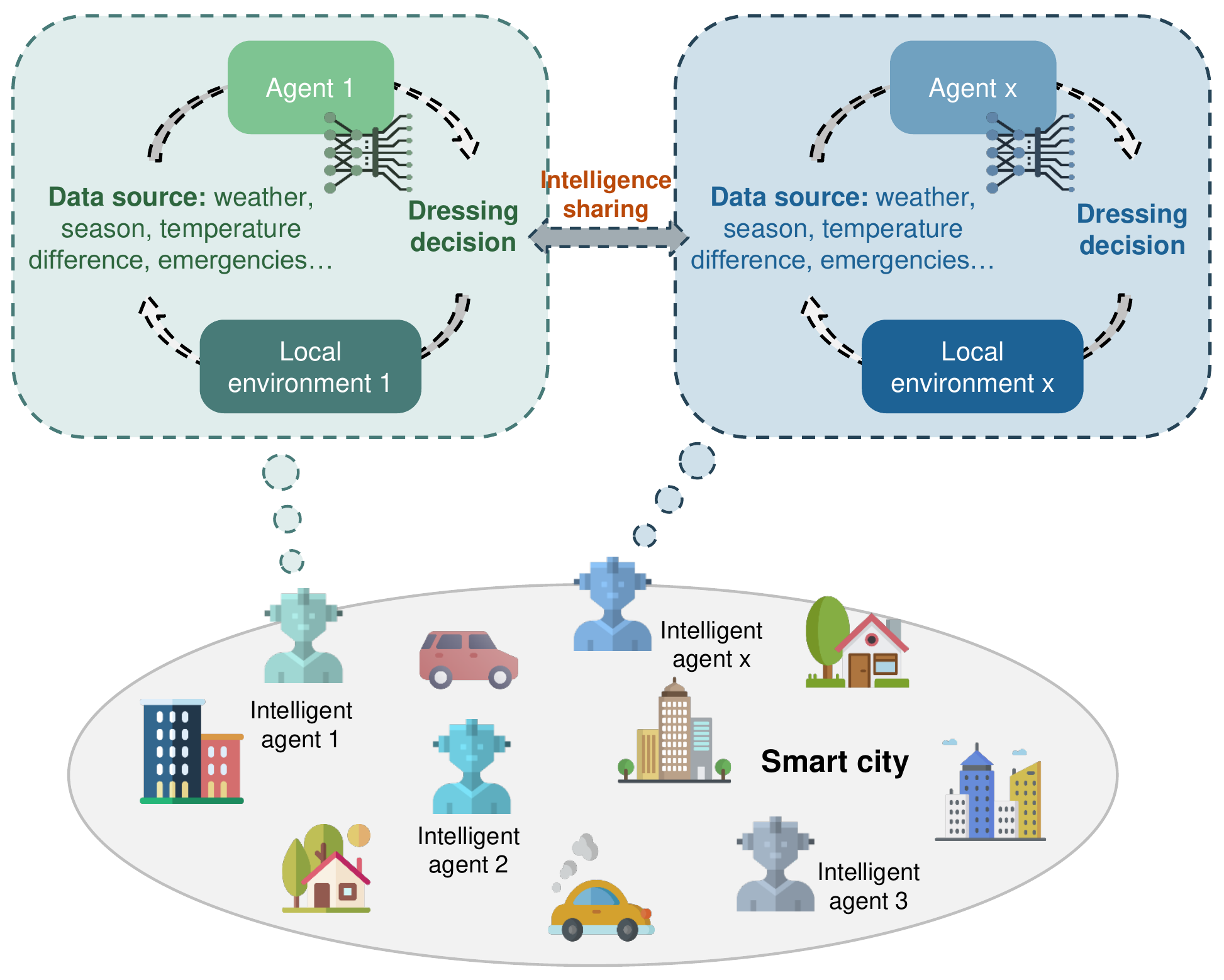}
	\caption{Smart city driven by the Internet of intelligence.}
	\label{Fig_SmartCity}
\end{figure}

\subsubsection{Efficient Data Analytics}
Hundreds of millions of intelligent devices and sensors in every corner of cities can automatically collect and monitor various data in real-time, including traffic, weather, energy, water consumption, shopping information, air quality, and so on \cite{7123563}. Through data aggregation, transmission, storage, organization, and analysis, what is happening and what may happen in the future can be understood.
However, to cope with such a huge amount of data, the current data processing of smart cities is of poor efficiency. In such a city-scale deployment of smart services, data is generated at a high rate, and only a small part of this massive data is typically used to improve the lives of urban residents \cite{8291121}.
Compared with the traditional information networking smart cities, the data processing efficiency can be greatly improved in intelligence networking smart cities through intelligence sharing.
Here, we take the way of dressing as an example.
When people migrate from one city to another, new ways of dressing need to be established according to the local weather. For traditional smart cities, ML may be used to learn the new environment. 
Based on the collected weather and dressing data, a new way of dressing can be established through a lot of learning and training. Humans are different from machines. When migrating to a new city, they will quickly obtain the correct way of dressing through the experience shared by others rather than random attempts. Therefore, as illustrated in Fig. \ref{Fig_SmartCity}, in a smart city driven by the Internet of intelligence, it is no longer to spend a lot of time, material and financial resources to re-training, but to refer to the way of human intelligence and establish the correct way of dressing through the intelligence sharing between intelligent agents.

\subsubsection{Improved Quality of Experience}
The Internet of intelligence plays a key role in improving the QoE of users in the digital and physical world of smart cities. Smart cities driven by the Internet of intelligence can connect and integrate physical environment and real-time events into a virtual system based on AI. Based on digital twins and various immersive technologies, the Internet of intelligence can integrate the physical world with the digital world, map the real subject to the virtual space, and gather all online and offline participants, so as to provide technical support for the construction of smart digital city scenes such as smart community, smart building and smart work area.
The realization of such a smart city system faces many challenges in terms of computing, storage and communication \cite{9430902}. The Internet of intelligence can make full use of intelligence to coordinate computing, caching, and transmission systems to achieve a ubiquitous immersive experience. The access network is built into the Internet of intelligence, and various big data analytics and intelligent algorithms will be deployed at the edge of the smart city network to provide users with a real-time and high-quality experience.

\subsubsection{Enhanced Security}
Ubiquitous data services have brought security challenges such as privacy, integrity, and trust to smart cities \cite{8447209}. The Internet of intelligence integrates blockchain to provide advanced security services, thus playing an important role in realizing secure and high-performance smart city scenarios. By constructing a decentralized security architecture, blockchain has the potential to provide authenticity, confidentiality, integrity and non-repudiation for smart cities \cite{9179702}. Moreover, with its high security, the blockchain proves the high efficiency of the Internet of intelligence in controlling the operation of smart cities in a distributed security way.

\subsection{Smart Healthcare}
Healthcare is another field that benefits from the Internet of intelligence. In terms of medical services, the Internet of intelligence can effectively integrate doctors, patients, hospitals and regulatory agencies to provide reliable customized and precise health management services. Medical and health services can continue to provide high-quality services throughout their life cycle.
Under the premise of satisfying privacy protection, the Internet of intelligence can store, protect, retrieve, analyze and share intelligence from patients, doctors, medical equipment, etc. Medical resources can also be shared between doctors and hospitals. Doctors or medical equipment can check patients remotely, and all shared intelligence can be retrieved from local or remote servers. As a result, the boundaries between hospitals and hospitals, and between regions and regions can be effectively eliminated, realizing the sharing of medical resources.
To sum up, in medical and health services enabled by the Internet of intelligence, the physical boundaries of hospitals and the geographic boundaries of cities can be eliminated, and the work efficiency of doctors and hospitals can be significantly improved, thereby helping to change health management, treatment and hospital management.
Some application domains in smart healthcare that can benefit from the Internet of intelligence technologies are summarized as follows.

\subsubsection{Doctor Recommendation}
Adopting medical appointment platforms in hospitals may pose challenges for patients in choosing the right doctor online. Specifically, people may not be able to understand who is the right doctor to solve their health problems without professional knowledge and experience \cite{9402991,7007848}. A smarter doctor recommendation system can be constructed based on the Internet of intelligence to cope with such issues. Patients can share the evaluations of the doctors they make appointments on the doctor recommendation system. Driven by the Internet of intelligence, the smart doctor recommendation system converts information collected from patients, doctor appointment platforms, patient comments, etc. into intelligence, and presents it to patients to achieve efficient doctor recommendations.

\begin{figure}[t]
	\centering
	\includegraphics[width=3.3in]{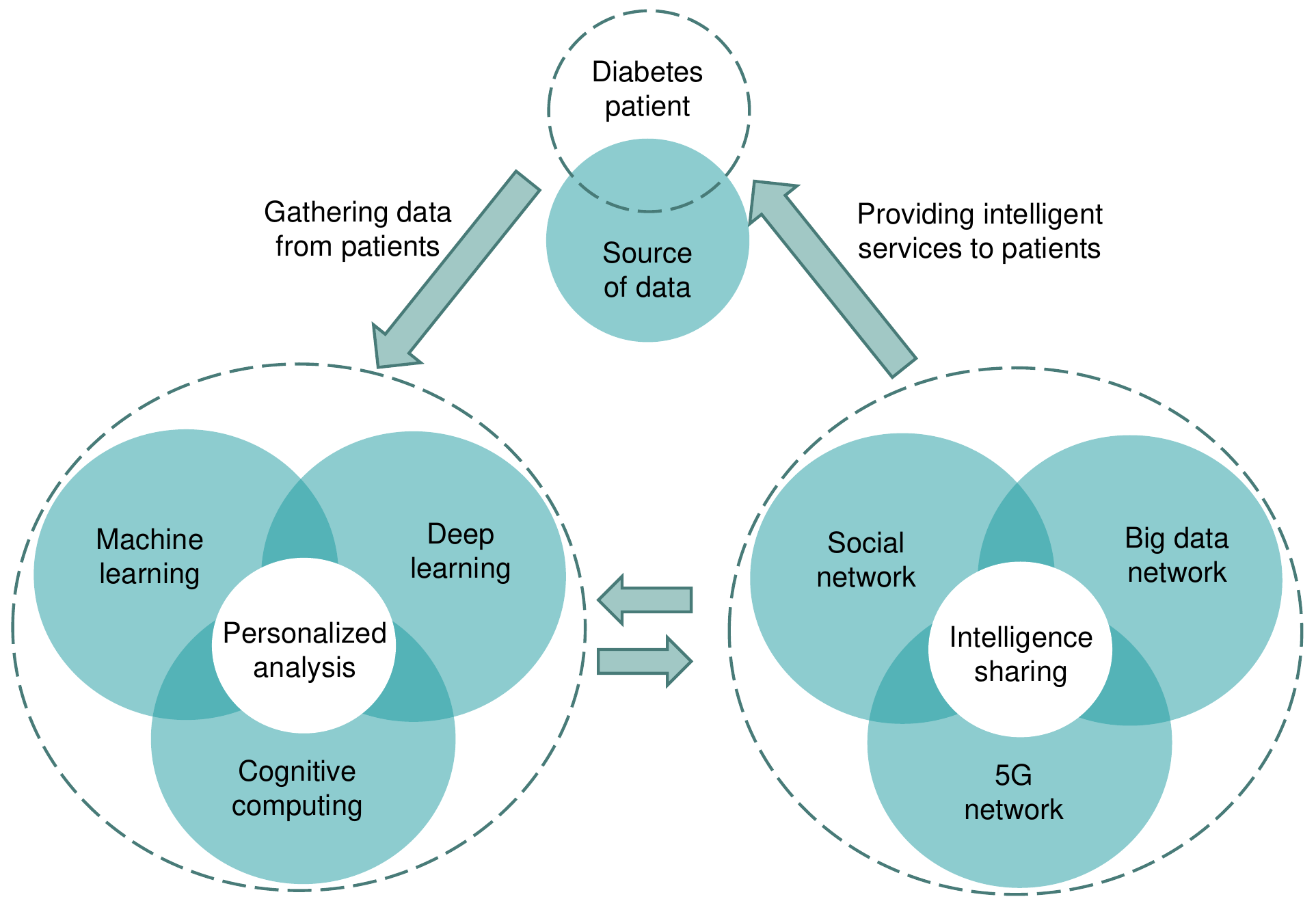}
	\caption{Intelligence sharing and personalized analysis model for Internet of intelligence-enabled diabetes.}
	\label{Fig_SmartHealthcare}
\end{figure}

\subsubsection{Personalized Diagnosis}
Based on the collected medical data and personalized physiological indicators, the Internet of intelligence can use various ML and cognitive computing algorithms to establish personalized diagnosis plans, so as to provide patients with personalized treatments to prevent and treat diseases \cite{schork2019artificial}. Taking diabetes as an example, through the collection, storage and analysis of personal diabetes information, the Internet of intelligence can extract the intelligence from the information and adjust treatment strategies in time according to changes in the patient's condition, as shown in Fig. \ref{Fig_SmartHealthcare}. In addition, effective intelligence sharing among patients, relatives, friends, personal health consultants, and doctors is established through the Internet of intelligence to maintain sustainability in diagnosing and treating intelligence-driven diabetes \cite{8337890}. Particularly, the Internet of intelligence mainly provides personalized diagnosis to patients in two ways. On the one hand, the Internet of intelligence enables patients to maintain a healthy lifestyle to prevent them from suffering from the disease at an early stage. On the other hand, the Internet of intelligence promotes out-of-hospital treatment, which can effectively reduce the cost of long-term hospitalization for patients. 

\subsubsection{Epidemic Prediction}
\begin{figure}[t]
	\centering
	\includegraphics[width=3.0in]{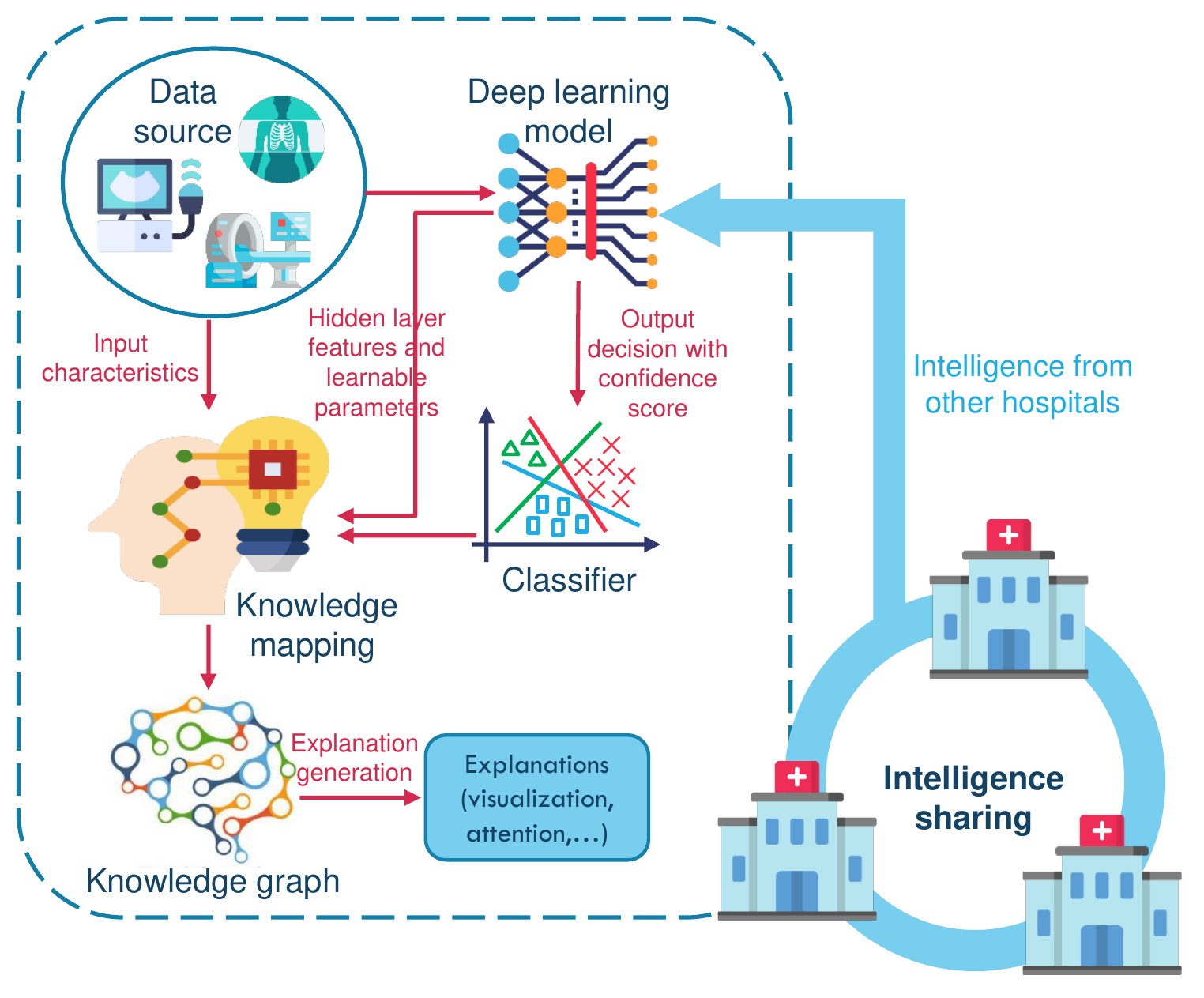}
	\caption{The process of the Internet of intelligence to combat COVID-19-like pandemics.}
	\label{Fig_SmartHealthcare2}
\end{figure}

A significant challenge for the current epidemic monitoring and control system is the need for a large amount of pathological, radiographic, genetic and other types of epidemic-related information to be absorbed, stored and processed. In the event of a sudden outbreak of an epidemic, such as COVID-19, it is challenging to conduct many tests manually on each test object due to time constraints \cite{9136589}. The Internet of intelligence will play an important role in such crises. In the Internet of intelligence, the hospital data center acts as an intelligent agent to train and establish a smart prediction system according to different types of physiological signals and hospital test signals, including chest X-ray images, computed tomography (CT) scanning images, protease sequences, eye surface images, cough sounds, body temperature, blood pressure, etc \cite{9136600}. Then, hospital data centers can extract intelligence about COVID-19 from the trained model, preserve the intelligence and share it with other hospital data centers. In this way, the Internet of intelligence can effectively process massive epidemic data and predict real-time epidemic crises. Fig. \ref{Fig_SmartHealthcare2} illustrates the process of the Internet of intelligence to combat COVID-19-like pandemics. Furthermore, the Internet of intelligence can help healthcare providers manage patients remotely. By allowing professionals outside the shock center to diagnose patients remotely, the tremendous pressure on front-line nursing staff can be reduced.

\subsection{Smart Education}
\begin{figure}[t]
	\centering
	\includegraphics[width=3.2in]{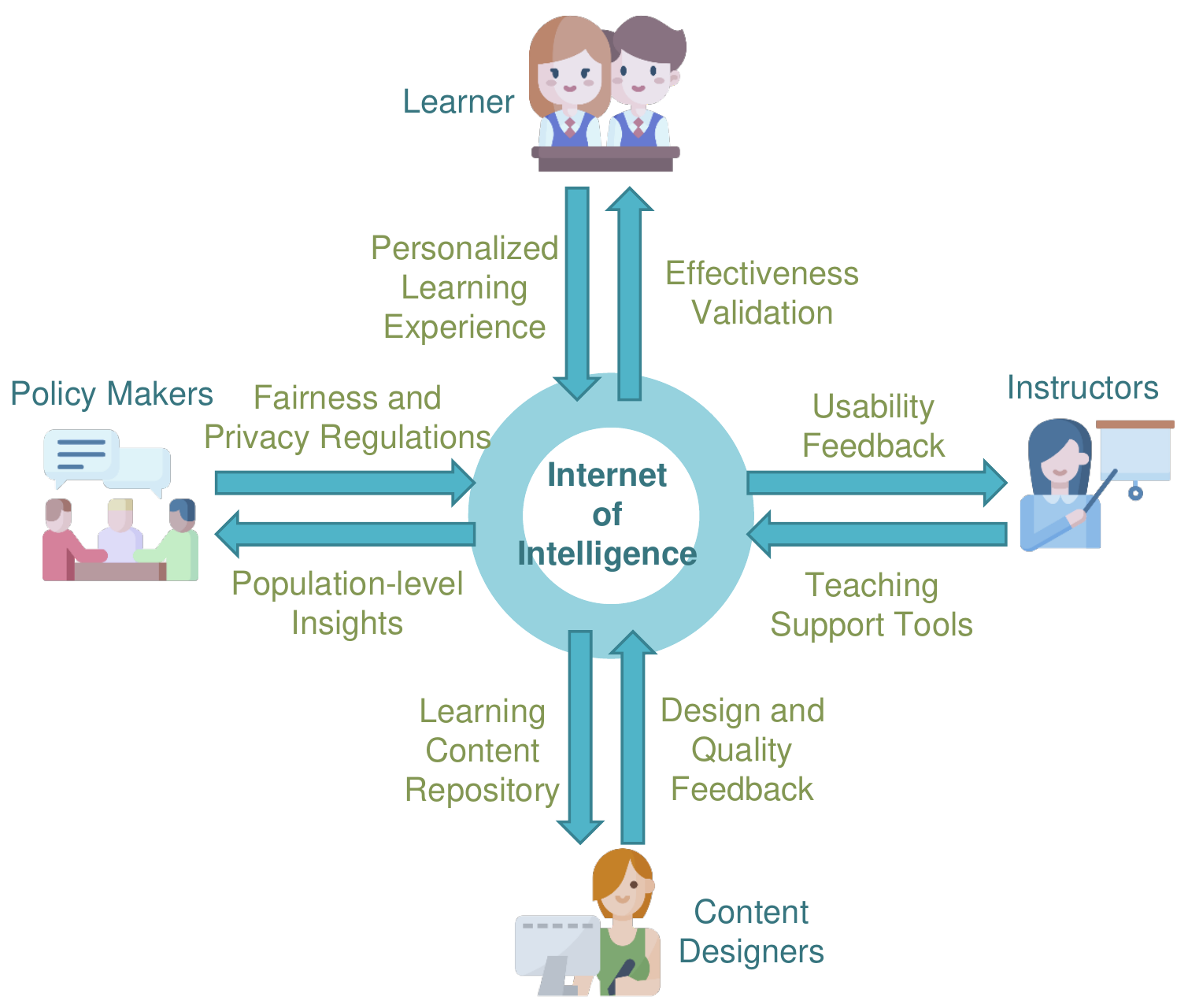}
	\caption{The baseline ecosystem of IoI-empowered smart education.}
	\label{Fig_SmartEducation}
\end{figure}

Traditional education requires more resources in terms of educational space, schedule, and human resources, making it easy to fail even with minor changes in conditions. For example, under unusual circumstances, such as the COVID-19 outbreak or natural disasters, traditional education may stagnate \cite{xie2020covid}. Therefore, emerging alternatives are inevitable. Enabling smart education is one of the most valuable values of the Internet of intelligence relative to education. The Internet of intelligence integrates students, teachers, schools, training institutions, universities, libraries and all learning resources, and has the potential for a revolutionary transformation from traditional education to modern concepts. 
Furthermore, by adapting to each learner's unique characteristics and expectations, such as background, goals, personality and talents, it greatly enhances the quality of education in several aspects.
Fig. \ref{Fig_SmartEducation} shows the baseline ecosystem of intelligence-empowered smart education. Several application domains of smart education can be benefited, as summarized below.

\subsubsection{Personalized Education}
\begin{figure}[t]
	\centering
	\includegraphics[width=3.2in]{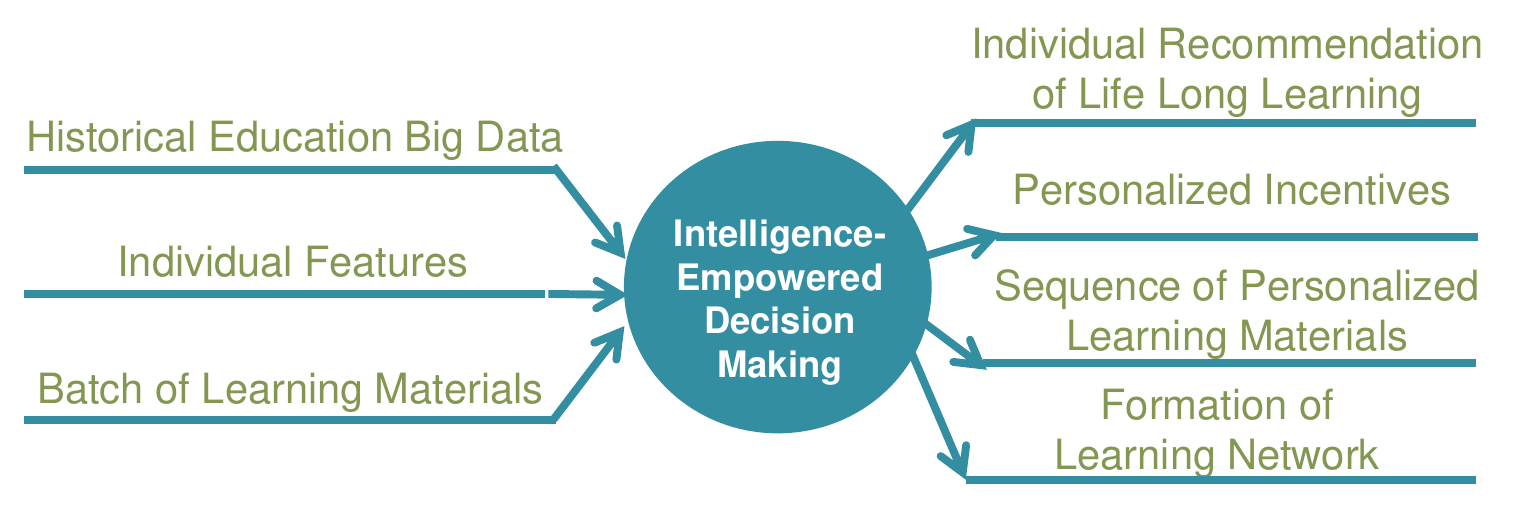}
	\caption{The basic concept of IoI-empowered personalized education.}
	\label{Fig_SmartEducation_2}
\end{figure}

Personalized education aims to design a compelling knowledge acquisition trajectory to achieve learners' expected goals by matching their strengths and bypassing their weaknesses \cite{9418572}. The Internet of intelligence introduces new perspectives and enhances personalized education by integrating advanced technologies such as AI and big data analytics technologies into smart education. Leveraging the advantages of online tools and personal tutoring, a new and more flexible type of learning technology has been created to adapt to students' learning and allocate resources on demand. As shown in Fig. \ref{Fig_SmartEducation_2}, smart education supported by the Internet of intelligence consists of a great quantity of decision-making strategies, which map available data and personal characteristics to various personalized educational materials and suggestions in the form of intelligence.

\subsubsection{Lifelong Learning}
There is a significant gap between the curriculum provided by schools and the job requirements. For example, soft skills such as communication and teamwork may be more important than some technical skills. Future research on lifelong learning may fill this gap. However, there are also some challenges that need to be resolved \cite{7524023}: i) Large decision-making space that increases as the number of courses increases; ii) The situation of attending multiple courses at the same time brings excellent flexibility; iii) Any static course is not the best because the knowledge, experience and performance of students are constantly developing and evolving during the learning process; iv) The backgrounds, knowledge and goals of students vary greatly. The Internet of intelligence can effectively combine the student's school records with possible long-term tracking of future employment results and other data sources (such as course outlines and job postings). 
Then, from the above educational information set, intelligence can be extracted, stored, shared and recreated to identify key skills that extend from the classroom to the profession. Educators can use this intelligence to dynamically adjust school courses and activities, so as to make fuller preparations for students' future.

\subsubsection{Diversity and Fairness Promotion}

In smart education, AI-driven personalization, such as student evaluation, feedback, and content recommendation, can effectively improve comprehensive learning achievement. However, the bias in the training data may lead to an imbalance in training results \cite{reich2017good}. This imbalance will jeopardize those underserved students since they usually do not have access to advanced digital education systems and rarely appear in the data sets collected by these systems \cite{doroudi2019fairer}. Intelligence can be extracted from the fairness characteristics between learners of different backgrounds by the Internet of intelligence, making smart education more inclusive for future generations. Furthermore, an intelligence-driven education ecosystem can be built, and a set of regulations can be formulated around intelligence ownership, sharing, continuous performance monitoring and verification issues to ensure the quality and diversity of the collected information. Meanwhile, algorithms with performance guarantees across different educational environments should be developed to identify misuse and implement fail-safe mechanisms.

\subsection{Smart Grid/Energy}
Smart grids are replacing traditional grids to provide consumers with reliable and efficient energy services. Smart grids introduced distributed energy generators to improve the utilization of distributed energy, introduced smart meters and two-way communication networks to realize customers and utility providers, and introduced electric vehicles to improve energy storage capacity and reduce carbon dioxide emissions. In this way, smart grids have demonstrated outstanding reliability, efficiency, security and interactivity \cite{7004067,6197400}.
The Internet of intelligence can effectively help smart grids achieve better performance in predictive dispatch, fault diagnosis, and network attack detection. 
Load forecasting is an important achievement of smart grid system. 
It forecasts the short-term, medium-term and long-term demand for electricity. 
By identifying better power distribution among consumers or customers (homes, industries, and corporate offices), power distribution companies will particularly benefit from smart grids.
Currently, different power utilization data is gathered together. Deep learning can use these data for load forecasting so that that power distributors can predict the power demand of users for power distribution.
However, the vast amount of data poses a challenge to the calculation time of deep learning. Through distributed intelligent sharing, the Internet of intelligence can effectively reduce ML training and computing time and improve feature selection.
For example, in \cite{9269337}, the authors proposed a distributed intelligent learning model for smart grid, which uses Hilbert Schmidt independence criterion (HSIC) bottleneck technology to provide higher precision learning, thus reducing the calculation time and cost in the case of limited data.
In addition, the Internet of intelligence can effectively manage its load scheduling while protecting the privacy of residents. Inspired by FL, the authors in \cite{9133518} proposed a distributed deep reinforcement learning approach, in which the action network is located in the distributed household, and the critic network is located in the aggregator from the trusted third party.
The security of smart grids under network attacks has aroused widespread concern. In order to improve the security of smart grids, the Internet of intelligence can detect DDoS attacks by analyzing the patterns of network data. Through the automatic analysis and detection method, it can respond in time, take preventive measures, greatly reduce the damage, and improve prediction accuracy.

In the energy field, the Internet of intelligence energy services can effectively complete the integration of power grids, thermal power grids and fuel grids, and realize the integration and complementarity of wind, solar and other energy sources. Energy Internet (also referred to as Smart grid 2.0) aims to become a smart grid with ultra-high voltage (UHV) grid as the backbone, variable clean energy transmission as the leader, and interconnection as the main feature \cite{yapa2021can}. Through the integration of the energy industry and intelligent technologies, comprehensive management of energy production, storage, transmission and consumption can be effectively realized \cite{huang2020application}. The Internet of intelligence can transform the traditional centralized AI into large-scale distributed intelligence networking systems, realizing self-organization, self-evolution and real-time intelligence. Distributed AI provides more accurate online real-time data training models to achieve smarter scheduling and control. Deterministic low latency and flexible network resource configuration guarantee a more real-time system response.
Meanwhile, the Internet of intelligence can effectively deal with energy inefficiency or energy failures in the energy Internet. For instance, by connecting the energy market and consumers, the Internet of intelligence can realize smart management of hybrid power energy consumption and production. When consumers have their own production capacity, the Internet of intelligence realizes two-way power transactions based on real-time electricity prices. That is, consumers sell their surplus electricity to the energy market at a satisfactory price during the peak period of electricity consumption, and they can buy the electricity in the market at a lower price during the low period of electricity consumption. This will not only reduce energy waste and promote the prosperity of the free energy trading market but also help promote clean energy to replace traditional energy.


\subsection{Smart Agriculture}

Under the influence of variable climate and other environmental driving factors, traditional agriculture faces the challenge of producing, storing and distributing nutritious and safe food from local to a global scale with higher efficiency.
The Internet of intelligence has the potential to transform agriculture and help meet these challenges.
Smart agriculture in the Internet of intelligence can integrate various emerging technologies to realize precision agriculture, visual management and intelligent decision-making in agricultural production. In addition, by connecting consumers, agricultural experts, farmers, distributors and other parties together, the efficiency and accuracy of agricultural production and circulation are greatly improved.
One of the challenging tasks for agriculture is to estimate the number of crops that will be produced in various regions before actual production. Changes in climate, weather or extreme events can have a negative impact on crop yields and lead to unpredictable losses and yields \cite{9098014}.
AI can help agriculture provide accurate and effective methods for crop nutrient management under these variable conditions and help capture the downstream effects of runoff and nutrient loss \cite{dharmaraj2018,bannerjee2018}.
Data on climate, soil and biological variables represent key processes in agroecosystems distributed across many geographic areas. Through AI and other analysis tools, large data sets can be integrated and analyzed in a long time and large space to obtain useful agriculture information and new insights \cite{talaviya2020}.
However, the use of AI technology also brings challenges of uncertainty, because the drivers of the local agroecosystem and the agroecosystem dynamics themselves may not reflect the past drivers.
These drivers exhibit intra- and inter-annual changes and changes in direction over time, and they can also interact with the food system that has its own feedback.
Through intelligence sharing between agricultural enterprises, various agricultural information can be fully utilized or integrated, and local, farm or ranch-scale information can be connected with regional to global Internet of intelligence.
Then, the global Internet of intelligence integrates AI technologies with transdisciplinary expert knowledge and harmonized information of diverse variables across large, spatially heterogeneous areas through time to address these complex agroecosystem problems.

\subsection{Smart Military}

Millions of sensors deployed on multiple military applications are used to provide services for command, control, communications, intelligence, surveillance, and reconnaissance systems \cite{9382385}. The situational awareness obtained through these sensors can help commanders conduct command and control, and help warfighters conduct strategic battles \cite{zheng2015leveraging}.
These data collected from sensors cannot be used directly but need to be manipulated, processed, and analyzed to obtain useful military information from them, bringing great challenges to the current military network. Manual analysis or even semi-automatic analysis to extract strategic knowledge is no longer a feasible solution since it requires a lot of human resources \cite{svenmarck2018possibilities}. Nowadays, the scale of threats is huge, and different hybrid security solutions are needed. Integrating the Internet of intelligence with technologies such as M2M communications, embedded systems, wireless sensor networks, edge/cloud computing, and mobile applications opened a new perspective for providing such solutions with human-grade intelligence and accuracy.
In the smart military scenario, the collected military information needs to be processed and analyzed before the military intelligence can be extracted \cite{cho2020priority}. Military intelligence can be used to visualize the battlefield situation, grasp the enemy's intentions and capabilities, identify the enemy's center, and support accurate future predictions to successfully execute military operations. The training was carried out both on public information and on specially developed information. Moreover, by constructing distributed smart military network systems in different military regions, real-time intelligence can be realized to provide more accurate online real-time training models. In general, the Internet of intelligence can effectively promote smart military by effectively utilizing military data, increasing decision-making speed, improving cyber defense, and reducing labor and costs.

\renewcommand{\arraystretch}{1.5}
\begin{table*}[!htb]
	\caption{Summary of Primary Application Scenarios in the Internet of Intelligence}
	\begin{center}
		\begin{tabular}{m{1.7cm}<{\centering}|m{2.4cm}<{\centering}|m{7.0cm}<{\centering}|m{3.4cm}<{\centering}|m{1.4cm}<{\centering}} 
			\hline
			\hline
			\textbf{Application} & \textbf{Use Case} & \textbf{Internet of Intelligence-based Solution } & \textbf{Advantages} & \textbf{References}\\ 
			\hline	
			\multirow{6}{*}{\shortstack{Smart\\ computer\\/communication\\ network}} &Joint optimized computing efficiency and accuracy& Intelligence sharing effectively improves the training efficiency and training accuracy without sampling and training large amounts of data & Improved efficiency and accuracy of ML; Reduced information redundancy& [179]-[184] \\
			\cline{2-5}
			& Global domain intelligence & Distributed intelligence and resources of intelligent modules are networked to enable global domain intelligence & Cooperative decision making; Global network optimization  & [21], [178], [185]-[187]\\
			\cline{2-5}
			& Improved network security and user privacy & Edge computing, blockchain and other technologies make ML decisions more decentralized and transparent to improve network security and user privacy & Decentralized and transparent decision making; Enhanced security & [16], [75], [188]-[193] \\
			\hline
			\multirow{6}{*}{\shortstack{Smart\\ transportation\\ /autonomous\\driving}} &Vehicular collective learning& The BCL framework is adopted so that each autonomous vehicle can share the learned local model as intelligence, and the blockchain is used to ensure the sharing process & Improved efficiency and accuracy of ML; Secure, automatic and transparent information sharing process & [14], [196]-[199] \\
			\cline{2-5}
			& Consumer-centric experience & Human characteristics are considered in the training, and the trained model can be stored, updated, and shared & High efficiency; Customized in-vehicle experience & [200]-[203] \\
			\cline{2-5}
			& Vehicular security and privacy & The blockchain is used to create a secure P2P network to achieve seamless communication between vehicles for service management, model sharing, and collaborative trust & Secure, trustworthy and decentralized decision making;
			High service quality & [204]-[207]\\
			\hline	
			\multirow{7}{*}{\shortstack{Smart \\ industry}} & Industry 4.0 & Collaborative learning and blockchain can be used to improve the efficiency and security of Industry 4.0 systems & High efficiency; Enhanced privacy
			& [212]-[214] \\
			\cline{2-5}
			& Robotics & Intelligence is assigned to robots instead of relying on remote servers for data processing & Strong security; High efficiency & [215]-[219] \\
			\cline{2-5}
			& Smart manufacturing & Distributed intelligence optimizes the manufacturing process and realizes intelligent information analysis & Reduced cost; Real-time decisions; Enhanced privacy & [220]-[222] \\
			\cline{2-5}
			& Customized production & The intelligence-oriented end-to-end production process is built to meet the needs of different companies and customers & Satisfied customization needs; Connected industrial production & [223] \\
			\hline	
			\multirow{5}{*}{\shortstack{Smart cities}} & Efficient data analytics & Intelligent agents quickly acquire intelligence through intelligence sharing without training and learning large amounts of data & Efficient processing of massive data; High accuracy; Less required data  & [226], [227] \\
			\cline{2-5}
			& Improved quality of experience & Intelligence is used to coordinate computing, caching, and transmission to achieve ubiquitous immersive experiences & Real-time and high-quality user experience & [13] \\
			\cline{2-5}
			& Enhanced security & Blockchain is used to build decentralized security architectures for smart cities & High security;
			High efficiency & [228], [229] \\
			\hline	
			\multirow{5}{*}{\shortstack{Smart\\ healthcare}} & Doctor recommendation & Smart doctor recommendation systems are constructed to convert information collected from patients, doctor appointment platforms, etc. into intelligence & Intelligent recommendation; High convenience; High efficiency
			& [230], [231] \\
			\cline{2-5}
			& Personalized diagnosis & Personalized diagnosis plan is established through intelligence sharing of social networks & Disease prevention; Personalized diagnosis & [232], [233] \\
			\cline{2-5}
			& Epidemic prediction & Intelligence is extracted by the hospital from the massive epidemic disease information, and can be saved and shared with other hospitals & Remote diagnosis; Relieved pressure on medical staff; Less required data  & [234], [235] \\
			\hline	
			\multirow{6}{*}{\shortstack{Smart\\ education}} & Personalized education & Educational data and personal characteristics will be mapped into personalized education suggestions in the form of intelligence & Personalized education; On-demand resource allocation & [237] \\
			\cline{2-5}
			& Lifelong learning & Educational information can be extracted as intelligence for long-term storage and recreation & Optimized decision-making; Adaptive education strategy adjustment
			& [238] \\
			\cline{2-5}
			& Promoting diversity and fairness & The fairness characteristics between learners of different backgrounds will be considered as intelligence by the education ecosystem & Increased educational diversity; High QoS & [239], [240] \\
			\hline	
			\multirow{4}{*}{\shortstack{Smart grid\\ /energy}} & Smart grid & Distributed intelligence sharing improves ML training and enhances feature selection & Reliable ML; High efficiency; High accuracy & [241]-[244] \\
			\cline{2-5}
			& Smart energy & Intelligent system realizes self-organization, self-evolution and real-time intelligence of the smart energy & Green energy; Improved energy utilization & [245], [246] \\
			\hline	
			\multirow{1}{*}{\shortstack{Smart\\ agriculture}} & - & Advanced intelligent technologies are combined with interdisciplinary expert knowledge and agricultural information across time and spatially heterogeneous regions & High resource utilization; High efficiency; Intelligent decision & [247]-[250] \\
			\hline	
			\multirow{1}{*}{\shortstack{Smart\\ military}} & - & Distributed and intelligent military is constructed to realize real-time military intelligence & Efficient use of data; Improved cyber defense & [110], [251]-[253] \\
			\hline
			\hline
		\end{tabular}
	\end{center}
	\label{tb3}
\end{table*}

\subsection{Lessons Learned: Summary and Insights}
This section is dedicated to providing a vision of the Internet of intelligence and discusses in detail the huge market potential of the Internet of intelligence in various applications, thus laying the foundation for its application in a wide range of fields. 
Table \ref{tb3} provides a brief summary of Internet of intelligence-based solutions for these applications.
Key lessons learned are as follows.

\begin{itemize} 
	
	\item Computing/communication technologies can effectively enable the Internet of intelligence, and in turn the development of the Internet of intelligence can promote further intelligence in computer/communication networks. The Internet of intelligence can jointly optimize computing efficiency and accuracy through direct intelligence sharing. 
	Moreover, by networking the decentralized intelligence and resources among intelligent modules, it can effectively achieve global domain intelligence in smart computer/communication networks. 
	The application of technologies such as edge computing and blockchain also makes ML decisions in smart computer/communication networks more decentralized and transparent, significantly improving network security and user privacy.
	
	\item The Internet of intelligence plays a vital role in CAV systems. First, on-board collective learning will effectively improve the efficiency, accuracy, and security of ML. Second, human characteristics and motivation are taken into account in the Internet of intelligence-driven CAVs to achieve the ultimate consumer-centric experience. Third, the Internet of intelligence has the potential to help build a secure, trustworthy and decentralized CAV ecosystem, thereby enhancing the security of vehicular communication and the QoS of vehicular services.
	
	\item The Internet of intelligence greatly enhances the intelligence and connectivity of industrial production, and can be widely used in many fields of the smart industry. First, the Internet of intelligence can effectively support the construction of distributed and intelligent industry 4.0 systems. Second, by distributing intelligence to robotic devices, the Internet of intelligence can alleviate the robotics industry's reliance on remote servers. Third, the Internet of intelligence can enhance and optimize the manufacturing process, enabling intelligent information analysis. Fourth, the personalized customization needs of users can be met through the intelligence interconnection between people and machines, machines and machines, and services and services.
	
	\item The Internet of intelligence can greatly improve the analysis efficiency of massive data in smart cities through intelligence sharing. In addition, the Internet of intelligence integrates the physical world with the digital world, thus supporting the construction of smart digital cities. By building access networks into the Internet of intelligence, various big data analysis and intelligent algorithms will be deployed at the edge of smart cities to provide users with a real-time and high-quality experience. The application of technologies such as blockchain can also enhance the security of smart cities to provide trusted services.
	
	\item In healthcare, the Internet of intelligence can transform the information collected from patients, doctors, and doctor appointment platforms into intelligence and present it to patients for efficient doctor recommendations. Based on the collected medical data and personalized physiological indicators, the Internet of intelligence can build personalized diagnosis plans for patients through various ML and cognitive computing algorithms.
	Besides, the intelligence networking paradigm can interconnect medical intelligence to efficiently process massive epidemic data and predict epidemic crises in real time.
	
	\item In the field of education, the Internet of intelligence can significantly improve the quality of education by integrating learning resources including students, teachers and schools. First, the Internet of intelligence personalizes education by mapping personal characteristics to various educational materials and recommendations in the form of  intelligence. Second, the Internet of Intelligence combines students' school records with possible future employment results and extracts valid intelligence from them to determine progress from the classroom to the profession.
	Third, the Internet of intelligence takes into account fairness characteristics among different learners as intelligence when training for education, thus making education more inclusive for future generations.
	
	\item In the grid/energy field, on the one hand, the Internet of intelligence can help smart grid achieve better performance in predictive scheduling, fault diagnosis and network attack detection through intelligence networking and trusted sharing. On the other hand, it can transform the traditional centralized AI into a large-scale distributed and intelligent network system for self-organization, self-evolution and real-time intelligence of smart energy.
	
	\item In agriculture, the Internet of intelligence can integrate various emerging agricultural technologies to enable precision agriculture, visual management and smart decision-making in agricultural production. By connecting local, farm or ranch-scale data to a regional to global Internet of intelligence, it can also integrate a variety of agricultural data, thereby greatly improving the efficiency and accuracy of agricultural production and distribution.
	
	\item The Internet of intelligence integrates with technologies such as WSNs, embedded systems, M2M communications to enable secure and efficient processing and analysis of large-scale military data to extract military intelligence with human-level intelligence and accuracy.
	Besides, it can support the construction of distributed and intelligent military network systems to achieve online and real-time intelligence.

\end{itemize}

\section{Research Challenges and Open Issues \label{sect_7}}

Despite the potential prospects of the Internet of intelligence, some remaining challenges and open issues need to be resolved before its widespread deployment, including modeling intelligence, incentives for intelligence sharing, intelligence discovery, protocol designs, massive data, scalability, and security and privacy. In this section, we summarize these research challenges and open issues to fully exploit the benefits of the Internet of intelligence.

\subsection{Modeling Intelligence}

It is essential to model the ``things'' that are networked in each networking paradigm. For instance, information modeling and energy modeling play key roles in the Internet and grid of energy, respectively. In particular, Shannon's information theory proposes that the use of ``entropy'' to quantify information is of vital importance to the success of the Internet. Therefore, intelligence modeling plays an essential role in the success of the Internet of intelligence. Turing test is widely used to test the ability of machines to perform identical or indistinguishable intelligent behaviors as humans \cite{saygin2000turing}. However, the Turing test does not have a quantitative measure of intelligence. How to model intelligence is still an open issue. It is worth pondering and studying whether it is possible to use a measure similar to entropy to quantify intelligence.

It can be seen from the evolution of networking paradigms that a higher level of networking paradigm provides a higher level of abstraction.
For instance, energy can be quantified as the speed of matter moving, and information can be quantified as the degree of energy spreading. Similarly, intelligence can be quantified to measure the amount of information distributed over time in the learning process. The mathematical equation can be expressed as: $\mathrm{d}L=\partial S/\partial R$, where $\mathrm{d}L$ is the change of intelligence, $\partial S$ is the similarity between the current order and the expected order, and $\partial R$ is a parameter in the general sense (such as time, data volume, etc.) \cite{9448012}. This kind of quantitative intelligence measurement can effectively describe how wide the spread of information after learning compared to its previous state, and is of significant importance to the development of the Internet of intelligence.

\subsection{Incentives for Intelligence Sharing}
The Internet of intelligence can reduce costs, improve resource utilization, and build a new foundation for socio-economic systems through trusted intelligence interaction. Tracing back to the successful development of the Internet, its most striking feature is spontaneity \cite{9277903}. Platform providers such as Amazon, Google are moving towards better performance platforms. The demand for various software promotes the development of software providers such as Oracle and Microsoft. Transmission pipelines put forward demands on network operators. We can observe that incentive mechanisms and business models are the main drivers for the development of the Internet. Therefore, the Internet of intelligence must provide appropriate incentives to promote practical collective intelligence between networks.

We always expect all participants in the Internet of intelligence to voluntarily contribute and share their intelligence and use their training data stored locally. However, in real-world scenarios, participants will not be interested in participating in the training of collective intelligence unless they can obtain satisfactory rewards. For example, due to concerns about privacy leakage and consumption of physical resources, participants in the Internet of intelligence may be unwilling to participate in the sharing of intelligence. In addition, for Internet of intelligence services, participants can be service consumers and data generators at the same time. How to cleverly price and reasonably distribute income among the participants of the Internet of Intelligence ecosystem based on the amount of intelligence contributed is a problem worthy of research. Accordingly, we need to ponder why participators are willing to contribute and share their intelligence, which is the essential shift from information networking to intelligence networking. Stimulating the driving force behind intelligence sharing and exchanges is the inner motivation for the Internet of intelligence.

Currently, much work has been done on motivating participants to contribute their own resources \cite{xiao2018spectrum,jin2015auction,kang2019incentive}, data \cite{xuan2020incentive,wang2013incentive,chen2019toward}, information \cite{wang2021information,gong2019research}, etc.
Game theory \cite{xiao2018spectrum}, auction theory \cite{jin2015auction}, contract theory \cite{kang2019incentive} and reputation mechanism \cite{gong2019research} are all commonly used optimization methods.
Blockchain has also been widely adopted to guarantee reliable and trustworthy incentives \cite{chen2019toward,wang2018blockchain,weng2019deepchain}.
The incentive mechanism design for the Internet of Intelligence has not been well studied.
Intelligence is a further abstraction of information and has different characteristics.
Therefore, on the basis of referring to the existing schemes, it is still challenging to build an efficient and secure incentive system with in-depth consideration of the modeling and quantification of intelligence.

\subsection{Intelligence Discovery}
In the Internet of intelligence, some operations such as distributed intelligence, intelligence sharing, etc., can only be carried out on the premise of understanding the distributed information of intelligence. For instance, the participants in the Internet of intelligence may not know which services are available in the network they are connected to. Besides, there are many participants in the Internet of intelligence, and they may not know with which participants they should be networked or share intelligence with. Therefore, intelligence discovery is another challenge in the Internet of intelligence. Since intelligent entities are distributed in different geographic locations of the Internet of intelligence, effective intelligence discovery mechanisms are of great significance for identifying and locating intelligence.
 
What exactly needs to be discovered in the Internet of intelligence? The participants need to discover the following information: i) the availability of the intelligence and ii) the naming conventions used for invoking intelligence. The first piece of information enables participants to discover the intelligence that provides the functionality they need along with intelligence-related metadata (e.g., intelligence description, version, complexity). The second one defines how participants can invoke the intelligence, for example, what are the input parameters of the intelligence, and what are the components required to form the name of the intelligence.

The publish-subscribe mechanism derived from ICN may be able to help the implementation of intelligence discovery \cite{8960319}.
In this mechanism, intelligence publishers are only responsible for publishing intelligence, and intelligence subscribers can subscribe to intelligence published by multiple publishers without knowing the valid sources of the intelligence. Such a mode effectively eliminates dependencies among participants, thereby enhancing the scalability of the network and enabling the communication infrastructure to adapt well to asynchronous and distributed environments \cite{8645643}.
Currently, publish-subscribe mechanisms have been widely used in resource discovery \cite{saghian2019efficient,meneguette2019vehicular,khalil2018comparative} and service discovery \cite{ngaffo2020information,sim2020study}.
However, existing solutions are not suitable for publishing/subscribing based on intelligence changes in the Internet of intelligence. It is challenging to improve and extend the existing solutions to enable intelligence discovery in the Internet of intelligence.

\subsection{Protocol Designs}
Designing new protocols for the Internet of intelligence to meet its service requirements is another crucial challenge.
The Internet of information has gone through the design paradigm of layering, cross-layer and cross-system \cite{yu2019internet}. When designing the Internet of intelligence, should we follow similar procedures or consider cross-system first? In the era of the Internet of information, the Internet is realized based on TCP/IP, where IP is the core of the entire TCP/IP protocol suite and the foundation of the Internet \cite{townes2012spread}.
The successful ``thin waist'' hourglass architecture of the Internet is centered on the universal network layer (i.e., IP), which realizes the major functionality of the information networking. 
Such an architecture enables the independent development of technologies at all layers, successfully driving the dramatic growth of the Internet of Information. 
Referring to the protocol design of the Internet of information, we can also conceive a ``thin waist'' hourglass architecture for the Internet of intelligence, which requires further study in the future.

\subsection{Massive Data}

The Internet of intelligence involves a great quantity of model learning and training that relies on generous and high-quality data to realize stable and good performance, leading to some challenges in the data aspect \cite{8666641}. 
Generally, the data sets for the Internet of Intelligence are scarce. It is challenging to generate synthetic data sets that can train models. Besides, the availability of these data sets is subject to laws, environments, and data owners' consent. To address these issues, one possible approach is to use public data sets \cite{8444669}, which is a common solution in several popular AI-enabled applications, such as image recognition \cite{dataset}. There is also a strong need to launch campaigns to encourage the publication of data
sets in the Internet of intelligence.

Meanwhile, due to the large-scale and heterogeneous nature of the Internet of intelligence, the generated data has multiple dimensions \cite{8951180}. Some data analysis models are needed to extract useful information and features from the data. 
A popular solution is multimodal learning, which aims to build a model from multiple modalities to process and correlate the information of multiple modalities and narrow the heterogeneity gap \cite{di2018signals,9068414}.
In such learning frameworks, how to represent, transform, align, fuse, and collaboratively learn the data considering the morphological characteristics of heterogeneous data in the Internet of intelligence is still challenging for future research.
Moreover, the data generated in the Internet of intelligence can be used for various purposes according to different application scenarios. In some scenarios, such as smart healthcare, smart military, etc., data privacy is important, and these data must be anonymized before being used for training \cite{9194237,9107229}. Therefore, the context is also of great significance in the collection of data from different domains.

\subsection{Scalability}
As supporting technologies for the Internet of intelligence, blockchain and AI systems are designed for special applications. It is difficult to achieve interoperability among these systems. For example, since the blockchain was initially designed for cryptocurrency, many important problems in the current blockchain system make it unable to be adopted as a universal platform for different services worldwide.

Blockchain offers the advantages of decentralization, security, and fault tolerance, but it sacrifices scalability. Especially in the Internet of intelligence, blockchain is still beyond devices with limited computing, storage, and network capabilities. 
Recently, a lot of approaches have been exploited to enhance the scalability of the blockchain, from SegWit for the Bitcoin blockchain to the sharding technology proposed by Ethereum \cite{8580364}. 
In addition, moving the processing and storage load out of the blockchain \cite{network2018cheap,dales2000swarm}, limiting the consensus scope of different components of the blockchain, and exploring communication for connecting multiple blockchains \cite{kwon2018network} are also promising solutions.

For AI systems, specific algorithms are suitable for specific applications \cite{9041689}. Moreover, deep-structured ML algorithms have been shown to be more powerful than shallow-structured ML in smart networks \cite{zhang2021deep,6786503,8395149}. High-complexity ML requires high computing capability, yet ML algorithms based on large-scale matrix operations may be limited by CPU-based hardware. This also leads to the scalability problem in applying AI to the Internet of intelligence. In terms of expanding the scalability of AI systems, how to effectively expand the storage space to include the training data of AI systems is an arduous challenge. The adoption of graphics processing unit (GPU)-accelerated hardware, such as scalable GPU servers \cite{8444694} and GPU-embedded soft-defined routers (SDRs) \cite{7935536}, may be a viable solution. Besides, distributed AI learning requires large training data
sets, which need extensive storage in the Internet of intelligence. The high-speed storage area network (SAN) framework may be promising, which can offer access to relevant data sets to train AI systems while moving unnecessary data sets to backup archive storage \cite{7932863}.

Therefore, considering the heterogeneous structure and dynamic topology of the Internet of intelligence, in the future, researchers need to design effective solutions to address the above scalability issues to meet the requirements of the Internet of intelligence services.

\subsection{Security and Privacy}
Due to security and privacy issues, users may have concerns when sharing intelligence. Therefore, security and privacy are critical issues in the Internet of intelligence. These issues are more essential in the Internet of intelligence compared to existing networking paradigms, because an action is usually involved intelligence and incorrect actions may cause more damage than incorrect information.

A large amount of data and model training are involved in the Internet of intelligence. 
The data collected from various sensors and smart devices in the Internet of intelligence are trained to build a model to take actions for the applications involved, such as user selection, resource allocation, and behavior prediction.
Attackers can inject false data or counter-sample input, thereby making the learning of the Internet of intelligence invalid. 
They can also manipulate the collected data, distort the model and change the output \cite{papernot2018sok}. 
For example, an attacker can tamper with the learning environment to incorrectly perceive the input information. Besides, by changing hardware settings or reconfiguring system learning parameters, an attacker can manipulate the implementation hardware \cite{9200330}. In these cases, ensuring the high security of the Internet of intelligence system is of great importance.

Privacy is another issue of increasing concern for Internet of intelligence systems. The distributed intelligence and intelligence sharing of the Internet of intelligence may cause serious privacy leakage of participants, making them unwilling to share their intelligence with others. Besides, in the context of big data, ubiquitous users and organizations can easily access large data sets and computing resources (such as GPU), which brings severe privacy issues to the Internet of intelligence, such as data loss or parameter tampering \cite{9069945}. Ensuring a high degree of privacy protection without affecting training performance is critical to be considered in the Internet of intelligence.

In this paper, we mentioned that blockchain could be adopted to cope with these issues. 
Much work has already been done using blockchain to enhance security and privacy in the ML process.
In \cite{9420742}, Rathore \emph{et al.} develop an ML and blockchain-driven security framework for 5G-enabled smart IoT to provide intelligent data and secure operations.
Liu \emph{et al.} in \cite{9311629} study a blockchain-enabled secure data sharing framework in MEC systems and propose an adaptive privacy protection mechanism based on this framework to protect the identity privacy of users in data sharing.
Chen \emph{et al.} in \cite{9523794} propose a blockchain-based decentralized privacy-preserving deep learning model for vehicle-based self-organizing networks (VANETs) to guarantee the privacy and data security during network transmission and data analysis.
However, since blockchain was initially developed for cryptocurrency, some critical issues, including interoperability, scalability, and other performance indicators, need to be discussed to apply it to the Internet of intelligence.

\subsection{Lessons Learned: Summary and Insights}

In this section, we provide some research challenges and open issues in the Internet of intelligence to motivate researchers in related research fields. A review of the past three networking paradigms reveals that the modeling of the ``things'' in networking paradigms is critical to their success.
Therefore, the foremost problem that needs to be solved before the widespread implementation and application of the Internet of intelligence is how to model intelligence and design appropriate protocols to meet the demand for Internet of intelligence services. On this basis, how to design efficient incentives and intelligence discovery mechanisms to motivate intelligence holders to share their intelligence and help requesters to discover intelligence is an open issue. Besides, due to the adoption of enabling technologies such as AI and blockchain of the Internet of intelligence, challenges such as massive data, system scalability, security and privacy also need to be addressed. The authors summarize and briefly outline these research challenges and open issues with a view to enlightening researchers for related research areas of the Internet of intelligence. Actually, the Internet of intelligence involves the fusion of all knowledge and the intersection of multiple disciplines. Using multidisciplinary knowledge to conduct research alternately is conducive to promoting the construction of the Internet of intelligence.

\section{Conclusions\label{sect_8}}

The current information networking paradigm (the Internet) is facing challenges such as information explosion, fake information, ``human-in-the-loop'' and the design of trustworthy, cost effective autonomous systems. The Internet of intelligence is expected to effectively help address these challenges and have a significant impact on socio-economic systems and human daily life.

In this survey, we provided an overview of the Internet of intelligence with emphasis on motivations, architecture, enabling technologies, applications, and challenges. We first reviewed the evolution of networking paradigms and AI, based on which we discussed the motivations of the Internet of intelligence by demonstrating that networking needs intelligence and intelligence needs networking. We then proposed the designed layered architecture of the Internet of intelligence, and recent advances of the enabling technologies in each layer are discussed. In addition, we introduced primary application scenarios of the Internet of intelligence to explain how the integration of Internet of intelligence technologies will change traditional industries such as transportation, cities, healthcare, industrial, energy, and so on. Finally, we put forward some research challenges and open issues, including intelligence modeling, incentives for intelligence sharing, intelligent discovery, protocol designs, massive data, scalability, security and privacy, to fully exploit the potential of future Internet of intelligence.

Currently, the research on the Internet of intelligence in industry and academia is still in its infancy.
This survey expects to provide a systematic and comprehensive manual to enable researchers to have a more straightforward and deeper understanding of intelligence networking paradigm and to deliver insightful guidelines to fully exploit the profound benefits of the Internet of intelligence.

\bibliographystyle{IEEEtran}
\bibliography{Reference}

\begin{IEEEbiography}[{\includegraphics[width=1in,height=1.25in,clip,keepaspectratio]{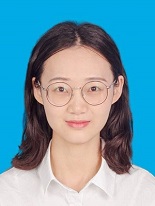}}]{Qinqin Tang} received the Ph.D. degree from Beijing University of Posts and Telecommunications, Beijing, China, in 2022. She is currently a Postdoctoral Research Fellow with Beijing University of Posts and Telecommunications. Her research interests include Internet of intelligence, edge/cloud computing, traffic offloading, and resource management.
\end{IEEEbiography}

\begin{IEEEbiography}[{\includegraphics[width=1in,height=1.25in,clip,keepaspectratio]{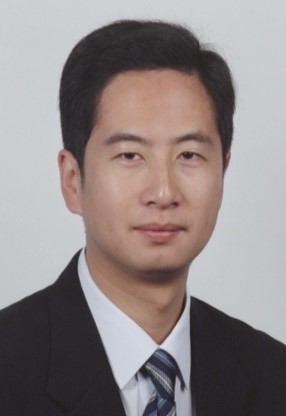}}]{F. Richard Yu} received the Ph.D. degree in electrical engineering from the University of British Columbia, Vancouver, BC,Canada, in 2003.	From 2002 to 2006, he was with Ericsson, Lund, Sweden, and a start-up in California, USA. He joined Carleton University, Ottawa, ON, Canada, in 2007, where he is currently a Professor. He is also with the Department of Future Networks, Purple Mountain Laboratories, Nanjing, China. His current research interests include blockchain, security, and green ICT. He serves on the Editorial Boards of several journals, and is a Co-Editor-in-Chief for Ad Hoc \& Sensor Wireless Networks, and is a Lead Series Editor for IEEE Transactions on Vehicular Technology and IEEE Communications Surveys \& Tutorials. He has served as a Technical Program Committee (TPC) Co-Chair of numerous conferences.
	
\end{IEEEbiography}

\begin{IEEEbiography}[{\includegraphics[width=1in,height=1.25in,clip,keepaspectratio]{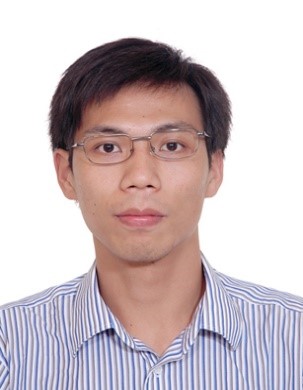}}]{Renchao Xie} is currently a Professor with Beijing University of Posts and Telecommunications. He is also with the Department of Future Networks, Purple Mountain Laboratories, Nanjing, China. He received the Ph. D. degree in the School of Information and Communication Engineering from Beijing University of Posts and Telecommunications in 2012. From July 2012 to September 2014, he worked as a Postdoctoral in China United Network Communications Group Co. From November 2010 to November 2011, he visited Carleton University, Ottawa, ON, Canada, as a Visiting Scholar. His current research interests include 5G network, edge computing, information centric networking, industrial Internet of things, and future network architecture. He has published more than 80 journal and conference papers, including 3 ESI highly cited papers. He has served as a Technical Program Committee (TPC) member of numerous conferences, including IEEE Globecom, IEEE ICC, EAI Chinacom, IEEE VTC-Spring and so on.
\end{IEEEbiography}

\begin{IEEEbiography}[{\includegraphics[width=1in,height=1.25in,clip,keepaspectratio]{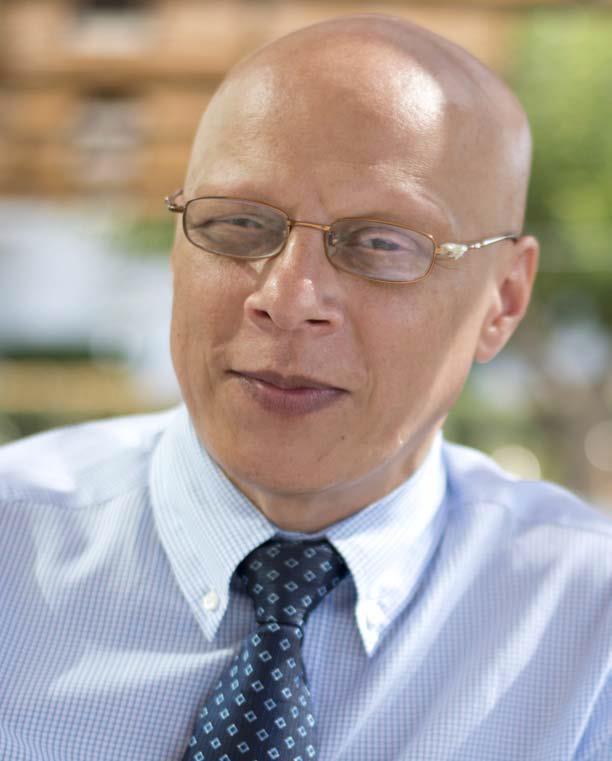}}]{Azzedine Boukerche} is currently the Distinguished University Professor and holds the Canada research chair tier-1 position with the University of Ottawa. He is also the Founding Director of the PARADISE Research Laboratory, the DIVA Strategic Research Center, and NSERC-CREATE TRANSIT, University of Ottawa. His current research interests include sustainable sensor networks, autonomous and connected vehicles, wireless networking and mobile computing, wireless multimedia, QoS service provisioning, performance evaluation and modeling of large-scale distributed and mobile systems, and large scale distributed and parallel discrete event simulation. He has published extensively in these areas and received several best research paper awards for his work. He is also a fellow of the Engineering Institute of Canada, the Canadian Academy of Engineering, and the American Association for the Advancement of Science. He has received the C. Gotlieb Computer Medal Award, the Ontario Distinguished Researcher Award, the Premier of Ontario Research Excellence Award, the G. S. Glinski Award for Excellence in Research, the IEEE Computer Society Golden Core Award, the IEEE CS-Meritorious Award, the IEEE TCPP Leaderships Award, the IEEE ComSoc ComSoft, the IEEE ComSoc ASHN Leaderships and Contribution Award, and the University of Ottawa Award for Excellence in Research. He also serves as the Editor-in-Chief for ACM ICPS and an associate editor for several IEEE transactions and ACM journals. He is also the steering committee chair for several IEEE and ACM international conferences.
\end{IEEEbiography}

\begin{IEEEbiography}[{\includegraphics[width=1in,height=1.25in,clip,keepaspectratio]{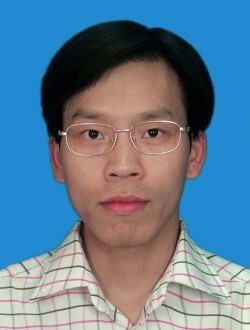}}]{Tao Huang}
	received his B.S. degree in communication engineering from Nankai University, Tianjin, China, in 2002, the M.S. and Ph.D. degree in communication and information system from Beijing University of Posts and Telecommunications, Beijing, China, in 2004 and 2007 respectively. He is currently a Professor with Beijing University of Posts and Telecommunications. He is also with the Department of Future Networks, Purple Mountain Laboratories, Nanjing, China. His current research interests include network architecture, network artificial intelligence, routing and forwarding, and network virtualization. 
\end{IEEEbiography}

\begin{IEEEbiography}[{\includegraphics[width=1in,height=1.25in,clip,keepaspectratio]{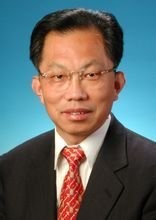}}]{Yunjie Liu}
	received the B.S. degree in technical physics from Peking University, Beijing, China, in 1968. He is currently an Academician with the China Academy of Engineering, Beijing; the Chief of the Science and Technology Committee of China Unicom, Beijing; and the Dean of the School of Information and Communication Engineering, Beijing University of Posts and Telecommunications, Beijing. He is also with the Department of Future Networks, Purple Mountain Laboratories, Nanjing, China. His current research interests include next generation network, and network architecture and management. 
\end{IEEEbiography}

\end{document}